\documentclass[a4paper,fleqn,usenatbib]{mnras}

\usepackage{newtxtext,newtxmath}

\usepackage[T1]{fontenc}
\usepackage{ae,aecompl}

\usepackage{graphicx}	
\usepackage{amsmath}	
\usepackage{amssymb}	
\usepackage{commath}    
\usepackage{enumitem}   
\usepackage{subcaption} 
\newcommand{\grad}{\gamma_{\mathrm{rad}}}
\newcommand{\eph}{\epsilon_{\mathrm{ph}}}

\newcommand{\zeltron}{\textsc{Zeltron}\,}
\newcommand{\acut}{3}
\DeclareMathOperator{\sech}{sech}
\newcommand{\tidx}{\zeta}

\newcommand{\nonthermal}[1][]{non-thermal#1 }
\newcommand{\lc}[1][]{light crossing#1 }
\newcommand{\ts}[1][]{time-scale#1 }
\newcommand{\quoted}[2][]{`#2'#1 }

\ifdefined \compilingRevision
\newcommand{\revtext}[1]{\textcolor{blue}{#1}}
\else
\newcommand{\revtext}[1]{}
\fi

\title[Kinetic Beaming and Gamma-Ray Flares]{Kinetic Beaming in Radiative Relativistic Magnetic Reconnection: A Mechanism for Rapid Gamma-Ray Flares in Jets}

\author[J. M. Mehlhaff et al.]{
J. M. Mehlhaff,$^{1}$\thanks{E-mail: john.mehlhaff@colorado.edu}
G. R. Werner,$^{1}$
D. A. Uzdensky,$^{1}$
M. C. Begelman$^{2, 3}$
\\
$^{1}$Center for Integrated Plasma Studies, Physics Department, 390 UCB, University of Colorado, Boulder, CO 80309, USA\\
$^{2}$JILA, University of Colorado and National Institute of Standards and Technology, 440 UCB, Boulder, CO 80309, USA\\
$^{3}$Department of Astrophysical and Planetary Sciences, 391 UCB, Boulder, CO 80309, USA
}

\date{Accepted XXX. Received YYY; in original form ZZZ}

\pubyear{2020}

\begin{document}
\label{firstpage}
\pagerange{\pageref{firstpage}--\pageref{lastpage}}
\maketitle

\begin{abstract}
  Rapid gamma-ray flares pose an astrophysical puzzle, requiring mechanisms both to accelerate energetic particles and to produce fast observed variability.
  These dual requirements may be satisfied by collisionless relativistic magnetic reconnection. On the one hand, relativistic reconnection can energize gamma-ray emitting electrons. On the other, as previous kinetic simulations have shown, the reconnection acceleration mechanism preferentially focuses high-energy particles -- and their emitted photons -- into beams, which may create rapid blips in flux as they cross a telescope's line of sight.
  Using a series of 2D pair-plasma particle-in-cell simulations, we explicitly demonstrate the critical role played by radiative (specifically inverse Compton) cooling in mediating the observable signatures of this \quoted{kinetic beaming}effect.
  Only in our efficiently cooled simulations do we measure kinetic beaming beyond one \lc[]time of the reconnection layer.
  We find a correlation between the cooling strength and the photon energy range across which persistent kinetic beaming occurs: stronger cooling coincides with a wider range of beamed photon energies.
  We also apply our results to rapid gamma-ray flares in flat-spectrum radio quasars, suggesting that a paradigm of radiatively efficient kinetic beaming constrains relevant emission models. In particular, beaming-produced variability may be more easily realized in two-zone (e.g. spine-sheath) set-ups, with Compton seed photons originating in the jet itself, rather than in one-zone external Compton scenarios.
\end{abstract}










\section{Introduction}

Powerful and rapid gamma-ray flares are observed in a variety of astrophysical objects, including pulsar wind nebulae \citep{agile11, fermi11, bsb12}, blazars \citep{magic07, hess07, magic11, fermi16a, fermi16b, magic19}, and radio galaxies \citep{veritas09, hess12, magic14}. In outbursts from all three types of systems, the gamma-ray flux can change faster than the \lc[]time of an important macroscopic scale: the size of the pulsar wind nebula itself, or, for radio galaxies and blazars, the radius of the central black hole's event horizon. Such extreme flares challenge theories of particle acceleration for two reasons. First, an acceleration mechanism is needed that is capable of energizing particles to emit in the GeV and TeV bands. Secondly, the short observed \ts[s]require a compact (causally limited) emitting region and an underlying physical process that yields rapid changes in observed flux.

One mechanism that may provide the requisite particle energization is relativistic magnetic reconnection \citep{bf94, lu03, l05}, a plasma process where an abrupt relaxation in the magnetic field topology can power relativistic single-particle and collective motion. Reconnection can be a promising alternative to shock acceleration, particularly in environments where an inferred high magnetization or relativistic shock front could suppress the latter \citep{ss09, ss11, spg15}. In contrast, it is precisely highly magnetized systems where reconnection may efficiently liberate stored magnetic energy, accelerating relativistic particles that subsequently produce the observed radiation.
Partly in view of these considerations, magnetic reconnection has been thought to power emission in many astrophysical objects and phenomena, both flaring and quiescent, including: pulsar magnetospheres \citep{l96, us14, cb14, cpp15, cps16, cmp16, ps18_philippov, c18, sgu19, hps19}, pulsar winds \citep{cp17}, pulsar wind nebulae \citep{ucb11, cub12, cwu12, cwu13, cwu14b, cwu14a, cl12, lks18}, giant magnetar flares \citep{l03, l06a, u11, sgu19}, gamma-ray bursts \citep{ds02_drenkahn, l06b, g06, gs07, u11, mu12}, black hole accretion disc coronae \citep{d98, dl05, dpk10, ug08, gu08, sdk15, kds15, b17, wpu19, sb20}, blazars \citep{gub09, g13, ngb11, nbc12, spg15, pgs16, wub18, zlg18, cps19, cps20, gu19, on20, hps20}, radio galaxies \citep{gub10}, and neutron star merger precursor emission \citep{ccd19, mp20}. Additionally, see the reviews by \cite{hl12} and \citet{u16}.

Beyond simply providing a means to access magnetic energy, collisionless relativistic reconnection has been shown to generate \nonthermal (power-law) distributions of energetic particles, which are required to produce the \nonthermal radiation spectra seen in many of the systems mentioned above. Evidence for reconnection-powered \nonthermal particle acceleration stems from a large number of particle-in-cell (PIC) simulation studies of both pair plasmas \citep{zh01, zh07, zh08, jtl04, bb07, bb12, ll08, cwu12, cwu13, cwu14b, cwu14a, gld14, gld15, gld19, ss14, sgp16, ps18, sb20, wuc16, wu17, wpu19, sgu19, on20, hps20} and electron-ion plasmas \citep{mwf14a, mwf14b, gll16, wub18, bso18}.

In this paper, we are specifically interested not in quiescent emission, but in abrupt high- and very high-energy flares, where the words \quoted{high}and \quoted{very high}are used in their technical senses to mean detected photons with energies in the ranges ${0.1-10 \, \rm GeV}$ and~${\gtrsim 0.1 \, \rm TeV}$, respectively \citep[e.g.][]{hess10, magic11, ms16}. When judging the feasibility of magnetic reconnection to power these extreme events, not only are the preceding energetic and spectral considerations important, but so too is the question of whether this plasma process can facilitate variability consistent with the dramatically short observed \ts[s.]

Aspects of this question have been addressed in connection with the Crab Nebula synchrotron flares. \citet{cwu12} showed that, near X-points (points in space where the magnetic field reconnects), particles are both accelerated and collimated, with the particles receiving more energy preferentially focused into tighter beams. This \quoted{kinetic beaming}effect naturally leads to rapid changes in the light an observer sees as beams of high-energy particles -- and their relativistically beamed synchrotron or inverse Compton (IC) emission -- sweep across the line of sight. Moreover, kinetic beaming predicts faster variability at higher energies, and is therefore observationally distinguishable from Doppler beaming \citep[e.g.][]{r66}, where relativistic bulk motion focuses emission achromatically. For the most energetic radiation, the light curve can change on \ts[s]as short as one-tenth the \lc time of the reconnection layer \citep{cwu13, cwu14b, cwu14a}.

Blazars, though very different from pulsar wind nebulae, also display evidence for kinetic beaming during rapid gamma-ray IC flares. For a~2010 outburst from PKS~1222+21 \citep{tst11, magic11}, kinetic beaming relaxes the energy density required to feed the emission region to a level accessible in the blazar's jet \citep{nbc12}. On the observational side, there have been at least two TeV events with approximately symmetric rise and decay times (consistent with a sweeping beam of light) and for which the TeV variability increases with photon energy \citep{magic07, hess10}. Yet another line of evidence, from numerical simulations, suggests that kinetic beaming at reconnecting current sheets may naturally occur within the turbulent environment of a magnetized blazar jet \citep{zuw20}.

Although compelling as an explanation for a diverse set of extreme flares, the kinetic beaming paradigm should perhaps be invoked with some caution. In particular, \citet{knp16} suggested that strong radiative cooling may be an important factor. They argued that weakly cooled particles will radiate most of their energy well after their momentum distribution -- initially collimated during X-point acceleration -- has been isotropized by the magnetic field external to the acceleration site. However, when strongly cooled, particles will dump their reconnection-acquired energy into energetic photons before dispersing. As one expects from this logic, the large simulations of \citet{ss14} and \citet{sgp16}, which, importantly, did not incorporate radiative losses, produced little to no anisotropy in the momentum distributions of high-energy particles confined to the largest magnetic islands. Additionally, \citet{ynz16} observed beaming to cease at late times in radiatively inefficient simulations. These studies do not cast doubt on kinetic beaming as an explanation for Crab Nebula flares -- for which this concept was originally proposed -- because in that case the gamma-ray emitting particles are subject to strong synchrotron losses. However, when considering other flaring systems, it seems quite likely that the observable signatures of kinetic beaming may be limited if the emitting particles do not radiate efficiently.

To summarize, kinetic beaming appears to be a generic by-product of the reconnection acceleration mechanism \citep{ucb11, cub12, cwu12, cwu13, cwu14b, cwu14a}, but it may not always be observationally relevant \citep{knp16, ss14, sgp16, ynz16}. The particles emitting, either via synchrotron or IC processes, at the energies of interest must do so efficiently. Otherwise, the observable signatures of kinetic beaming (i.e. rapid light curve variability) may become washed out by particle isotropization. This simple picture -- despite rapidly progressing numerical work on radiative relativistic reconnection \citep[e.g.][]{jh09, cwu13, cwu14b, cwu14a, ynz16, nyc18, wpu19, sgu19, hps19, sb20, on20} -- has yet to be made rigorous by a systematic study of radiatively cooled kinetic beaming. The need for such a study is underscored by its astrophysical implications: it would reveal whether and under what conditions a kinetic beaming scenario may viably explain fast gamma-ray flares observed from many types of sources.

Therefore, in this work, we perform the first systematic investigation into the impact of radiative cooling on observable kinetic beaming. By analysing a series of 2D PIC simulations of relativistic pair-plasma reconnection with varying IC cooling strength, we answer the following related questions:
\begin{enumerate}
\item \label{en:q1} For weak radiative cooling, does observable kinetic beaming disappear as the reconnection layer evolves? That is, does particle anisotropy vanish as reported by \citet{ss14}, \citet{sgp16}, and \citet{ynz16}, just as one would expect from \citet{knp16}?
\item \label{en:q2} If so, does strong cooling restore observable kinetic beaming?
\end{enumerate}
We note that our specialization to IC cooling is mostly for definiteness. We expect our results to carry over, in a qualitative sense, to systems where other emission mechanisms (most notably synchrotron) dominate the radiative output.

Our findings indicate affirmative answers to both questions~\ref{en:q1} and~\ref{en:q2}. In our simulations where radiative cooling is extremely weak or absent, no significant anisotropy is retained by the distribution of particles beyond a single \lc[]time of the reconnection layer. However, as cooling becomes more efficient, a persistent beaming effect emerges across an increasing range of particle (and photon) energies. When persistent, kinetic beaming occurs only among the highest energy particles -- those roughly within a decade of their radiatively imposed maximum possible energy.

This last result translates to a powerful constraint on rapid astrophysical gamma-ray flares, adding a radiative efficiency requirement to models attributing the observed variability to reconnection-driven kinetic beaming. As an illustration of the potential applicability of this result, after presenting our numerical findings, we specialize to the case of rapid flares in flat-spectrum radio quasars (FSRQs). Via simple analytical estimates, we show that invoking kinetic beaming constrains possible models for the PKS~1222+21 TeV flare \citep{magic11}. We find that the most appropriate radiative scenario may be inherently two-zone \citep[for example, spine-sheath,][]{gtc05, srb16}, where, as opposed to more traditional one-zone external Compton models, the photons seeding TeV IC emission come from inside the jet.

In Section~\ref{sec:sims}, we describe our simulations, including our self-consistent incorporation of IC cooling and how it limits particle acceleration. In Section~\ref{sec:beams}, we develop a quantitative language to describe beaming as manifested in distributions of particles and radiation. Section \ref{sec:kinbeam} applies this language to our simulations to answer questions \ref{en:q1} and~\ref{en:q2}. In Section~\ref{sec:fsrqflares}, we demonstrate the utility of our findings by analysing flaring FSRQs. We conclude in Section~\ref{sec:conclusions}.

\section{Simulations}
\label{sec:sims}
\subsection{Set-up}
We present relativistic pair-plasma simulations run using the radiative electromagnetic PIC code \zeltron \citep{cwu13}. The simulation domain is a 2D box of size~$L_x \times L_y$ with~$L_y = 2L_x = 2L$ and periodic boundary conditions enforced in both dimensions. Spatial dependence is limited to the~$(x,y)$ dimensions, but vectorial quantities, including velocities and field components, are fully three-dimensional.

We initialize the simulations with standard double Harris current sheets \citep{ks03} of half-thickness~$\delta$ carrying anti-aligned currents in the $\pm z$-directions and centred on the planes~$y_1=L_y / 4$ and~$y_2=3L_y / 4$. This double-sheet configuration is chosen because it is consistent with the periodic box boundaries. Namely, the currents establish an in-plane magnetic field~$B_x(y) = \pm B_0 \tanh \left[ (y - y_{1,2}) / \delta \right]$ that reverses twice -- once at each sheet -- and is therefore periodic in~$y$. Additionally, we add a uniform initial guide field~$\pmb{B}_{\rm g} = (B_0 / 4) \hat{\pmb{z}}$. This serves chiefly as a numerical device to support magnetic islands against radiative cooling-induced contraction to the point where the Debye length becomes unresolved. The value~$B_{\rm g} = B_0 / 4$ does not substantially alter \nonthermal particle acceleration \citep{wu17, wpu19}.

The current in the Harris layers is carried by a drifting plasma component. In each layer, drifting electrons and positrons begin the simulation counterstreaming with bulk velocities~$\pm c \beta_{\rm d} \hat{\pmb{z}}$ and a combined drifting lab-frame number density~$n_{\rm d}(y) = n_{\mathrm{d},0} \sech^2 \left[ (y - y_{1,2}) / \delta \right]$. The resultant current profile~$J_z(y) = \pm c \beta_{\rm d} e n_{\rm d}(y)$ is precisely that necessary to generate the field~$B_x(y)$ as dictated by Amp\`ere's Law. In addition to the current-governing mean velocity~$\beta_{\rm d}$, we initialize the counterstreaming species with relativistic comoving temperature~$\theta_{\rm d} = T_{\rm d} / m_{\rm e} c^2 = 1050 \gg 1$ to support the current layers against the upstream magnetic pressure.

With the Harris equilibrium satisfied by the drifting particles, each simulation also contains an initially stationary uniform background plasma of combined density~$n_{\rm b}$ and relativistically hot temperature~$\theta_{\rm b} = T_{\rm b} / m_{\rm e} c^2 = 25 \gg 1$ that provides the inflow material for the reconnection layer. Two important dimensionless quantities associated with the background plasma are the cold and hot upstream magnetizations,~$\sigma \equiv B_0^2 / 4 \pi n_{\rm b} m_{\rm e} c^2$ and~$\sigma_{\rm h} \equiv B_0^2 / 4 \pi w$, respectively. Here, the enthalpy density~$w$ is given in the~$\theta_{\rm b} \gg 1$ limit by~$w = 4 n_{\rm b} T_{\rm b} = 4 \theta_{\rm b} n_{\rm b} m_{\rm e} c^2$, and, as a result, the hot magnetization becomes one-half inverse plasma-beta:~$\sigma_{\rm h} = B_0^2/16 \pi \theta_{\rm b} n_{\rm b} m_{\rm e} c^2 = 1 / (2 \beta_{\rm plasma})$. The two $\sigma$-parameters have the following physical interpretations. The cold magnetization characterizes the available magnetic energy per upstream particle; for~$\sigma \gg 1$, individual particles may acquire energy far in excess of their rest mass, and we set~$\sigma = 10^4$. The hot magnetization, on the other hand, decides whether the energy inflow to the reconnection layer is dominated by the magnetic field or by the particles. Taking~$\sigma_{\rm h} = 10^2$, we operate in the magnetically dominant~$\sigma_{\rm h} \gg 1$ limit, which also implies a relativistic Alfv\'en speed and places us in the relativistic regime of reconnection.

To minimize system-size effects, the simulation box must considerably exceed the largest kinetic scale in the problem: the typical Larmor radius of energetic particles \citep[see][]{wuc16}. Because the magnetic energy per upstream particle is~$\sigma m_{\rm e} c^2 / 2$, an average particle accelerated through the reconnection layer emerges with Larmor radius of order~$\sigma \rho_0$ where~$\rho_0 = m_{\rm e} c^2 / e B_0$ is a nominal Larmor radius. We conduct simulations in the large system regime~$L_x \gg 40 \sigma \rho_0$ identified by \citet{wuc16} and set~$L_x / \sigma \rho_0 = 320$. To confirm the insensitivity of our results to our simulation box size, we also run a series of simulations at different~$L_x$.

At the small-scale end, our cell size is~$\Delta x = \Delta y = \sigma \rho_0 / 24$, which is just smaller than the Debye length:~$\lambda_{\rm D} = 1.2 \Delta x$. With the cell size set, we employ a corresponding time-step satisfying the Courant--Friedrichs--Lewy condition~$c \Delta t = 0.7 \Delta x < \Delta x / \sqrt{2}$. Our initial number of simulation particles per grid cell is~$80$.

We incorporate radiative cooling into our simulations via inverse Compton scattering of a background (\quoted[)]{seed}radiation field that is static, homogeneous, and isotropic. The photons comprising this field are not tracked simulation entities, but, in the Thomson limit, give rise to a continuous radiative drag force~$\pmb{f}_{\mathrm{IC}}$ that enters self-consistently into the PIC particle push \citep{tpd10}. For a particle of 4-velocity~$(c \gamma, c \gamma \pmb{\beta})$, the expression for the drag force is~$\pmb{f}_{\mathrm{IC}} = - (4/3) \sigma_{\rm T} U_{\mathrm{ph}} \gamma^2 \pmb{\beta}$ where~${\sigma_{\rm T} = 8 \pi e^4 / 3 m_{\rm e}^2 c^4}$ is the Thomson cross section \citep[cf.][]{bg70, rl79, pss83, u16, wpu19, sb20}. From this, one sees that particle cooling depends only on the total energy density of background photons~$U_{\mathrm{ph}} \equiv \int d \epsilon \, u(\epsilon)$, not on their spectral distribution. We therefore adopt, without loss of generality, a simple monochromatic spectral energy density~$u(\epsilon) = U_{\mathrm{ph}} \delta (\epsilon - \eph)$. The Thomson limit is satisfied if the photon energies encountered by a particle of Lorentz factor~$\gamma$ in its rest frame are small compared with its mass:
\begin{align}
  \label{eq:kndef}
  \frac{\gamma \epsilon_{\rm ph}}{m_{\rm e} c^2} \ll 1 \, .
\end{align}
This condition allows the recoil in any single scattering event to be neglected and justifies our continuous treatment of radiative losses \citep{bg70}. When equation~(\ref{eq:kndef}) is not satisfied, IC cooling transitions to the discrete Klein--Nishina regime, with particles delivering an order-unity fraction of their energies to single photons.

A convenient dimensionless parameter that quantifies the IC cooling strength is~$\grad$, the Lorentz factor of a particle whose acceleration force due to the reconnection electric field matches its radiation-reaction force \citep[cf.][]{u16, wpu19, sb20}
\begin{equation}
  \label{eq:gradintuit}
  e E_{\mathrm{rec}} = \frac{4}{3} \sigma_{\rm T} U_{\mathrm{ph}} \grad^2 \, .
\end{equation}
(Here we have assumed~$\beta \simeq 1$.)
If one takes the reconnection electric field as~$E_{\rm rec} = \beta_{\mathrm{rec}} B_0$ with a standard value of the relativistic reconnection rate~$\beta_{\mathrm{rec}} = 0.1$, then~$\grad$ can be defined \citep[again, see][]{u16, wpu19} as
\begin{equation}
  \label{eq:graddef}
  \grad \equiv \sqrt{3(0.1)eB_0/(4 \sigma_{\rm T} U_{\mathrm{ph}})} \, .
\end{equation}
By construction,~$\grad$ is an upper bound on the particle energy distribution, because particles of higher energy would radiate more power than is delivered them via the reconnection electric field. This suggests the quantity~$\grad / \sigma$ as a quantitative measure of the IC cooling strength. For~$\grad / \sigma \gg 1$, the particle energy distribution may develop a hard power-law tail as in the non-radiative regime \citep[e.g.][]{ss14, gld14, wuc16, ps18, wpu19}, with \nonthermal particles having typical Lorentz factors of order~$\sigma$ and subject to relatively weak cooling. However, in the case~$\grad / \sigma \lesssim 1$, particles with energy~$\sigma m_{\rm e} c^2$ are strongly cooled and near their upper limit, controlled now by radiative cooling rather than by the available magnetic energy.

In this work, our main results come from a series of simulations scanning across~$\grad / \sigma \in [1/2, 1, 2, 4, 6, 8, 16, 32, 64, \infty]$ at the fiducial box size~$L_x / \sigma \rho_0 = 320$. Note that \textit{higher}~$\grad / \sigma$ corresponds to \textit{weaker} IC cooling. In particular, the case~$\grad / \sigma = \infty$ implies~$U_{\mathrm{ph}} = 0$: no cooling. In addition, we run a few simulations with a uniform radiative efficiency~$\grad / \sigma = 4$ but differing~$L_x / \sigma \rho_0 \in [80, 120, 160, 240, 320]$ as a first step towards characterizing the system-size dependence of our findings. Table~\ref{table:params} summarizes these values and those of the other parameters discussed so far.
\begin{table*}
  \centering
\begin{tabular}{|| l l l l l ||}
  \hline
  Parameter & Symbol (=definition) & Value \\
  \hline \hline
  Upstream magnetic field & $B_0$ & & \\
  Nominal gyroradius & $\rho_0 = m_{\rm e} c^2 / e B_0$ & & \\ 
  Radiation-limited Lorentz factor & $\gamma_{\mathrm{rad}}$ & $(1/2, 1, 2, 4, 6, 8, 16, 32, 64, \infty) \times \sigma$ \\
  System size & $L_x=L$ & $(80, 120, 160, 240, 320) \times \sigma \rho_0$ \\
  \quoted{Cold}magnetization & $\sigma = B_0^2 / 4 \pi n_{\rm b} m_{\rm e} c^2$ & $10^4$ \\
  \quoted{Hot}magnetization & $\sigma_{\rm h} = B_0^2 / 16 \pi n_{\rm b} T_{\rm b}$ & $10^2$ \\
  Background temp. & $\theta_{\rm b} = T_{\rm b} / m_{\rm e} c^2$ & $25$ \\
  Guide field & $B_z$ & $B_0 / 4$ \\
  Peak drift over background density & $\eta = n_{\rm d} / n_{\rm b}$ & $5$ \\
  Harris layer drift velocity & $\beta_{\rm d} c$ & $0.3 c$ \\
  Harris layer (comoving) temp. & $\theta_{\rm d} = T_{\rm d} / m c^2$ & $1050$ \\
  Harris layer half-thickness & $\delta = \sigma \rho_0 / \beta_{\rm d} \eta$ & $0.67 \sigma \rho_0$ \\
  Cell size & $\Delta x , \, \Delta y$ & $\sigma \rho_0 / 24$ \\
  Time step & $\Delta t$ & $0.7 \Delta x / c$ \\
  Macroparticles per cell & & $80$ \\  
  \hline
\end{tabular}
\caption{Simulation parameters used in this study. Note that we do not scan across all combinations of~$L_x$ and~$\grad$. Instead, we conduct a series of simulations exploring all~$\grad$ values in the table at a fixed system size~$L_x = 320 \sigma \rho_0$ and a second series across all~$L_x$ values at a fixed radiation-reaction strength~$\grad = 4 \sigma$.
}
  \label{table:params}
\end{table*}

Our wide-ranging scan in~$\grad$ is limited on the strong-cooling end by radiative losses in the upstream region, which can cause the background plasma feeding the reconnection layer to change in time. In order to avoid this effect, one should require that the upstream IC cooling time~$t_{\rm cool}$ exceed the duration of our simulations~$3L_{\rm x} / c$. Using the average Lorentz factor~$\langle \gamma \rangle = 3 \theta_{\rm b}$ of the background plasma, the ratio~$t_{\rm cool} / (3 L_x / c)$ can be written as
\begin{align}
    \frac{t_{\rm cool}}{3 L_x / c} &= \frac{\langle \gamma \rangle m_e c^2 / P_{\rm IC}(\langle \gamma \rangle)}{3 L_x / c} = \frac{10}{3} \frac{\sigma \rho_0}{L_x} \frac{\grad}{\langle \gamma \rangle} \frac{\grad}{\sigma} \notag \\
    &= \frac{40 \sigma_{\rm h}}{9} \frac{\sigma \rho_0}{L_x} \left( \frac{\grad}{\sigma} \right)^2 \notag \\
    &\simeq 1.4 \left( \frac{\sigma_{\rm h}}{100} \right) \left( \frac{L_x / \sigma \rho_0}{320} \right)^{-1} \left( \frac{\grad}{\sigma} \right)^{2} \, .
    \label{eq:upstreamcool}
\end{align}
Here, we used~$P_{\rm IC}(\gamma) = |c\pmb{\beta} \cdot \pmb{f}_{\rm IC}| = (4/3) c \sigma_{\rm T} U_{\rm ph} \gamma^2 \beta^2$ and the relativistic limits~$\beta = 1$ and~$1 \ll \theta_{\rm b} = \sigma / 4 \sigma_{\rm h}$.
Let us go one step farther, employing equation~(\ref{eq:upstreamcool}) along with~$\dif \gamma / \dif t = -P_{\rm IC}(\gamma) / m_e c^2$ to estimate the amount by which the upstream plasma cools during a simulation. The temperature~$\theta_{\rm b,f}$ reached at~$t = 3 L_{\rm x} / c$ may be as low as
\begin{align}
    \frac{\theta_{\rm b,f}}{\theta_{\rm b}} \simeq \frac{1}{1 + (3 L_{\rm x} / c) / t_{\rm cool}} \, .
    \label{eq:upstreamcoolfrac}
\end{align}

Evidently, our simulation with~$\grad = \sigma / 2$ is problematic, with the upstream plasma cooling in time~$t_{\rm cool} \sim 1 L_x / c$ and potentially falling to~$\simeq 26$ per cent of its initial temperature by the end of the run. This simulation also exhibits the worst energy conservation, with the energy error peaking at about~$3.6$ per cent (all our other simulations have per cent level or better error). Thus, the results of this most strongly radiative run should not be taken as definitive on their own. Fortunately, the conclusions we draw from our~$\grad$ scan (Section~\ref{sec:kinbeamgrad}, Figs~\ref{fig:epscut},~\ref{fig:epsiso}, and~\ref{fig:epsCutDivEpsIso}), do not depend on whether we include or exclude this simulation from our quantitative analysis. We have therefore chosen to include it as a tentative endpoint on the data generated by the rest of our~$\grad$-varying~($L_x$-constant) simulation series. We also note that our benchmark radiative case~$\grad = \sigma$ is not completely free of the upstream cooling issue. However, according to equation~(\ref{eq:upstreamcoolfrac}), the background temperature may decrease by less than a factor of~$2$ by the end of that simulation. Because such a discrepancy is within the error bars on our main kinetic beaming quantities measured in Section~$\ref{sec:kinbeamgrad}$, we view our~$\grad = \sigma$ simulation as marginally acceptable.

Finally, we would like to point out that upstream radiative losses only increase our effective~$\sigma_{\rm h}$. Because~$\sigma_{\rm h}$ is quite large to begin with, raising it by order-unity factors preserves (indeed, enhances) the asymptotically large-$\sigma_{\rm h}$ limit. Hence, we do not expect the acceleration and beaming of high-energy particles to be significantly impacted, even in our most strongly cooled~($\grad/\sigma = 1/2,1$) simulations.

\subsection{Evolution of the reconnection layer}
\begin{figure*}
  \centering
  \includegraphics[height=0.95\textheight]{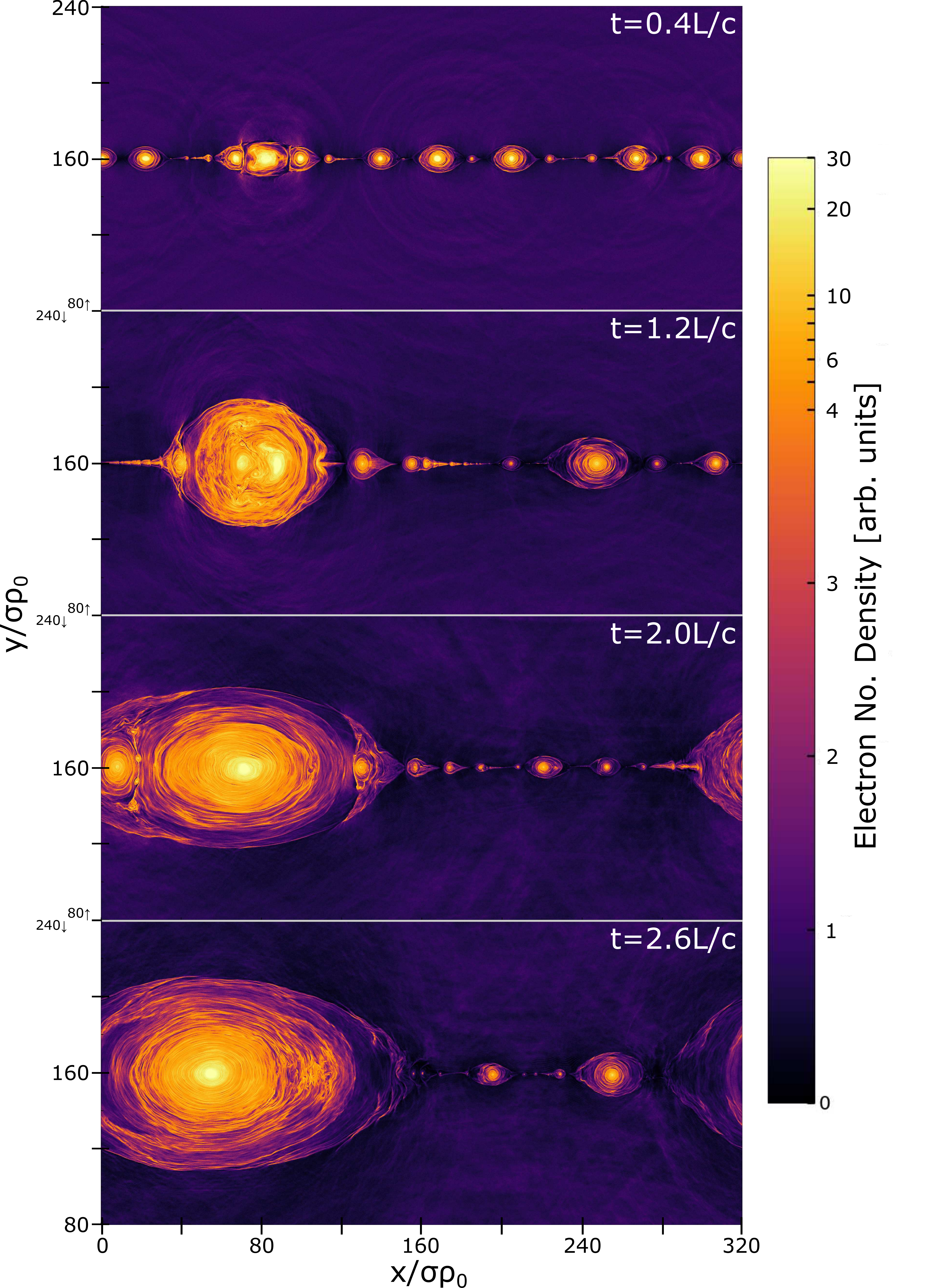}
  \caption{\revtext{(Figure updated/changed from original manuscript.)} Total (drift plus background) electron number density pictured in the lower reconnection layer at key moments during our~$\grad = \sigma$ simulation. The full simulation width~($L = L_x = 320 \sigma \rho_0$) and a restricted height range~($L_y / 4 \pm 80 \sigma \rho_0$) are displayed. See text for a description of the various phases of the time evolution.}
  \label{fig:reconnLayerInTime}
\end{figure*}

Having described the set-up of our simulations, let us now move on to how they evolve in time. In every run, we trigger magnetic reconnection with a small (1 per cent) perturbation to the initial magnetic field. The current sheet then tears into a number of magnetic islands or \quoted{plasmoids}which begin to merge with one another (Fig.~\ref{fig:reconnLayerInTime},~$t=0.4L_x/c=0.4L/c$). Initially, the plasmoids are all about the same size, but eventually -- in Fig.~\ref{fig:reconnLayerInTime} at about~$t=1.2L/c$ -- a single largest plasmoid begins to dominate the reconnection layer. This primary plasmoid proceeds to consume the others that have also accumulated to considerable but smaller sizes, culminating in a spectacular merger between the largest and next-to-largest plasmoids (Fig.~\ref{fig:reconnLayerInTime},~$t=2.0L/c$). After this most dramatic merger, additional small plasmoids are continually born from the main X-point and venture across the box to be consumed by the large primary plasmoid (Fig.~\ref{fig:reconnLayerInTime},~$t=2.6L/c$).

\begin{figure*}
  \centering
  \begin{subfigure}{0.49\textwidth}
    \centering
    \includegraphics[width=\linewidth]{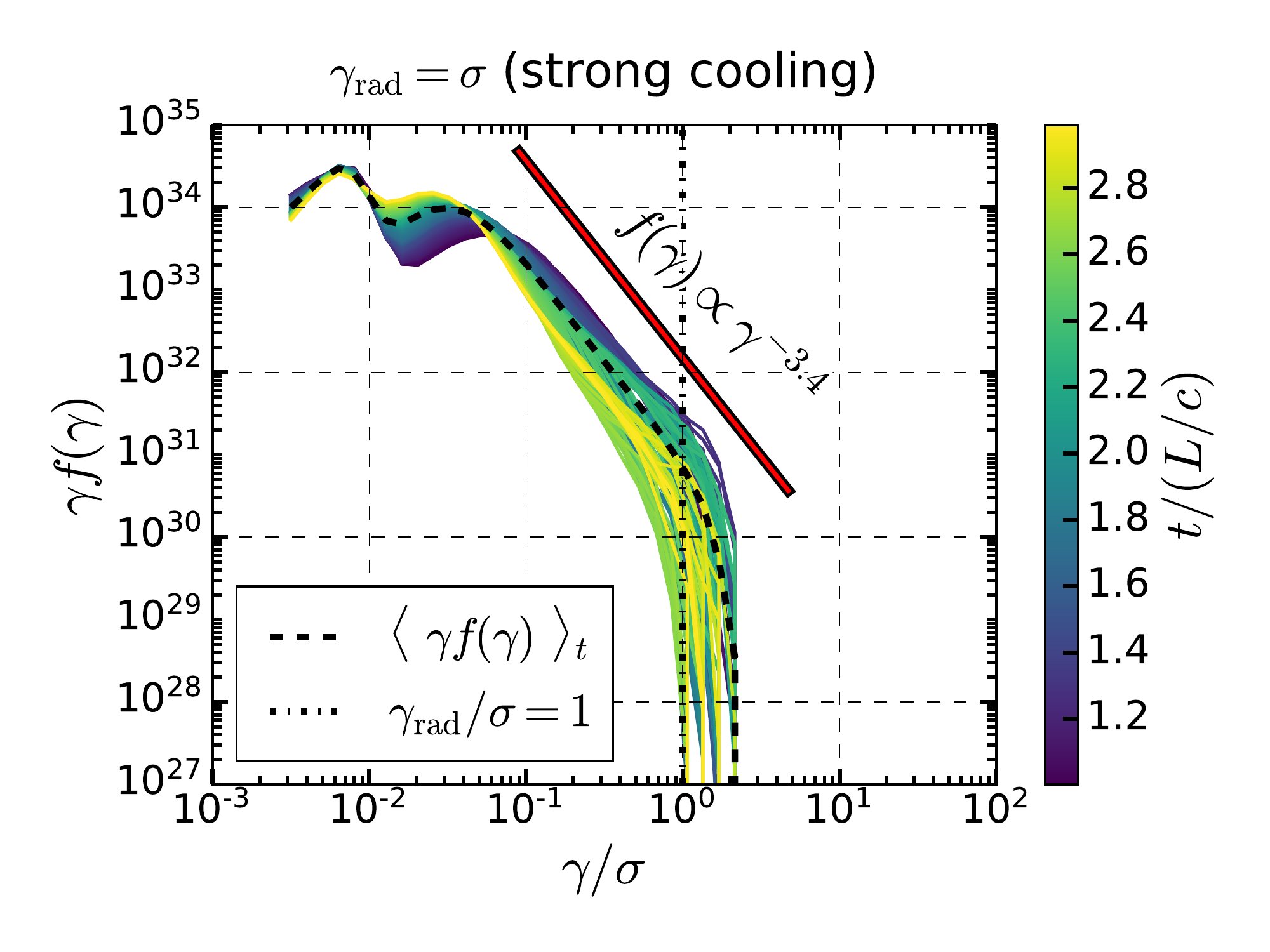}
  \end{subfigure}
  \begin{subfigure}{0.49\textwidth}
    \centering
    \includegraphics[width=\linewidth]{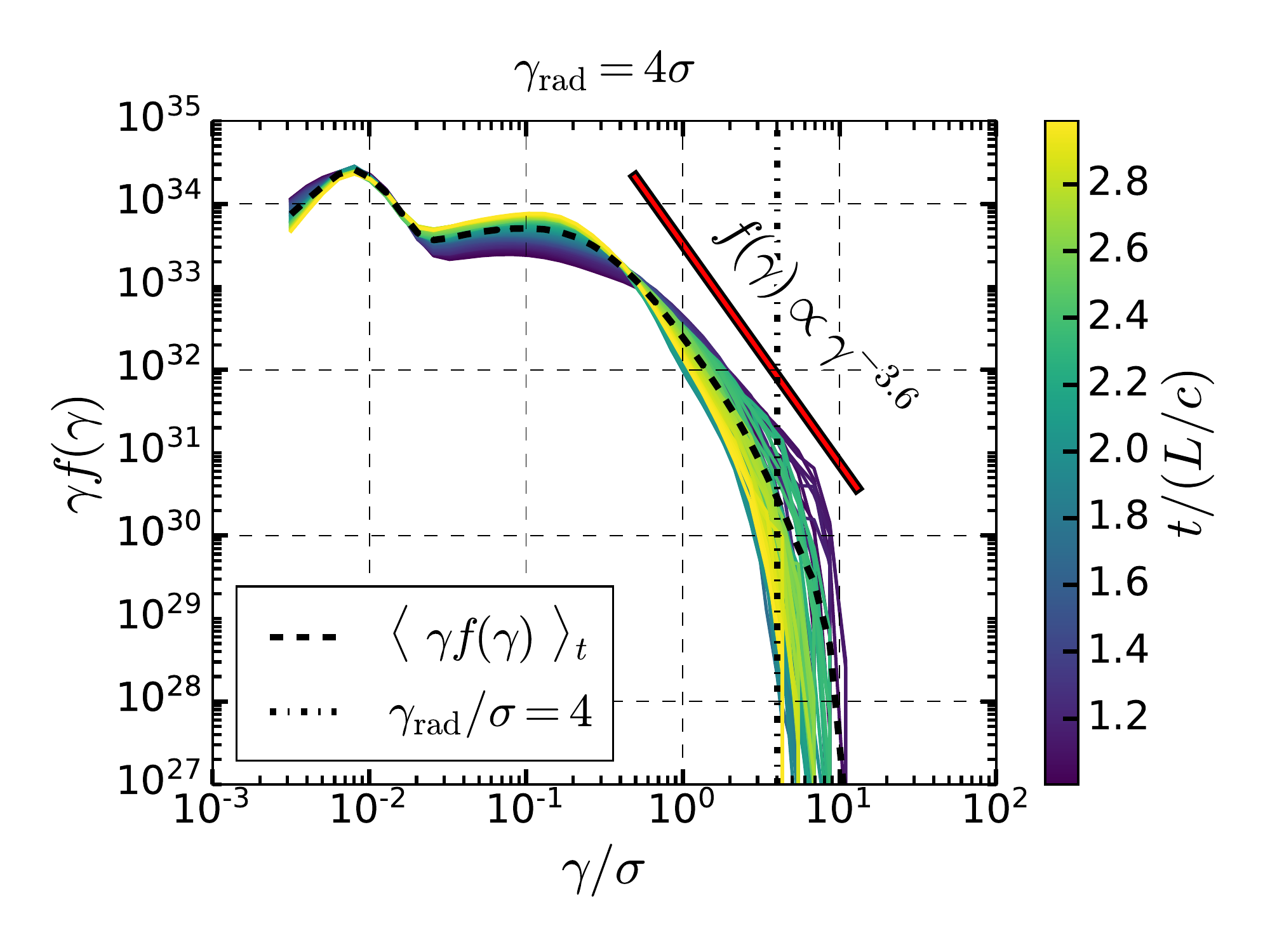}
  \end{subfigure} \\
  \begin{subfigure}{0.49\textwidth}
    \centering
    \includegraphics[width=\linewidth]{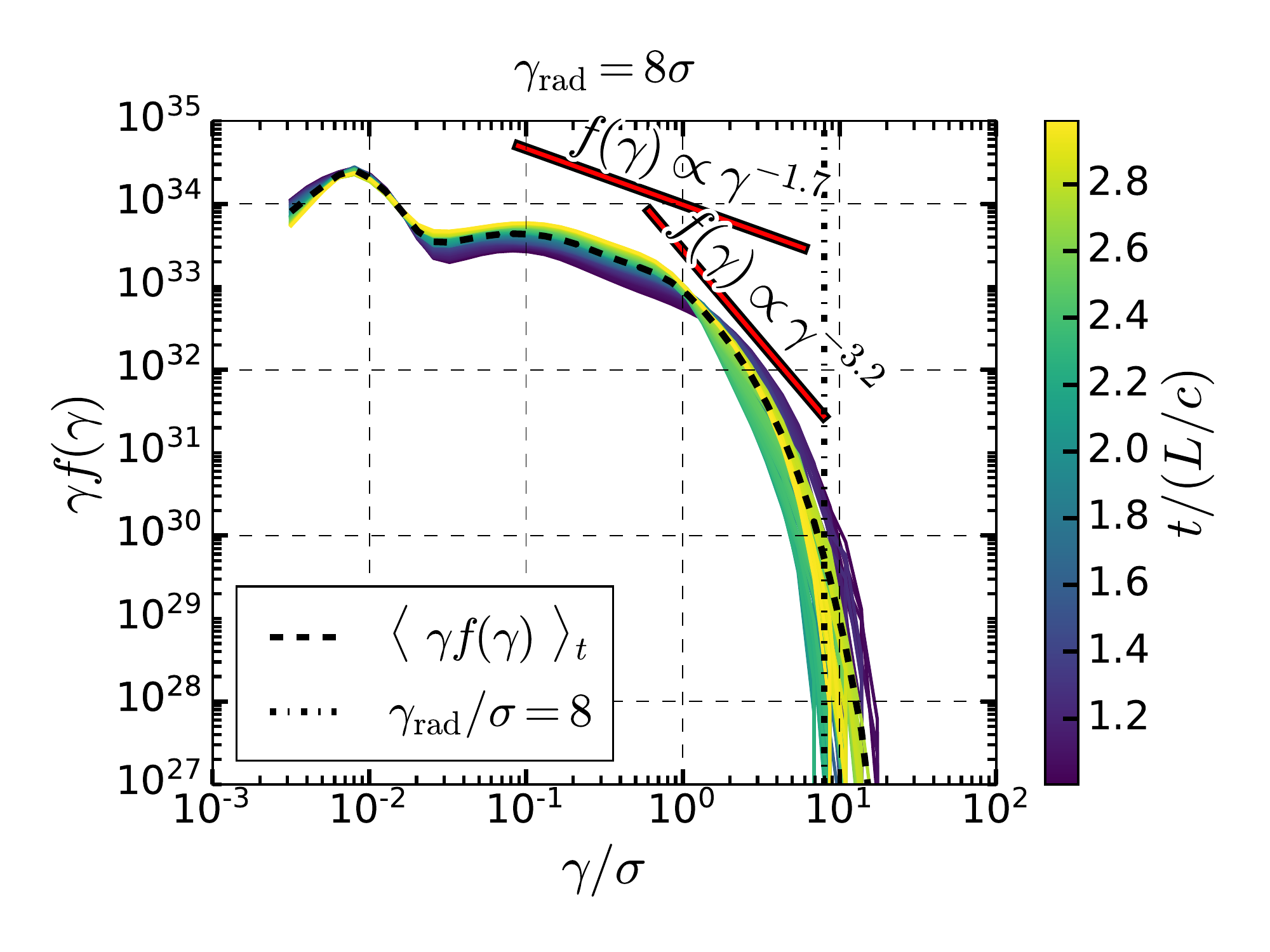}
  \end{subfigure}
  \begin{subfigure}{0.49\textwidth}
    \centering
    \includegraphics[width=\linewidth]{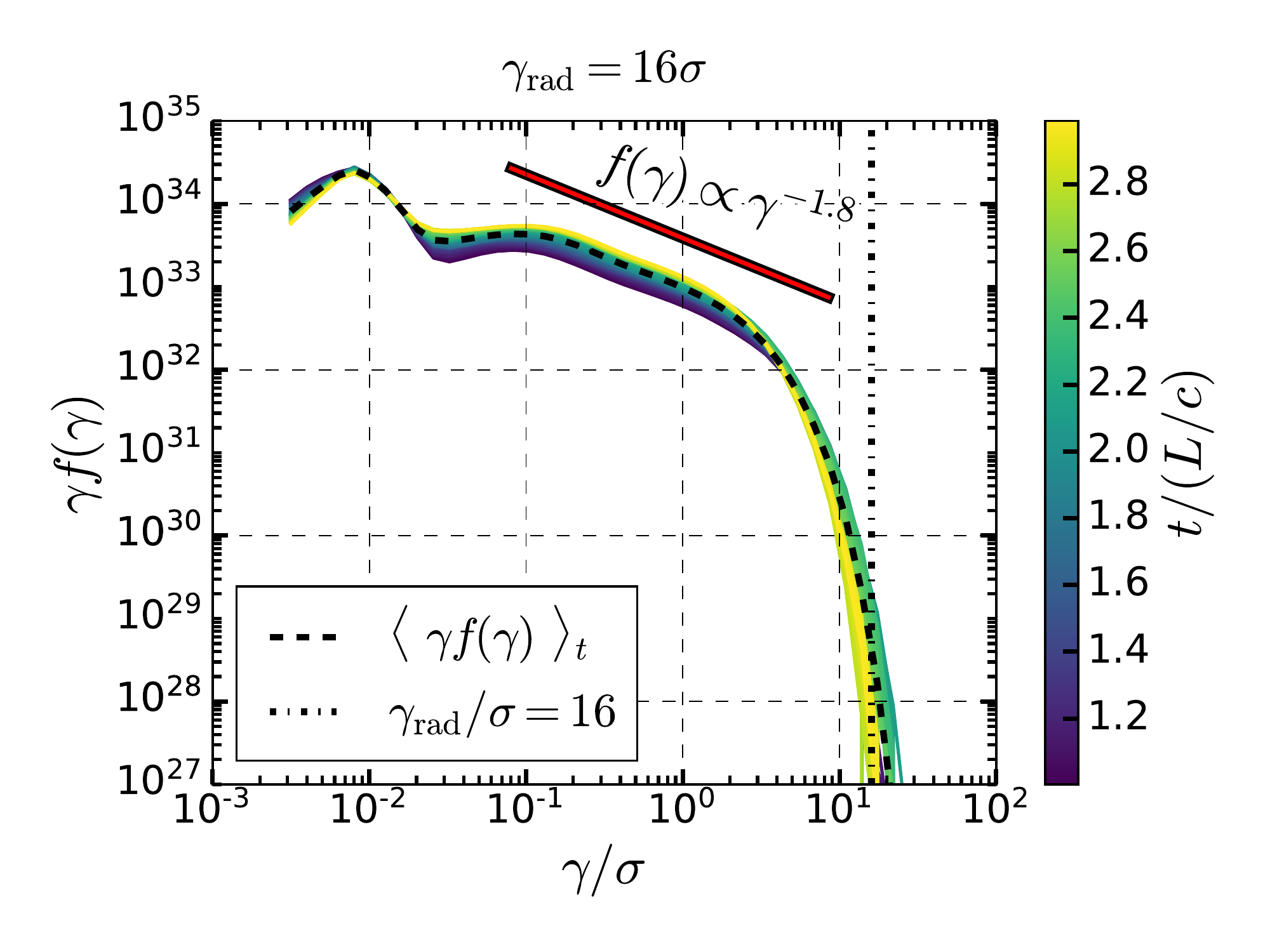}
  \end{subfigure}
  \caption{\revtext{(Figure updated/changed from original manuscript.)} Lower layer electron energy distribution as a function of time between~$L/c$ and~$3L/c$ for a subset of the simulations presented in this work. Coloured solid lines denote the distribution at different times. Black dashed lines represent the time-averaged (over~$1 \leq ct/L \leq 3$) distribution. Vertical black dot--dashed lines label~$\grad$. Red bars indicate power-law segments in the time-averaged particle distributions. The qualitative features are consistent with those reported in \citet{wpu19}. Namely, the simulations with strongest cooling~($\grad = 1, 4 \sigma$) exhibit only a steep power law~$f(\gamma) \propto \gamma^{-p}$ with variable index~$p \gtrsim 3$; the simulation with intermediate cooling~($\grad = 8 \sigma$) exhibits both soft variable~($p \gtrsim 3$) and hard steady~($p \lesssim 2$) power-law segments; and the weakly cooled simulation~($\grad = 16 \sigma$) only contains a hard steady~($p \lesssim 2$) power law.}
  \label{fig:tdpdists}
\end{figure*}

We observe significant \nonthermal particle acceleration during our simulations. At late times, this energization is bursty: merging plasmoids sporadically punctuate ongoing reconnection from the main X-point with short intense episodes of particle acceleration. In our radiatively efficient runs, following these episodes, the high-energy particles rapidly cool, leading to a steepening of their \nonthermal energy distribution. This effect was analysed by \citet{wpu19} \citep[see also][]{sb20}, and we illustrate it in Fig.~\ref{fig:tdpdists}, which presents time-dependent particle energy distributions from the lower reconnection layer in a subset of our simulations.
In the limit of weak cooling (e.g. $\grad / \sigma = 16$ in Fig.~\ref{fig:tdpdists}), the distribution develops a shallow power law. However, due to long periods of continuous IC losses interrupted by bursts of plasmoid merger-initiated magnetic reconnection, the particle distributions for the simulations with stronger cooling (e.g. those with $\grad / \sigma \leq 4$) all exhibit steeper, more variable power laws at late times. Additionally, for all the displayed simulations, the cut-off particle energy is well approximated by~$\grad$, indicating that radiative losses control this limit (even when too weak to steepen the \nonthermal power-law tail).

\subsection{A view in angular space}
\label{sec:angmaps}
Up until now, we have described the evolution of our magnetic reconnection simulations from spatial and energetic viewpoints. We displayed several snapshots of the electron number density $n(x, y)$ in Fig.~\ref{fig:reconnLayerInTime}. Then, in Fig.~\ref{fig:tdpdists}, we described the electron energy distribution~$f(\gamma)$ and how its evolution is impacted by radiative cooling. These pictures represent different ways of viewing the master distribution function in phase space~$f(x, y, \gamma, \pmb{\Omega}; t)$. At a given time~$t$, this master distribution is five-dimensional, containing two spatial and three velocity dimensions, the latter of which we decompose into a Lorentz factor~$\gamma = E / m c^2$ and a direction labelled by the solid angle~$\pmb{\Omega}$. In terms of the master distribution, the number density and energy distribution are~$n(x, y; t) = \int \dif \gamma \dif \Omega\, f(x, y, \gamma, \pmb{\Omega}; t)$ and~$f(\gamma; t) = \int \dif x \dif y \dif \Omega\, f(x, y, \gamma, \pmb{\Omega}; t)$, respectively.

\begin{figure}
  \centering
  \includegraphics[width=\linewidth]{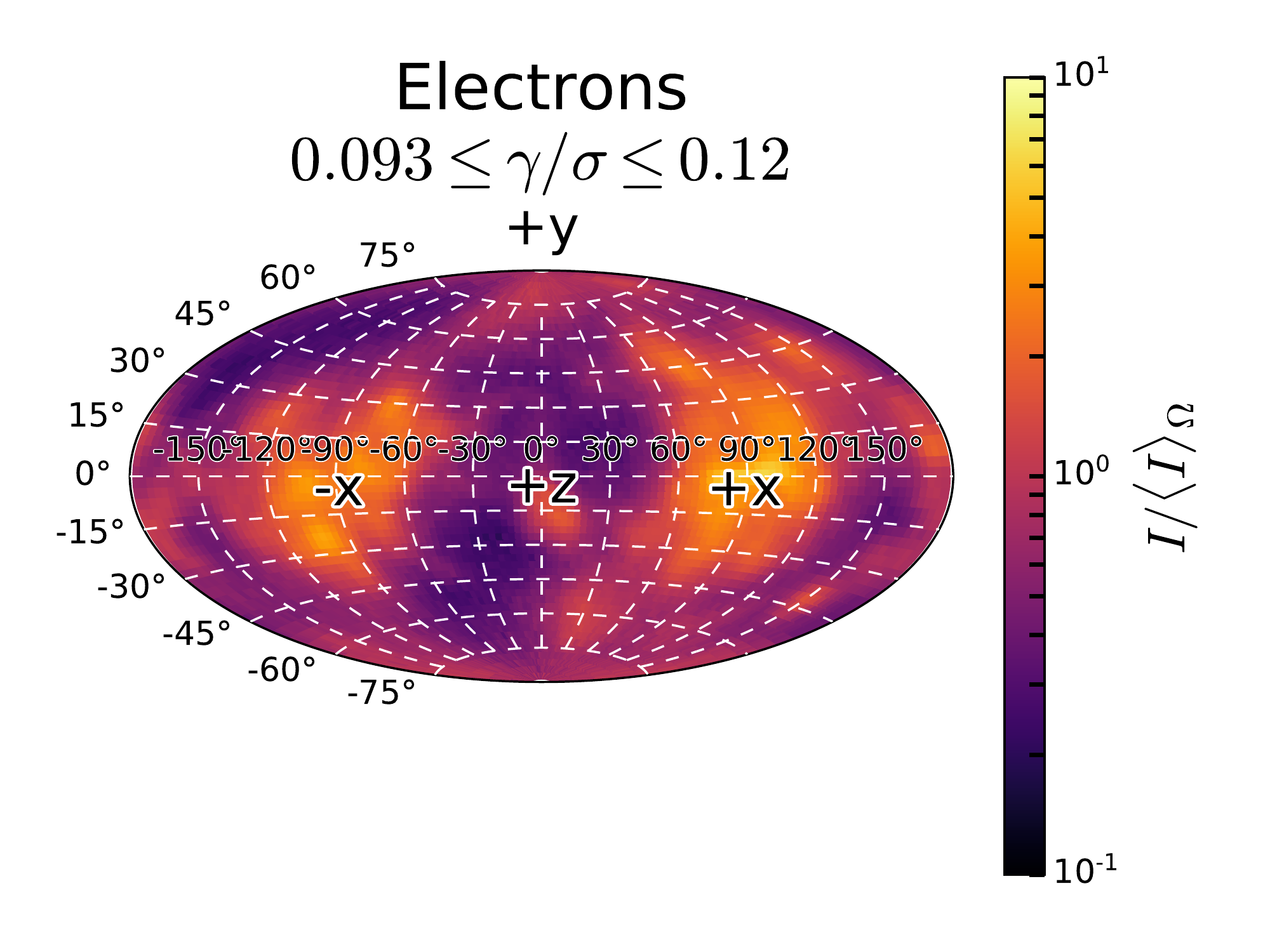}
  \caption{\revtext{(Figure updated/changed from original manuscript.)} An angular map for the~$\grad = \sigma$ simulation displaying the angular intensity~$I = \dif N_{t}/ \dif \gamma \dif \Omega$ of lower layer electrons at~$2.0L/c$. This is a low-energy map -- in the sense that the electron Lorentz factors~$\gamma$ are a decade below~$\grad$ -- and exhibits only mild beaming.}
  \label{fig:hm1nc}
\end{figure}

\begin{figure}
  \centering
  \includegraphics[width=\linewidth]{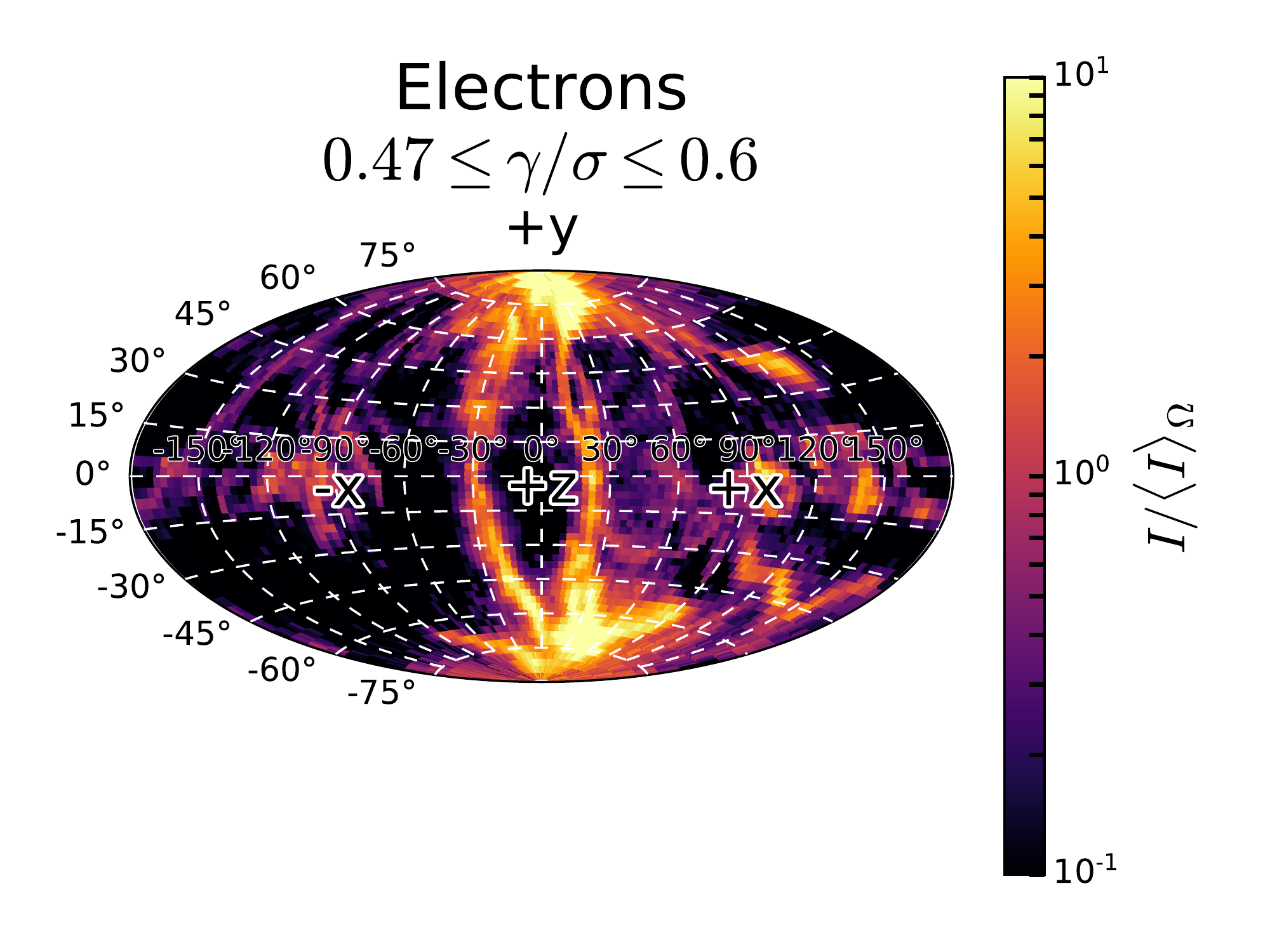}
  \caption{\revtext{(Figure updated/changed from original manuscript.)} An angular map for the~$\grad = \sigma$ simulation displaying the angular intensity~$I = \dif N_{t}/ \dif \gamma \dif \Omega$ of lower layer electrons at~$2.0L/c$. This is a higher energy map relative to Fig.~\ref{fig:hm1nc} (the electron Lorentz factors are closer to~$\grad$) and, as a result, exhibits more intense beaming patterns.}
  \label{fig:hm2nc}
\end{figure}

In this work, we are also interested in how particle momenta (and emitted photons) are distributed directionally. As a result, we must keep the angular information in the distribution function~$f(x, y, \gamma, \pmb{\Omega}; t)$, as was first done by \citet{cwu12}. Furthermore, because we are interested in \textit{kinetic} beaming -- beaming as a function of particle or photon energy -- we must preserve correlations between~$\gamma$ and~$\pmb{\Omega}$. To visualize all three velocity dimensions of the distribution function, we separate the energy information from the angular information, viewing the entire angular distribution at a single energy. Examples of this view are the angular maps (also \quoted{intensity maps}or \quoted[)]{heatmaps}of Fig.~\ref{fig:hm1nc} and Fig.~\ref{fig:hm2nc}. These display the spatially integrated angular particle distribution~$\dif N_t/\dif \gamma \dif \Omega = \int_{y < L_y / 2} \dif x \dif y \, f(x, y, \gamma, \pmb{\Omega}; t)$ at fixed~$\gamma$ and~$t$ using the Aitoff projection. A particle contributes to latitude~$\varphi \in [-90^\circ, 90^\circ]$ and longitude~$\lambda \in [-180^\circ, 180^\circ]$ on a map if its velocity vector parallels the unit vector
\begin{align}
    \hat{\pmb{n}} = \cos(\lambda) \cos(\varphi) \hat{\pmb{z}} + \sin(\lambda) \cos(\varphi) \hat{\pmb{x}} + \sin(\varphi) \hat{\pmb{y}} \, .
    \label{eq:philamdef}
\end{align}
To isolate a single reconnection layer, Fig.~\ref{fig:hm1nc}, Fig.~\ref{fig:hm2nc}, and all subsequent angular plots are generated using only particles (or, later, photons emitted from particles) located in the lower half of the simulation box.

Let us now describe the physical origins of the basic features in Fig.~\ref{fig:hm1nc} and Fig.~\ref{fig:hm2nc}. These features can be neatly decomposed into two broad categories: mild horizontal beaming stemming from bulk motion along the primary reconnection current sheet and extreme beaming arising near X-points (not only in the main current layer, but also in secondary ones between merging plasmoids). In the primary current layer near the main X-point, the reconnection electric field points in the~$+z$-direction~($\varphi = 0^\circ$,~$\lambda = 0^\circ$), resulting in electron acceleration in the~$-z$-direction~($\varphi = 0^\circ$,~$\lambda = \pm 180^\circ$). As electrons are ejected towards~$-z$, they begin to be deflected by lines of reconnected magnetic field, which causes them to disperse towards~$\pm x$~($\varphi = 0^\circ, \lambda = \pm 90^\circ$). This results in the mild concentration of particles along the equator in Fig.~\ref{fig:hm2nc}. As these particles radiatively cool, they are simultaneously deposited into plasmoids, and plasmoid bulk motion along the reconnection layer induces a gentle low-energy momentum anisotropy along the~$\pm x$-directions as in Fig.~\ref{fig:hm1nc}.

We move now to the more extreme beaming. Generally, such pronounced anisotropy occurs only among the higher energy particles, a trend that Fig.~\ref{fig:hm2nc} illustrates nicely. Less universal, but still common, is the fact that the strong beaming patterns in that figure result from plasmoid mergers, evidenced by the prominent vertical swaths in the angular particle distribution. This comes about because, between merging plasmoids, a secondary reconnection layer forms approximately parallel to the~$zy$-plane. The reconnection electric field in this secondary layer points along~$-z$ and accelerates electrons along~$+z$. Owing to the rotated orientation of the reconnecting magnetic field, these electrons begin to fan out towards~$\pm y$ -- towards the poles -- rather than~$\pm x$ as in the primary current sheet.

The fan shapes in Fig.~\ref{fig:hm2nc} are not perfectly vertical because plasmoids with unequal sizes and speeds are merging. At this time, there are actually two ongoing mergers -- one on either flank of the large primary plasmoid (Fig.~\ref{fig:edensforhm12nc}) -- both actively accelerating particles. In the merger on the right-hand side, a smaller and faster left-moving plasmoid creates a secondary current sheet that bends and moves to the left, biasing the accelerated particles towards the~$-x$-direction. The opposite is true for the merger on the left involving a small/fast right-moving plasmoid. The combined result is that the swaths of high-energy particles shown in Fig.~\ref{fig:hm2nc} do not extend along a single meridional plane running through~$\lambda = 0^\circ$, but through two slightly offset planes intersecting~$\lambda \simeq \pm 30^\circ$: one for each current sheet created at asymmetric plasmoid mergers.

\begin{figure}
  \centering
  \includegraphics[width=\linewidth]{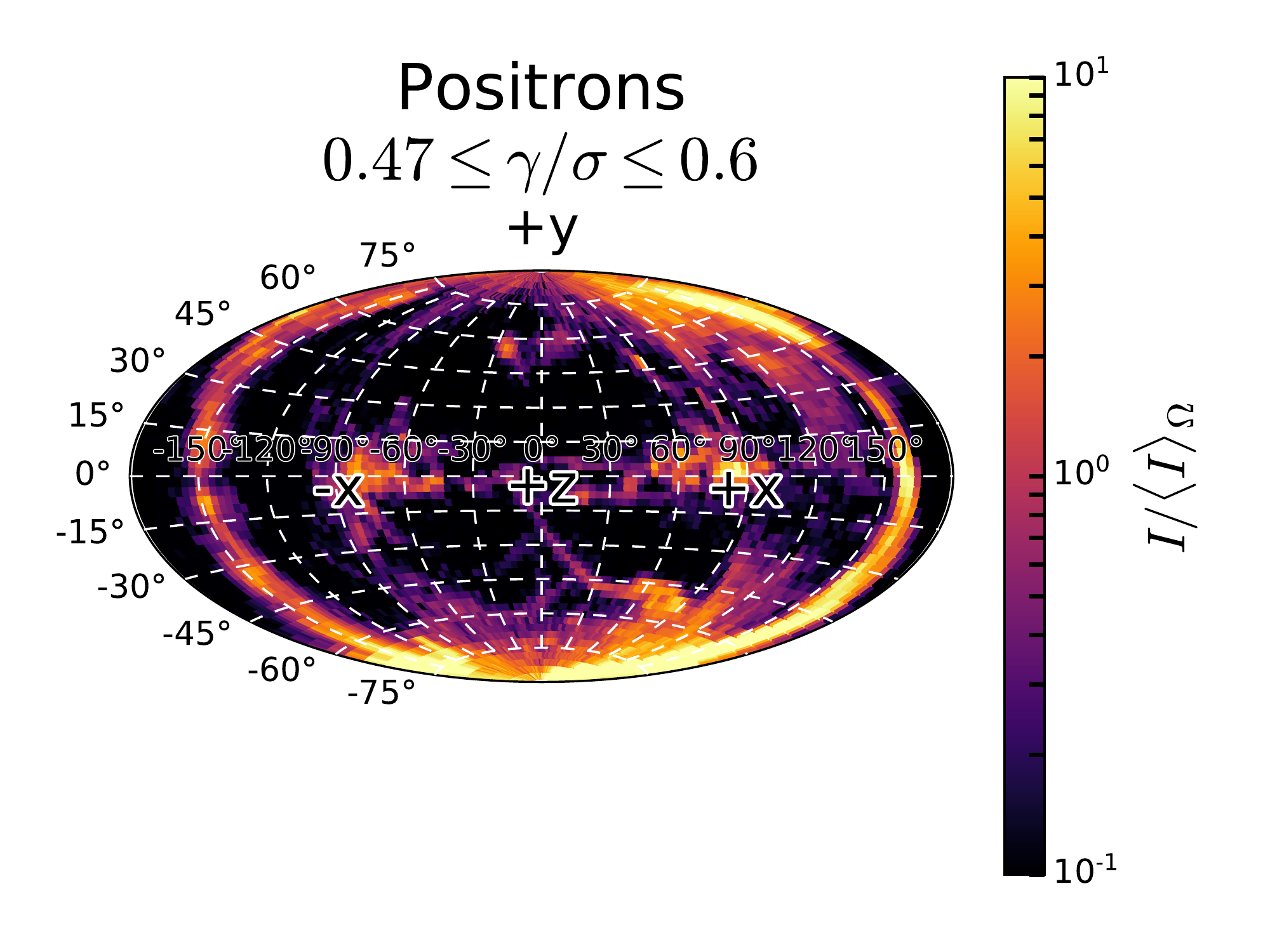}
  \caption{\revtext{(Figure updated/changed from original manuscript.)} A positron intensity map drawn for the same energy band and at the same time as the electron map of Fig.~\ref{fig:hm2nc}. Positrons feel the same reconnection electric field as electrons, but experience acceleration in the opposite direction as a result.}
  \label{fig:hm2ionsnc}
\end{figure}

\begin{figure}
  \centering
  \includegraphics[trim={0 0 160 0},clip,width=\linewidth]{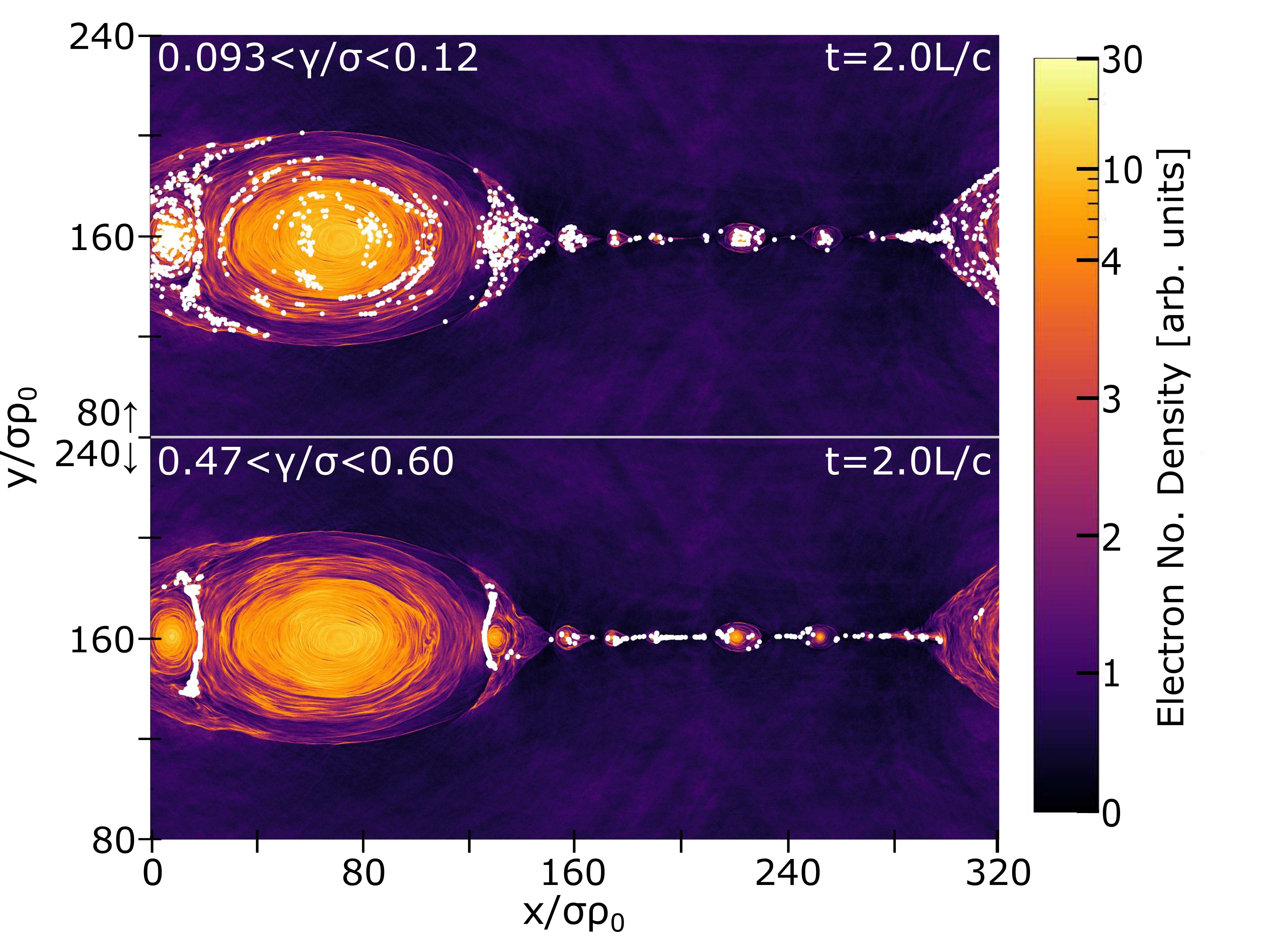}
  \caption{\revtext{(Figure updated/changed from original manuscript.)} Electron density for the same time ($2.0L/c$) used to generate the angular maps of Fig.~\ref{fig:hm1nc}, Fig.~\ref{fig:hm2nc}, and Fig.~\ref{fig:hm2ionsnc}. In the top (bottom) panel, white ovals indicate the locations of a random subset of the electrons contributing to the angular map in Fig.~\ref{fig:hm1nc} (Fig.~\ref{fig:hm2nc}) with Lorentz factors in the range~$0.093 \sigma < \gamma < 0.12 \sigma$~($0.47 \sigma < \gamma < 0.60 \sigma$). The higher energy particles reside in the hearts of the primary and inter-plasmoid current sheets -- near X-points -- and the lower energy particles in plasmoids.}
  \label{fig:edensforhm12nc}
\end{figure}

The features in Fig.~\ref{fig:hm1nc} and Fig.~\ref{fig:hm2nc} are nicely mirrored by those in the corresponding positron angular maps. As an example, Fig.~\ref{fig:hm2ionsnc} displays the angular distribution of positrons in the same energy band and at the same time as the electrons in Fig.~\ref{fig:hm2nc}. Owing to their opposite response to the reconnection electric field, the positrons yield the same X-point-generated beaming configurations found among the electrons in Fig.~\ref{fig:hm2nc} but reflected about the~$xy$-plane.

To tie the beaming features on angular maps to their spatial origins, Fig.~\ref{fig:edensforhm12nc} shows the spatial electron number density at the time for which Figs~{\ref{fig:hm1nc}--\ref{fig:hm2ionsnc}} were drawn. Also shown are the locations of randomly chosen subsets of electrons with Lorentz factors in the low-energy range of Fig.~\ref{fig:hm1nc} and the high-energy range of Figs~\ref{fig:hm2nc} and~\ref{fig:hm2ionsnc}. The positional clustering explicitly demonstrates the dichotomy described above: low-energy particles are confined to plasmoids whose bulk motion governs their momentum anisotropy while high-energy particles exhibit more extreme beaming shaped by reconnection X-points.
Fig.~\ref{fig:edensforhm12nc} also shows that X-point acceleration and collimation is ongoing in the primary reconnection layer even after the plasmoid chain has fully developed. As a result, it is not always the case (as it is in the angular maps shown previously) that the strongest beaming signatures are vertical. In fact, vertical fan shapes tend to be only intermittently prominent: when plasmoids -- particularly large ones like in Fig.~\ref{fig:edensforhm12nc} -- are actively merging. In between these episodes, strong beaming arising in the primary current sheet can still be significant (e.g. Fig.~\ref{fig:heatmapintime_GradSigma1}).

At this point, we would like to clarify that, of the two beaming origins discussed in this section, it is only the one operating near X-points that was associated with kinetic beaming in the works that originally introduced the concept \citep{ucb11, cub12, cwu12}. As discussed by those authors, the configuration of electromagnetic fields at these locations is particularly suited to accelerate and collimate high-energy particles: the reconnection electric field delivers energy while the reconnecting magnetic field focuses particles into beams. The particles remaining near X-points longer are consequently more energized and more focused. As we have seen already (Fig.~\ref{fig:hm2nc}), this mechanism is responsible for the most severe beaming at the highest particle energies; as we shall see later, it also yields the most energy-\textit{dependent} beaming. By comparison, plasmoid motion-generated anisotropy (Fig.~\ref{fig:hm1nc}) is milder and tends, because it derives from fluid level motion, to be more achromatic. (Plasmoids, along with their associated bulk motion and Doppler beaming, provide the basis for the \quoted{minijets}model of \citealt{gub09, gub10}.)

Thus, when we use the term \quoted[,]{kinetic beaming} we are not referring to just any energy-dependence in the degree of particle or photon collimation. We refer specifically to the most extreme and energy-dependent anisotropy at the highest energies generated near X-points. In this sense, the very different signatures of beaming evident in Figs~\ref{fig:hm1nc} and~\ref{fig:hm2nc}, despite occurring at different particle energies, do not illustrate kinetic beaming. Instead, they portray two separate beaming mechanisms that merely dominate at different energy scales. We will illustrate kinetic beaming as we and previous authors apply the term -- which involves energy-dependent anisotropy sourced only by the X-point mechanism -- after we develop a more quantitative description of beaming in the next section.

\section{Quantifying beaming}
\label{sec:beams}
In this section, we present two quantitative notions of \quoted{beaming}as manifested in angular maps such as Figs~{\ref{fig:hm1nc}--\ref{fig:hm2ionsnc}}. Before proceeding, it will be helpful to introduce some additional terminology with which to describe the information on these maps: the angular distribution of particles~$\dif N / \dif \Omega$, of the instantaneous radiated power~$\dif P / \dif \Omega$, and quantities derived from these.\footnote{We temporarily omit to explicitly write the dependence on time~$t$, as well as the~$\gamma$-dependence of~$\dif N / \dif \Omega$ and the spectral dependence of~$\dif P / \dif \Omega$, while establishing our nomenclature.} The angular distribution~$\dif P/\dif \Omega$ is ordinarily called \quoted[~$I$,]{intensity}and the power~$P$ radiated into a finite solid angle is~$P = \int I \dif \Omega$. In analogy with light, we shall frequently call the angular distribution of particles~$\dif N / \dif \Omega$ by the name \quoted[,]{intensity}as well as borrow the symbol~$I$. Furthermore, we will use the word \quoted{power}to refer to the total number of particles travelling within a finite angular patch. This language enables us to describe beaming in generic terms. Whether we mean a beam of particles or a beam of photons will be clear from the context.

\subsection{Two notions of beaming}
As demonstrated by the intensity maps of Fig.~\ref{fig:hm1nc} and Fig.~\ref{fig:hm2nc}, the \zeltron particle distributions do not necessarily exhibit what one typically imagines as a beam: a spot of high intensity that is nearly symmetric about some axis. Rather, the high intensity regions on angular maps can be quite extended and complicated in shape, particularly at higher energies. Any quantitative definition of beaming one adopts must therefore be sufficiently versatile to handle the diverse set of momentum-space configurations attained by the particles (or photons, but for concreteness this section confines the discussion to the particle distribution).

\begin{figure}
  \centering
  \includegraphics[width=\linewidth]{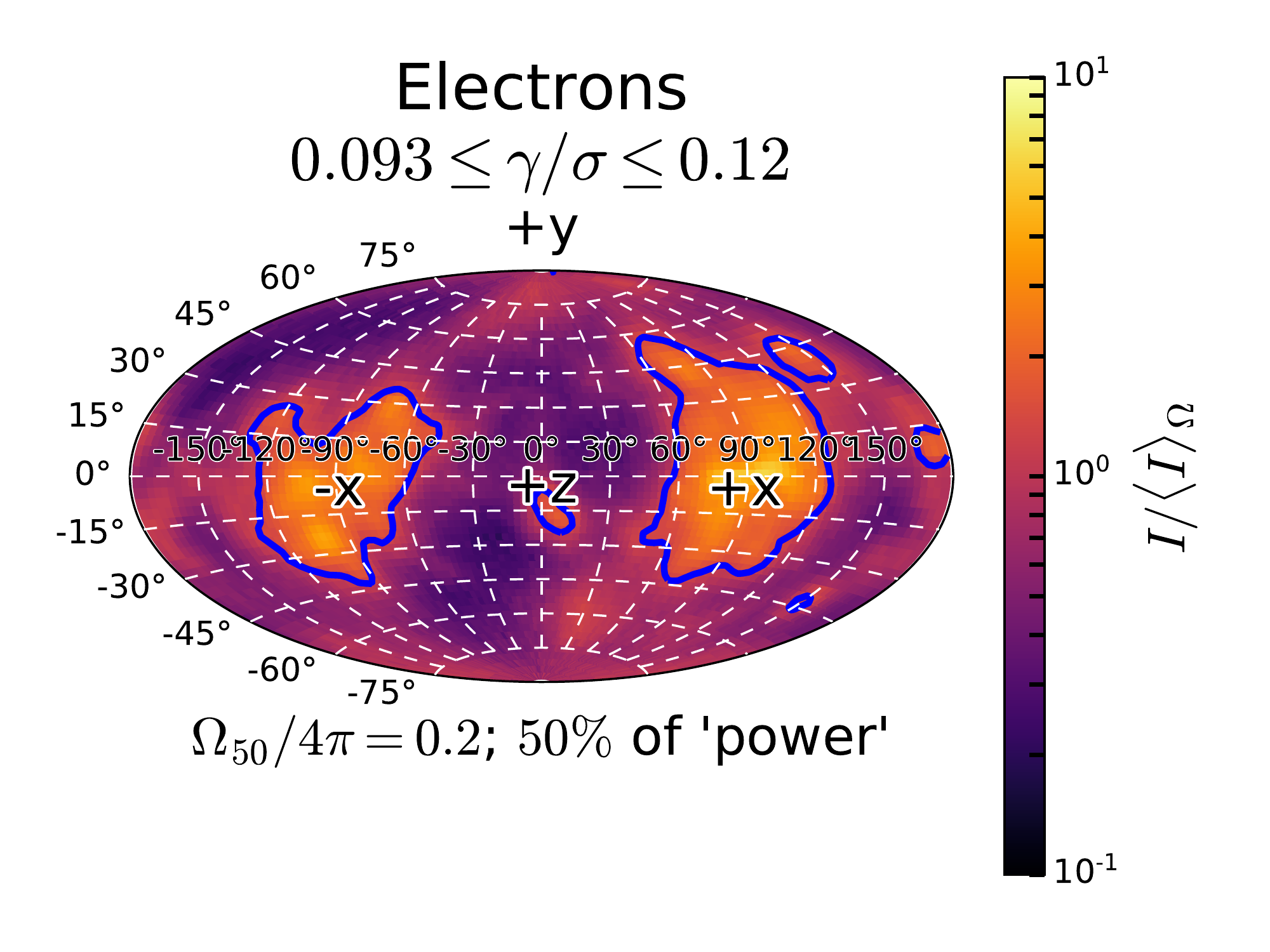}
  \caption{\revtext{(Figure updated/changed from original manuscript.)} The same angular map as Fig.~\ref{fig:hm1nc}, but with a blue contour outlining the smallest solid angle that contains half of the total power -- i.e.,~$\Omega_{\rm 50}$, which in this case is~$20 \rm \, per \, cent $ of $4\pi$.}
  \label{fig:hm1}
\end{figure}

\begin{figure}
  \centering
  \includegraphics[width=\linewidth]{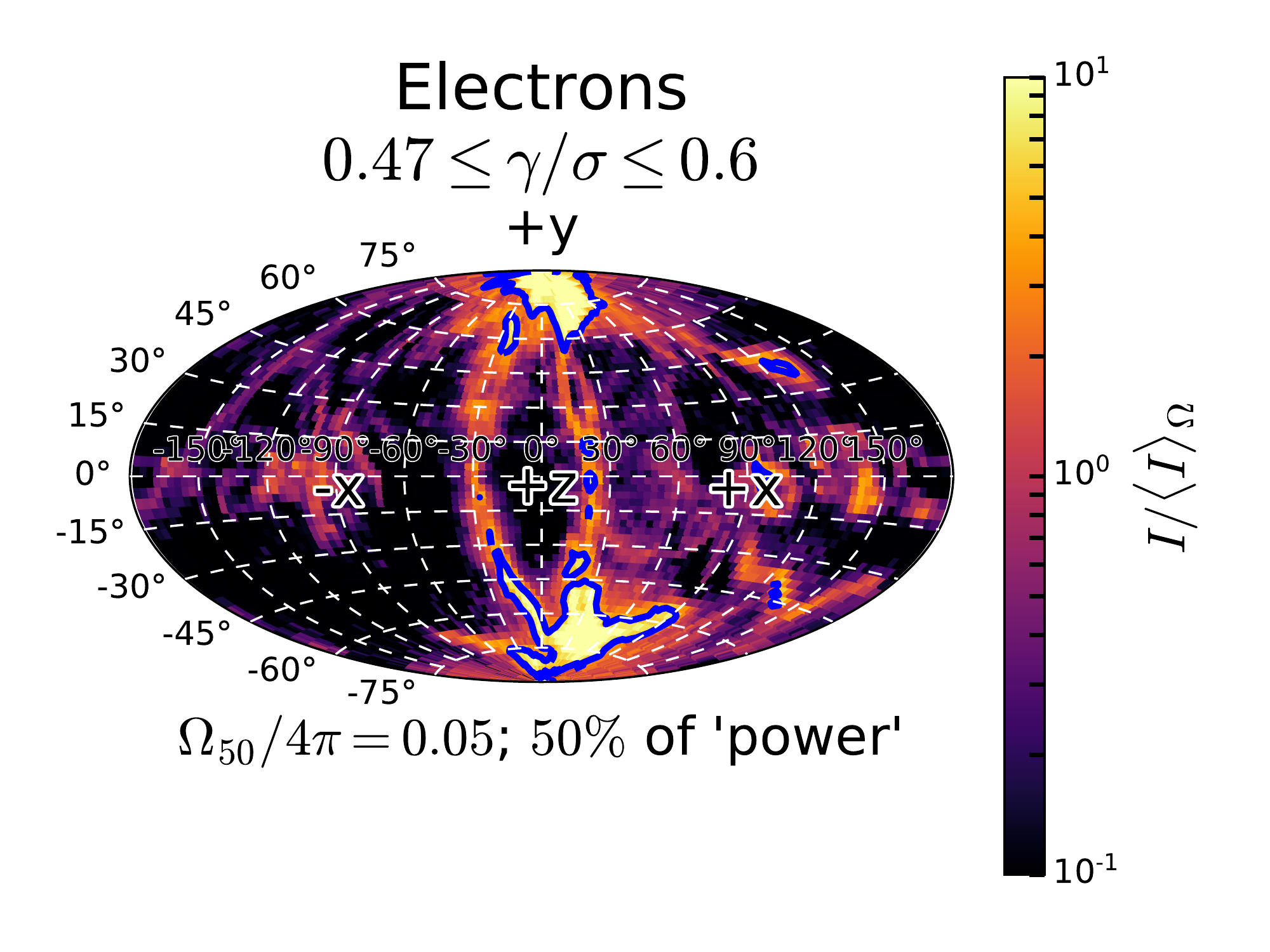}
  \caption{\revtext{(Figure updated/changed from original manuscript.)} The same intensity map as in Fig.~\ref{fig:hm2nc}, but with the~$\Omega_{50}$ contour labelled. In this case~$\Omega_{50}$ is~$5 \rm \, per \, cent $ of~$4\pi$.}
  \label{fig:hm2}
\end{figure}

To meet this challenge, we employ two complementary measures of beaming. The first was originally introduced by \citet{cwu12}, who parametrized beaming by~$\Omega_{50}$: the smallest total (possibly non-contiguous) solid angle containing half of the power on an angular map (within the given energy bin). This quantity is illustrated in Fig.~\ref{fig:hm1} and Fig.~\ref{fig:hm2}. A smaller value indicates more extreme beaming because a smaller fraction of the sphere contains an order unity fraction of the power.

\begin{figure}
  \centering
  \includegraphics[width=\linewidth]{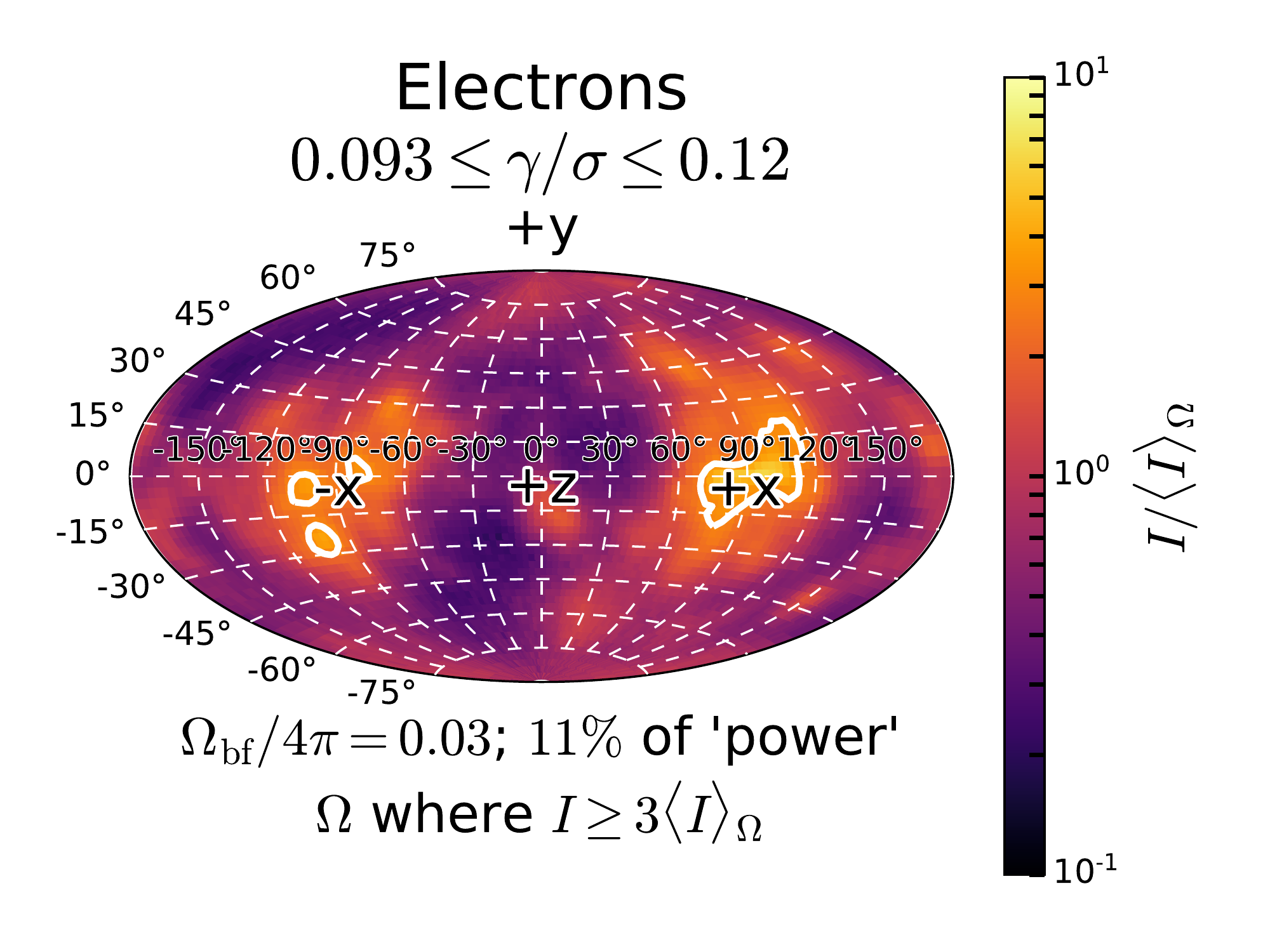}
  \caption{\revtext{(Figure updated/changed from original manuscript.)} The same angular map as in Fig.~\ref{fig:hm1nc}, but with a white contour outlining the high intensity region (where the intensity exceeds three times its angle-average). The fraction of the heatmap power contained in this contour is the \quoted{beamed fraction}and in this case is equal to~$11 \rm \, per \, cent $. The solid angle footprint of the high intensity region is~$3 \rm \, per \, cent $ of the sphere (note that the solid angle~$\Omega_{bf}$ enclosed by the contour is \textit{not}~$\Omega_{50}$). The beamed fraction provides an alternative measure of beaming to~$\Omega_{50}$.}
  \label{fig:hm1bf}
\end{figure}

\begin{figure}
  \centering
  \includegraphics[width=\linewidth]{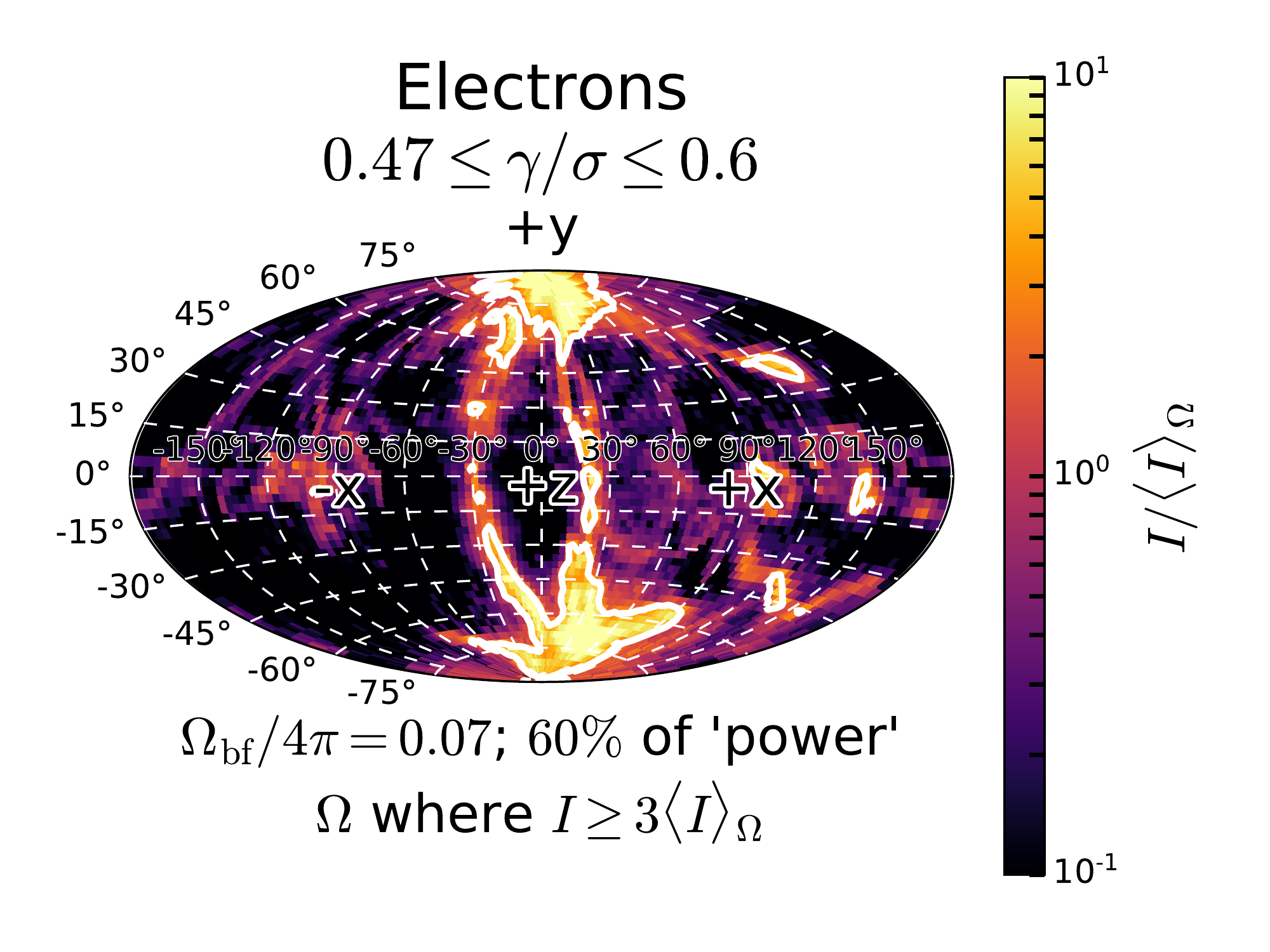}
  \caption{\revtext{(Figure updated/changed from original manuscript.)} The same map as Fig.~\ref{fig:hm2nc}, but with a white contour outlining the high intensity region~($I \ge \acut \langle I \rangle_{\Omega}$). Here the beamed fraction is equal to~$60 \rm \, per \, cent $ and is contained in~$7 \rm \, per \, cent $ of the available solid angle.}
  \label{fig:hm2bf}
\end{figure}

The second measure of beaming characterizes the angular regions where the intensity exceeds three times the angle-averaged intensity (denoted~$\langle I \rangle_\Omega$ in the figures; again within a single energy bin). We define the \quoted[,]{beamed fraction}which we abbreviate as~$bf$, to be the fraction of the total power contained within these regions. Extraction of the beamed fraction from the angular maps in Figs~\ref{fig:hm1nc} and~\ref{fig:hm2nc} is demonstrated in Figs~\ref{fig:hm1bf} and~\ref{fig:hm2bf}.

Both measures of beaming --~$\Omega_{50}$ and~$bf$ -- have advantages and disadvantages. The beamed fraction does not rely on regions of extreme intensity being confined to small fractions of the sphere. On the other hand,~$\Omega_{50}$, when small, is perhaps a more convincing indicator of beaming because it means that the corresponding angular map region contains high power and occupies a small solid angle; large beamed fraction indicates only high power. We will use both tools in order to give a more compelling account of kinetic beaming.

One thing that these metrics have in common is that they are insensitive to the shapes and continuity of beams. Although one may conceive of more detailed and observer-centric measures of beaming, perhaps characterizing the morphologies of individual contiguous beams, this would greatly complicate the analysis. Leaving that for a future work, we find that our more coarse-grained measures are sufficient to illustrate a number of intriguing properties of the global system-wide beaming produced by magnetic reconnection.

\begin{figure*}
  \centering
  \begin{subfigure}{0.33\textwidth}
    \centering
    \includegraphics[width=\linewidth]{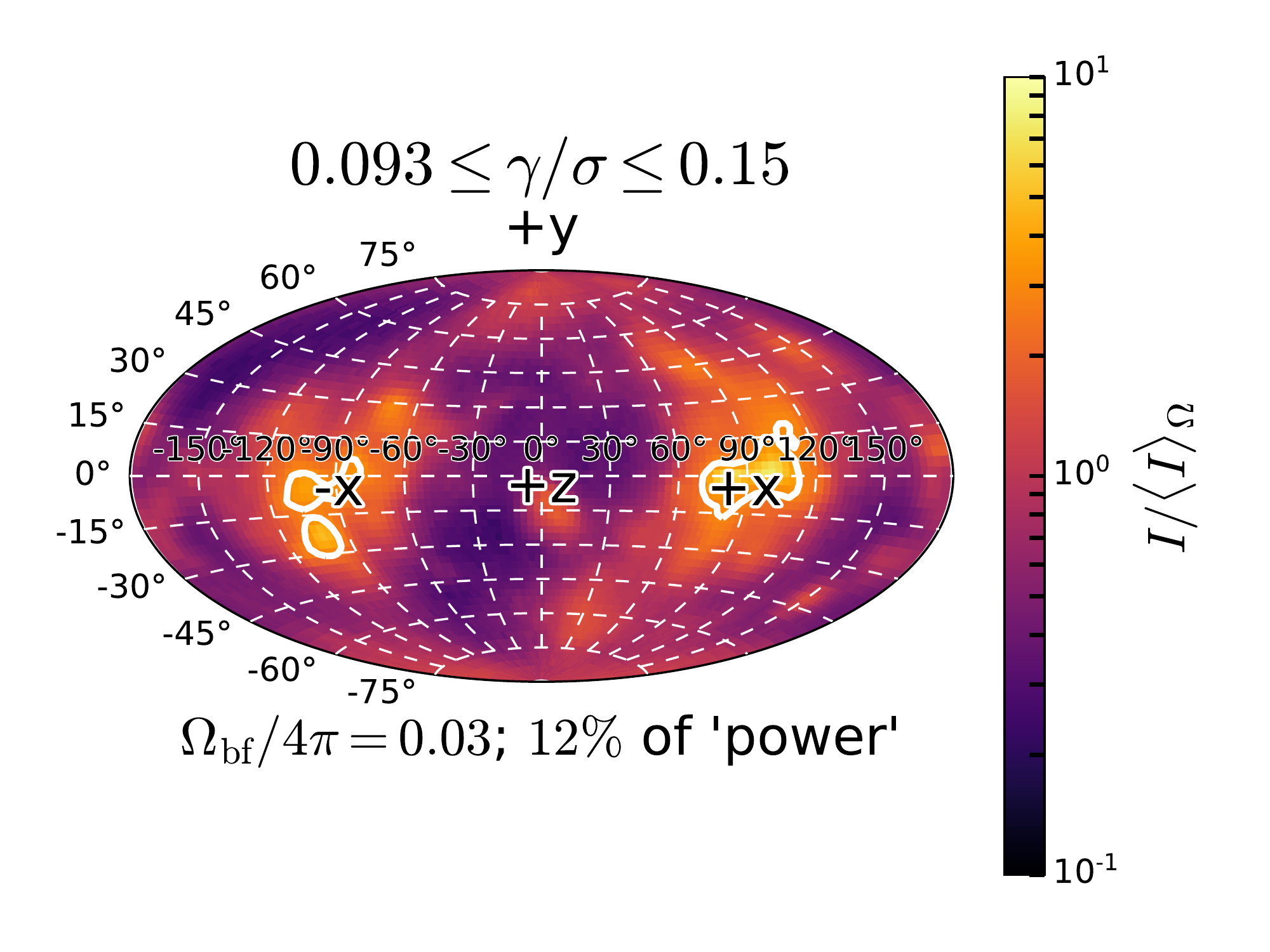}
    \label{sfig:hm1}
  \end{subfigure}
  \begin{subfigure}{0.33\textwidth}
    \centering
    \includegraphics[width=\linewidth]{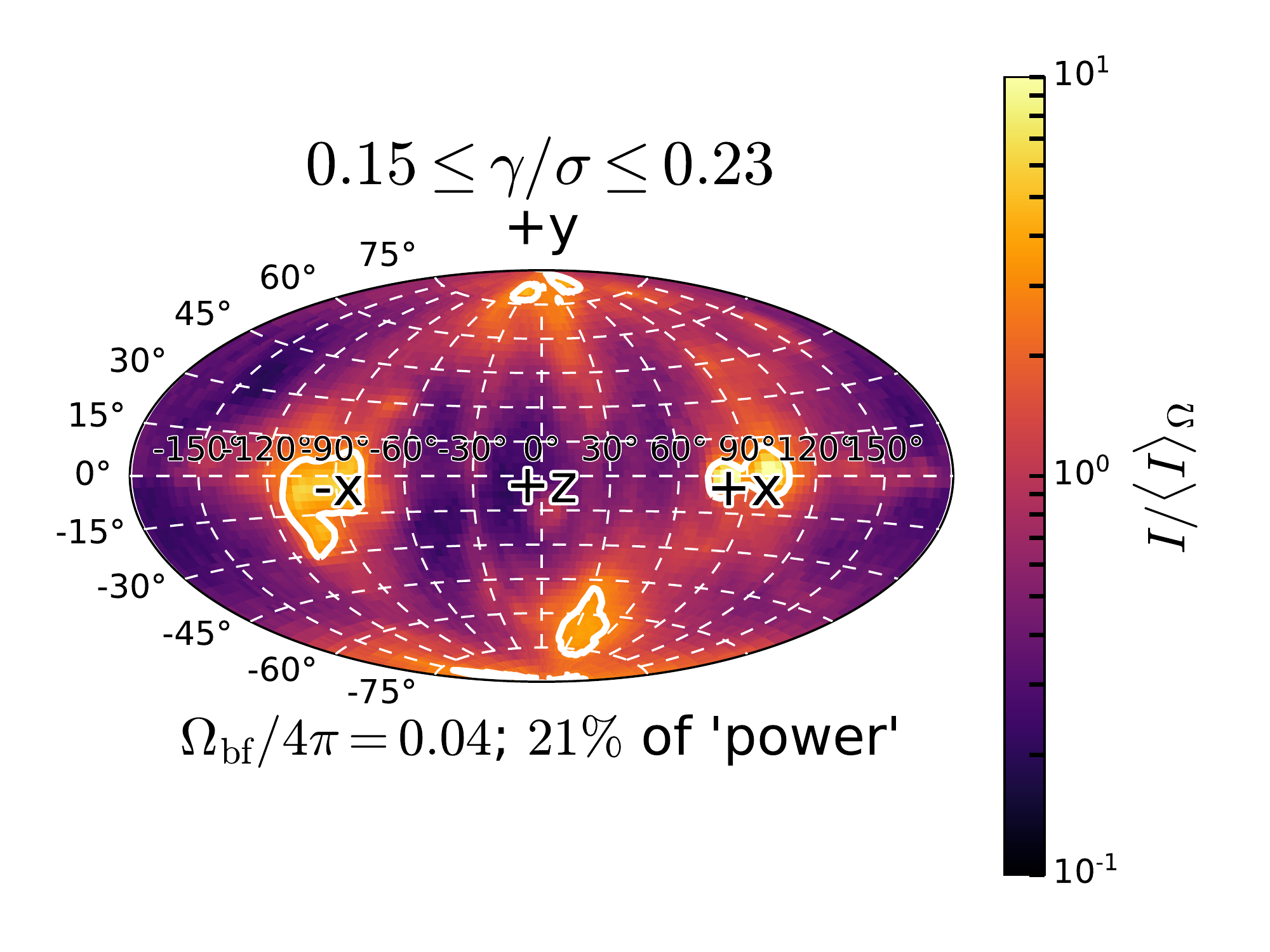}
  \end{subfigure}
  \begin{subfigure}{0.33\textwidth}
    \centering
    \includegraphics[width=\linewidth]{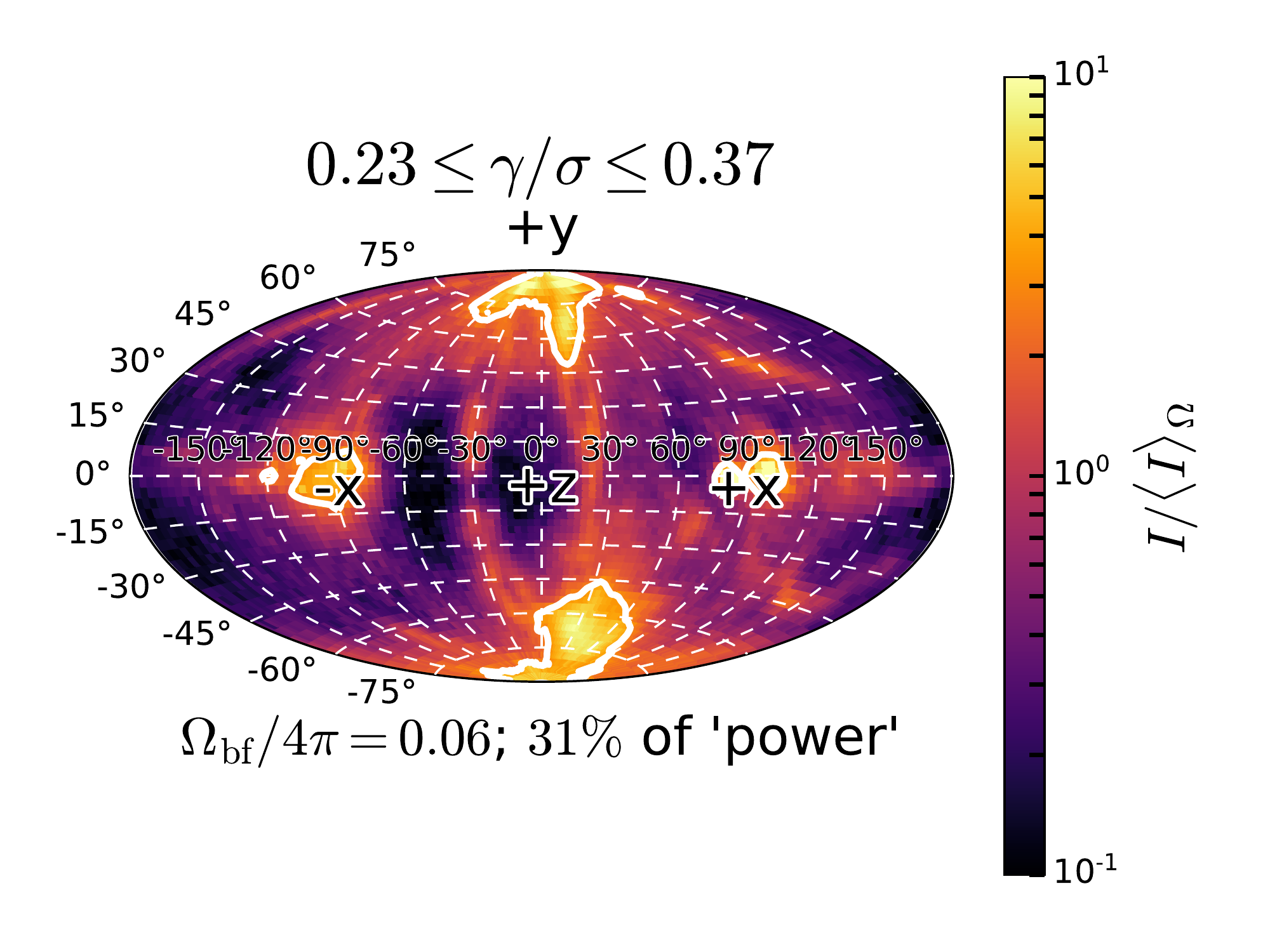}
  \end{subfigure} \\
  \begin{subfigure}{0.33\textwidth}
    \centering
    \includegraphics[width=\linewidth]{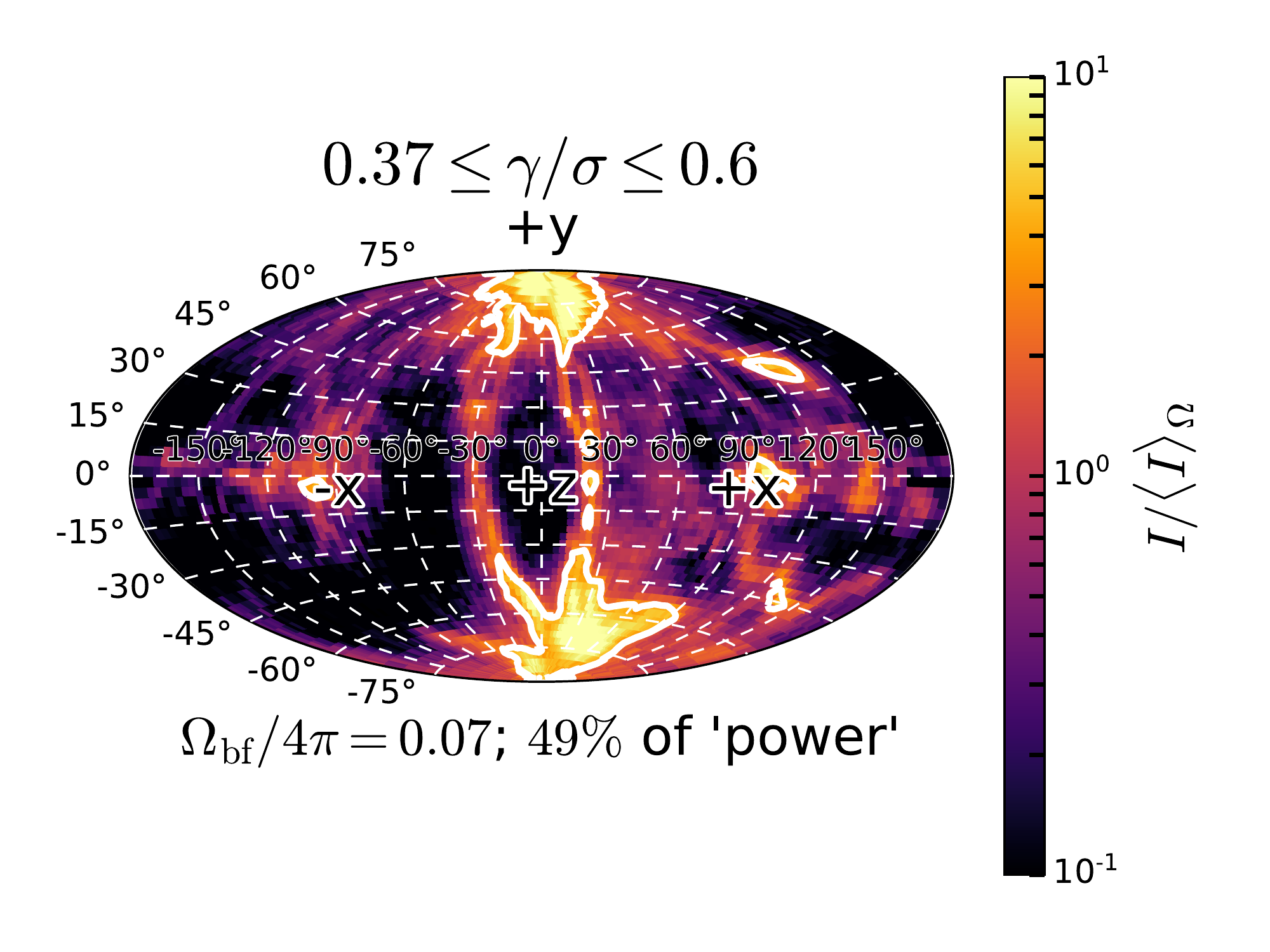}
  \end{subfigure}
  \begin{subfigure}{0.33\textwidth}
    \centering
    \includegraphics[width=\linewidth]{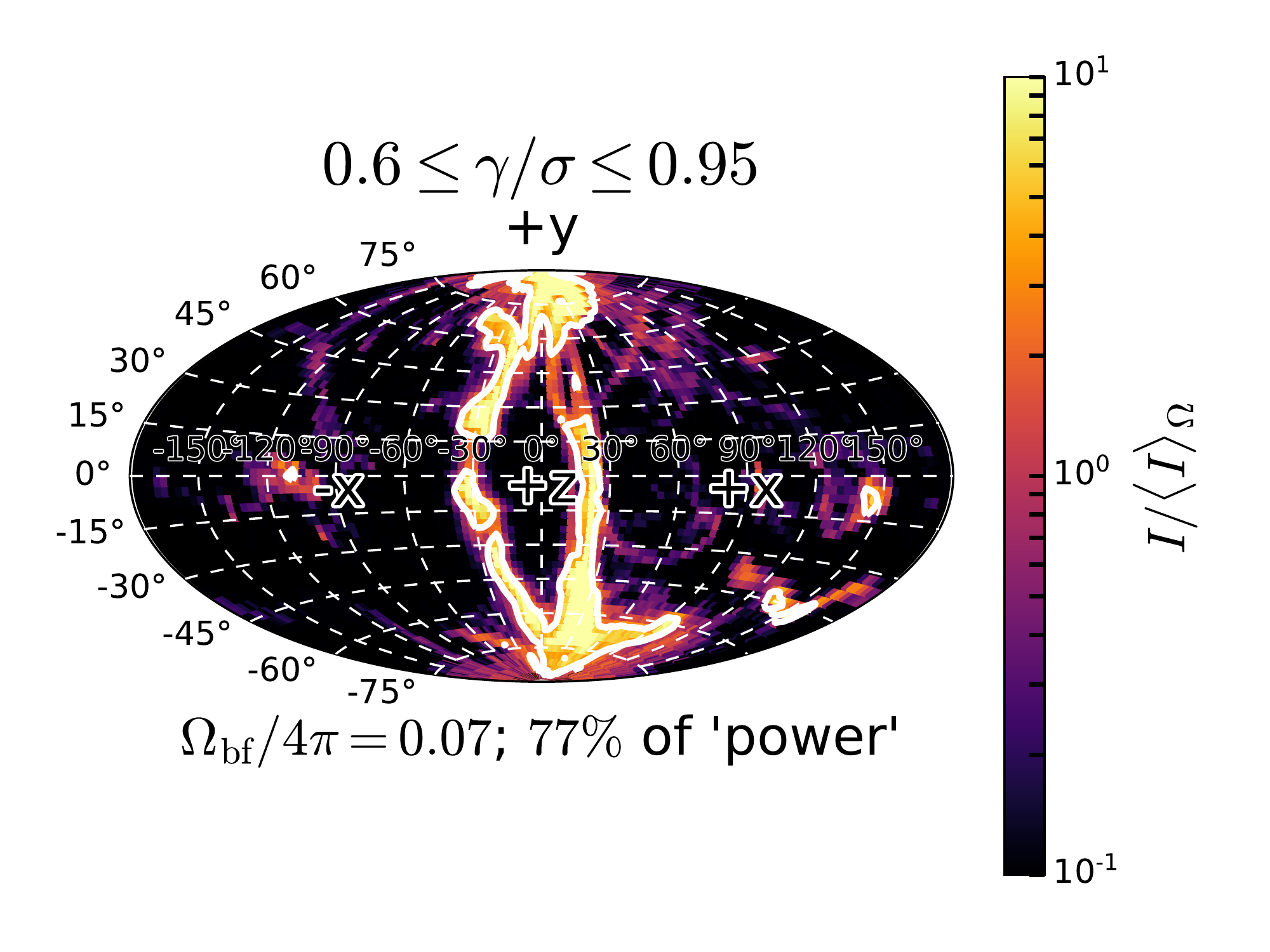}
    \label{sfig:badstats1}
  \end{subfigure}
  \begin{subfigure}{0.33\textwidth}
    \centering
    \includegraphics[width=\linewidth]{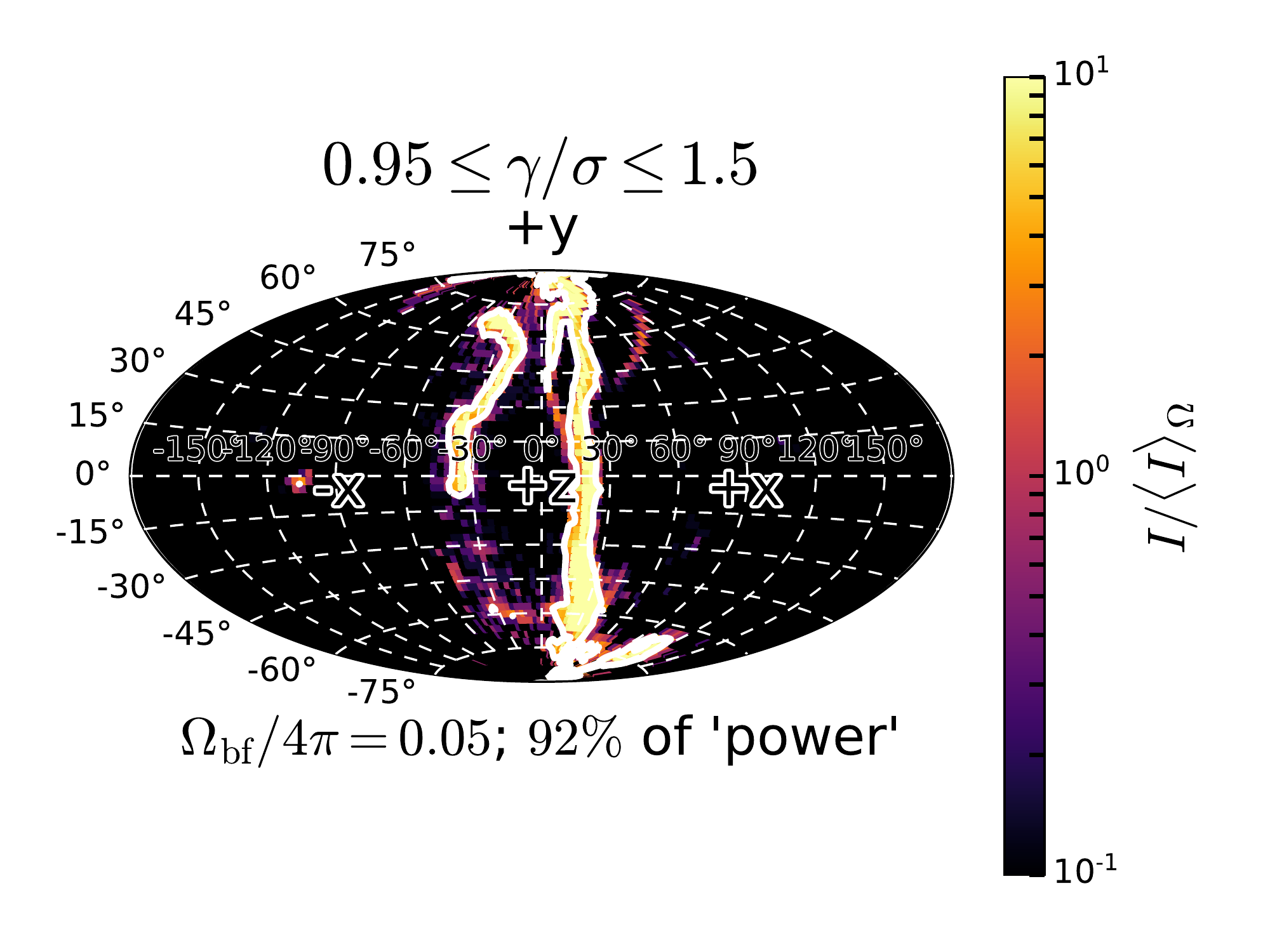}
    \label{sfig:badstats2}
  \end{subfigure}
  \caption{\revtext{(Figure updated/changed from original manuscript.)} Electron intensity maps from our~$\grad = \sigma$ simulation, with white contours outlining the \quoted{beamed fraction}(where the intensity is more than three times the average), for a series of particle energy bins at a single time~$t = 2.0L/c$. Higher energy particles are more strongly beamed and, in this case, originate from reconnection sites between merging plasmoids.}
  \label{fig:heatmapscan}
\end{figure*}

\begin{figure}
  \centering
  \includegraphics[width=\columnwidth]{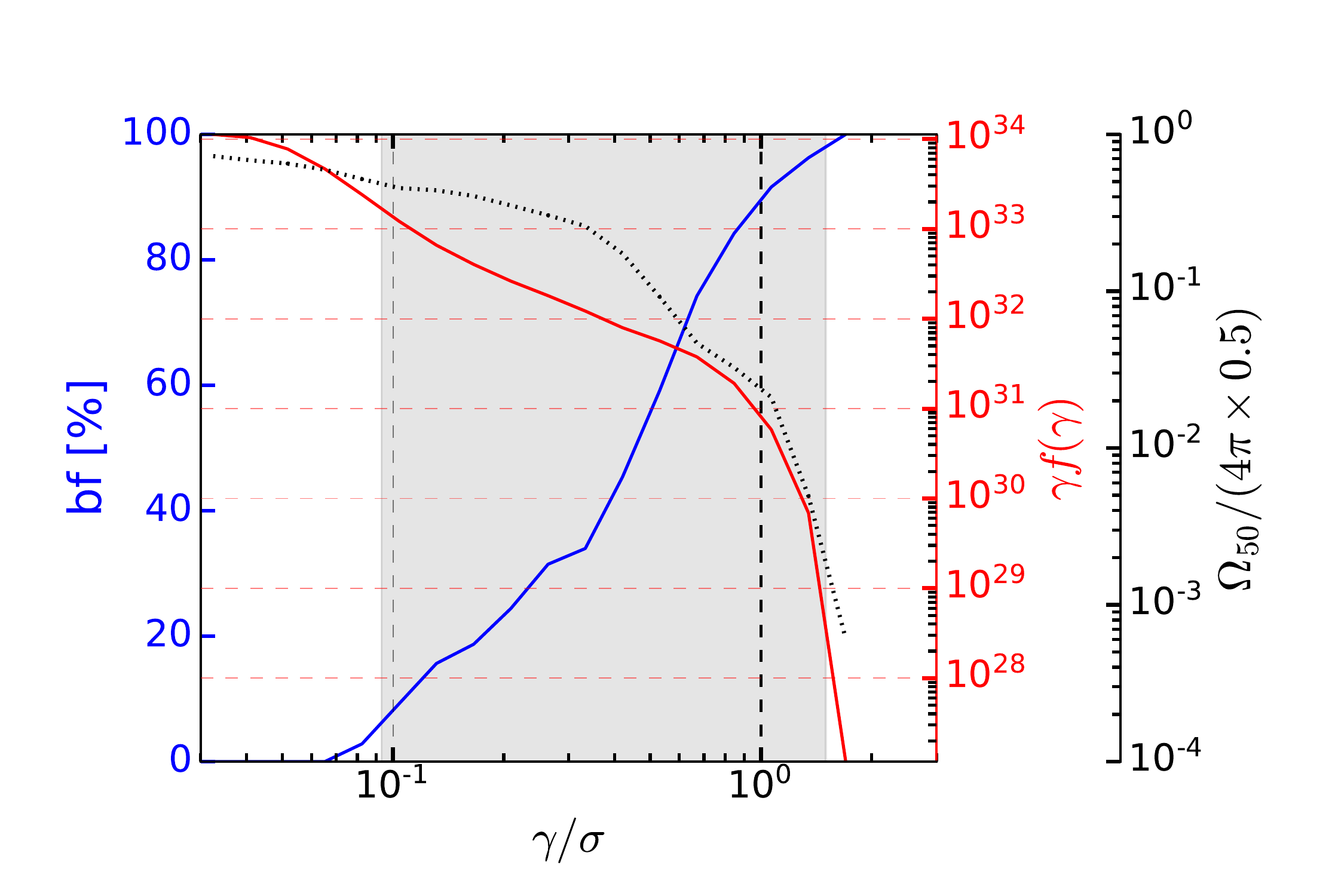}
  \caption{\revtext{(Figure updated/changed from original manuscript.)} Electron beamed fraction (solid blue),~$\Omega_{50}$ (dotted black), and energy distribution (solid red) as a function of electron energy at~$t = 2.0L/c$ in the~$\grad = \sigma$ simulation. The Lorentz factor~$\gamma = \grad = \sigma$ is denoted by a vertical dashed line. The~$\Omega_{50}$ curve is normalized such that a perfectly isotropic angular map produces the value~$\Omega_{50} / (4 \pi \times 0.5) = 1$. In the same limit (attained at low energies), the beamed fraction tends to zero because the intensity is everywhere less than three times the isotropic intensity. The shaded region indicates the energy range shown in Fig.~\ref{fig:heatmapscan}. The coincident sharp rise in the beamed fraction and precipitous drop in~$\Omega_{50}$ demonstrate pronounced kinetic beaming at the highest energies.}
  \label{fig:hmtabbeaming}
\end{figure}

Having developed two notions of beaming, we are now in a position to analyse \textit{kinetic} beaming, which necessarily involves many maps across the particle energy spectrum. In this effort, the chief utility of the~$\Omega_{50}$ and~$bf$ measures is to enable a reduction of the data contained on any given heatmap to two meaningful numbers, which we may then plot as a function of particle or photon energy. This procedure is illustrated in Fig.~\ref{fig:heatmapscan} and~Fig.\ref{fig:hmtabbeaming}. In the first figure, we display a collection of electron intensity maps spanning a decade in particle energy at a given instant in our~$\grad = \sigma$ simulation (strongly radiative). Each map in that figure is distilled to two numbers, its~$\Omega_{50}$ and its beamed fraction, which are then plotted as a function of particle energy in Fig.~\ref{fig:hmtabbeaming}. The latter figure depicts the pronounced energy-dependence of beaming more concisely and dramatically, and we will make use of many similar plots throughout the remainder of this work.

Before moving on to concentrate more exclusively on the succinct energy-centric view of beaming afforded by plots like Fig.~\ref{fig:hmtabbeaming}, we would like to pause and emphasize, once more, the connection between the angular configurations realized in Fig.~\ref{fig:heatmapscan} and their underlying physical mechanisms (discussed previously in Section~\ref{sec:angmaps}). Namely, Fig.~\ref{fig:heatmapscan} demonstrates:~(1) mild beaming in the~$\pm x$-directions~($\varphi = 0^\circ, \lambda = \pm 90^\circ$) due to bulk plasmoid motion; and~(2) dramatic beaming originating near X-points -- in this case, X-points between merging plasmoids -- and extending from the~$+z$-direction~($\varphi = 0^\circ, \lambda = 0^\circ$) towards the poles~($\varphi = \pm 90^\circ$). The former mechanism is most prominent at lower energies but the latter takes precedence at higher energies and gives rise to the steepest energy-dependence in Fig.~\ref{fig:hmtabbeaming}. As a reminder, it is this more extreme beaming that we call \quoted{kinetic beaming} and to which we shall devote the majority of our analysis in the next section.

\section{Kinetic beaming and radiative cooling}
\label{sec:kinbeam}
In this section, we apply the quantitative measures of beaming described above -- $\Omega_{50}$ and beamed fraction ($bf$) -- to answer questions~\ref{en:q1} and~\ref{en:q2} posed in the Introduction. First, we consider the question of observable kinetic beaming for two extreme cases: no radiative cooling and strong radiative cooling. After examining these scenarios in detail, we conduct a higher level analysis that makes use of our full parameter scan in~$\grad$ to create a more complete picture of the dependence of kinetic beaming on cooling efficiency.

\subsection{No cooling: $\grad / \sigma = \infty$}
\label{sec:kbnocool}
\begin{figure}
  \centering
  \includegraphics[width=\columnwidth]{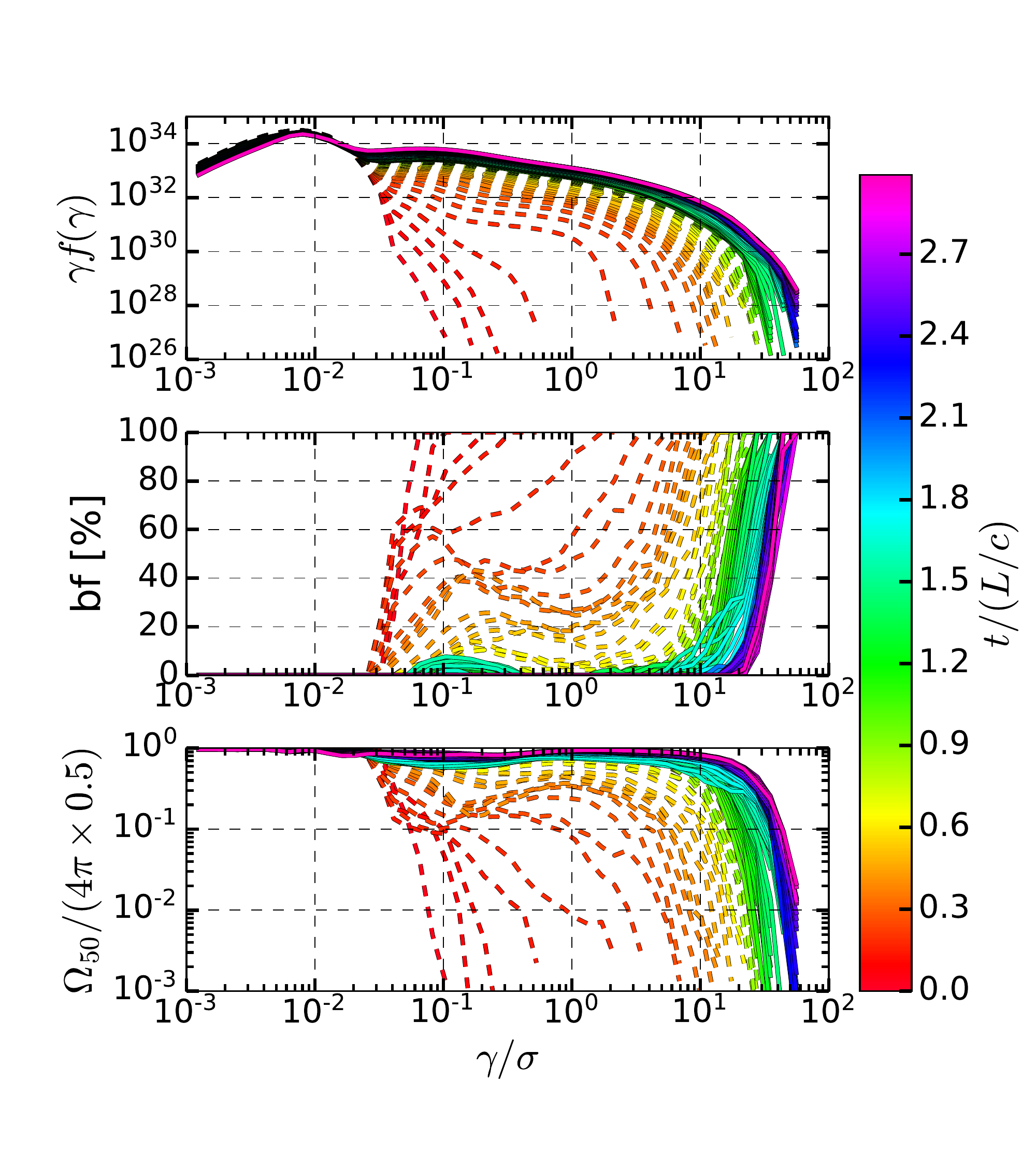}
  \caption{The evolution of the electron energy distribution (top panel), beamed fraction (middle panel), and~$\Omega_{50}$ (bottom panel) for our simulation with~$\grad / \sigma = \infty$ (no cooling). Dashed lines indicate data prior to the first \lc time and solid lines data between~$1$ and~$3$ \lc[s.] The solid lines show that, at late times, beaming vanishes at all but the highest (recently populated) energies: the beamed fraction tends to zero and~$\Omega_{50}$ to~$0.5 \times 4 \pi$.}
  \label{fig:BeamFracVsT_norad}
\end{figure}

For our simulation without IC cooling ($\grad / \sigma = \infty$), the time evolution of three quantities as a function of particle Lorentz factor is displayed in Fig.~\ref{fig:BeamFracVsT_norad}. From top to bottom, these are the electron energy distribution, electron beamed fraction, and electron $\Omega_{50}$.
In the figure, transient behaviour in all three quantities persists through about one \lc time. During this early stage, beaming is both present and energy-dependent, with $bf$ rising sharply and $\Omega_{50}$ falling steeply at high Lorentz factors: we observe clear kinetic beaming.

Here, we are restricting our discussion to the highest particle energies, ignoring the non-monotonic behaviour in~$bf$ and~$\Omega_{50}$ that takes place at lower energies (and primarily at early times). This behaviour stems chiefly from a competition between the two sources of anisotropy in the particle distribution discussed previously in Section~\ref{sec:angmaps}: plasmoid motion, which induces mild beaming among the low-energy particles, and collimation near reconnection X-points, which has a much more dramatic beaming effect primarily at  high particle energies. At intermediate energies, the contributions from both plasmoids and X-points to the global (spatially integrated) distribution of particles can be approximately equal, causing bright regions to cover a larger portion of the angular map and, hence, making it appear more isotropic. By focusing on the highest particle (and, later, photon) energies, where beaming is monotonically increasing~[$\dif \, (bf) / \dif \gamma > 0$ and~$\dif \, \Omega_{50} / \dif \gamma < 0$], we isolate the contribution from X-points, the true underlying agents of \quoted{kinetic beaming}as defined here and in previous works \citep[][see also Section~\ref{sec:angmaps}]{ucb11, cub12, cwu12}.

\begin{figure*}
  \centering
  \begin{subfigure}{0.33\textwidth}
    \includegraphics[width=\linewidth]{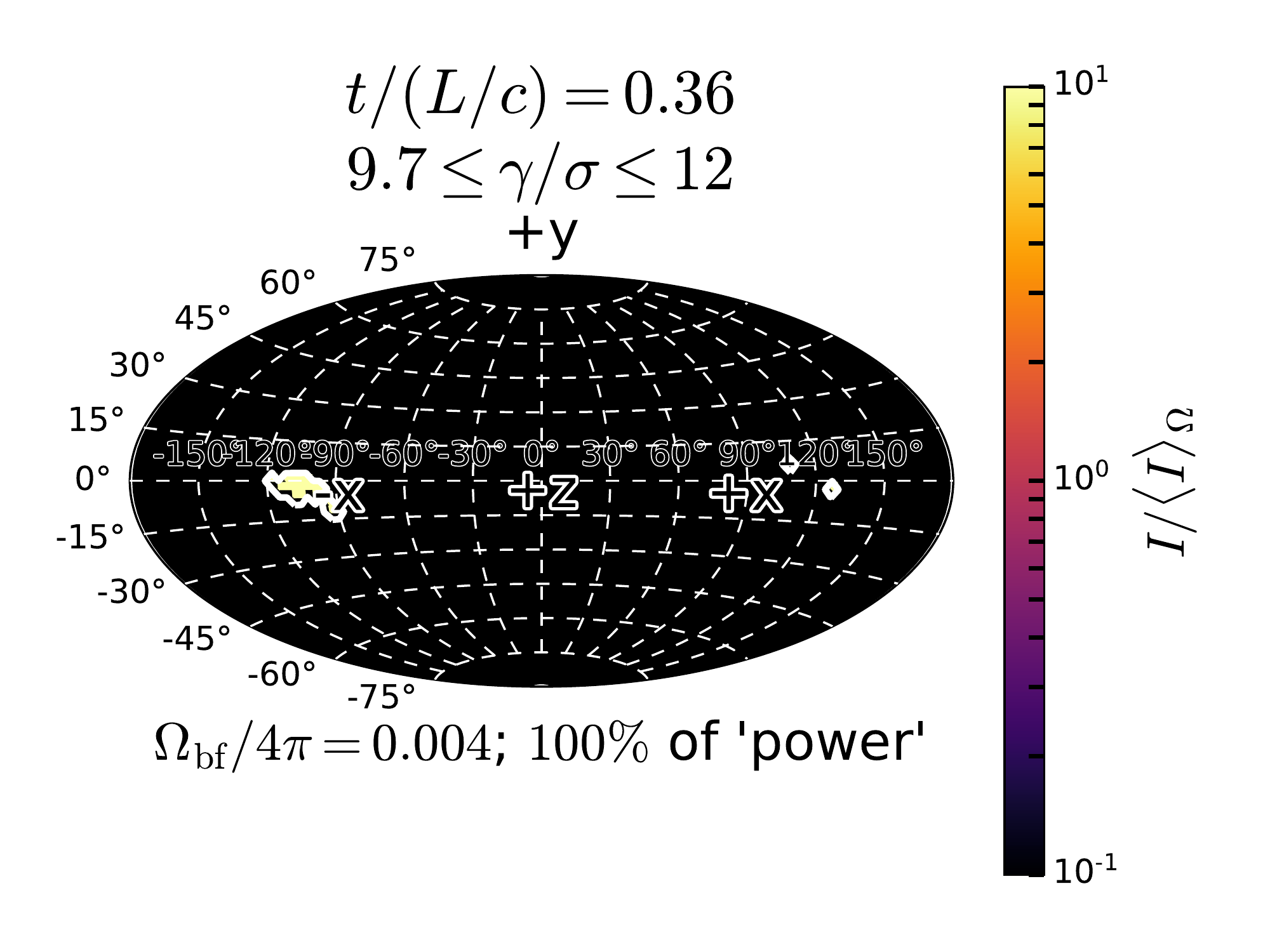}
  \end{subfigure}
  \begin{subfigure}{0.33\textwidth}
    \includegraphics[width=\linewidth]{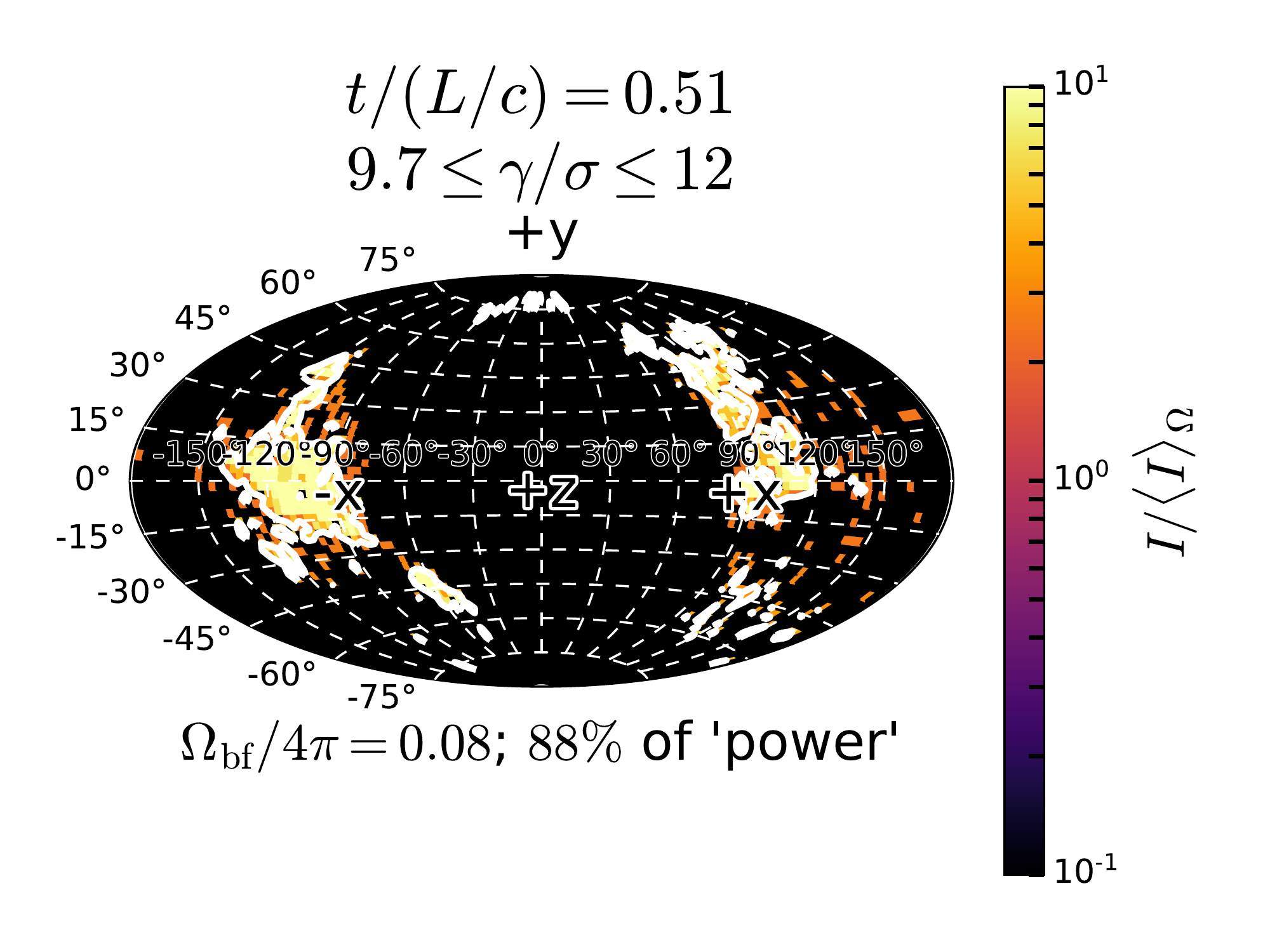}
  \end{subfigure}
  \begin{subfigure}{0.33\textwidth}
    \includegraphics[width=\linewidth]{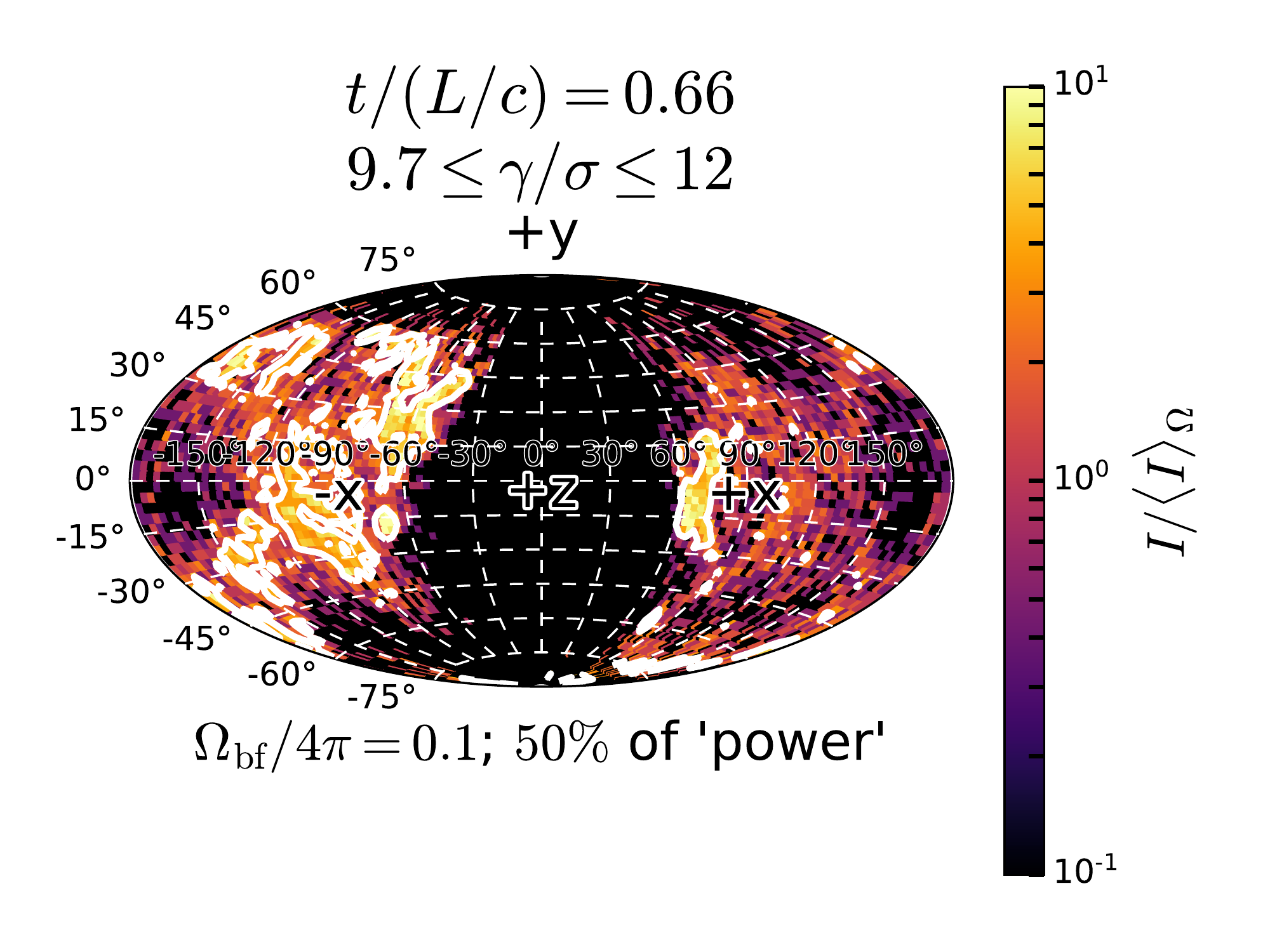}
  \end{subfigure} \\
  \begin{subfigure}{0.33\textwidth}
    \includegraphics[width=\linewidth]{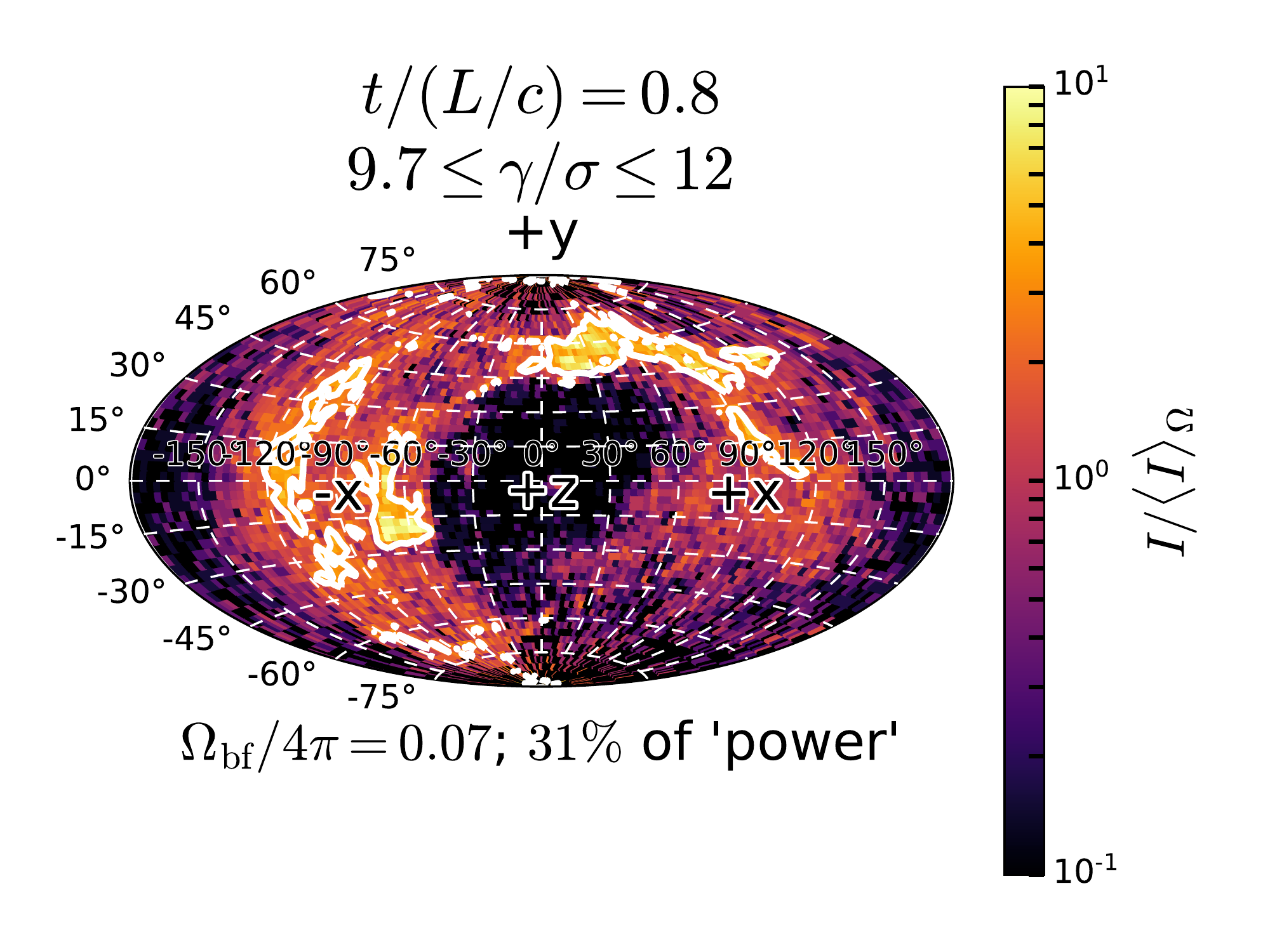}
  \end{subfigure}
  \begin{subfigure}{0.33\textwidth}
    \includegraphics[width=\linewidth]{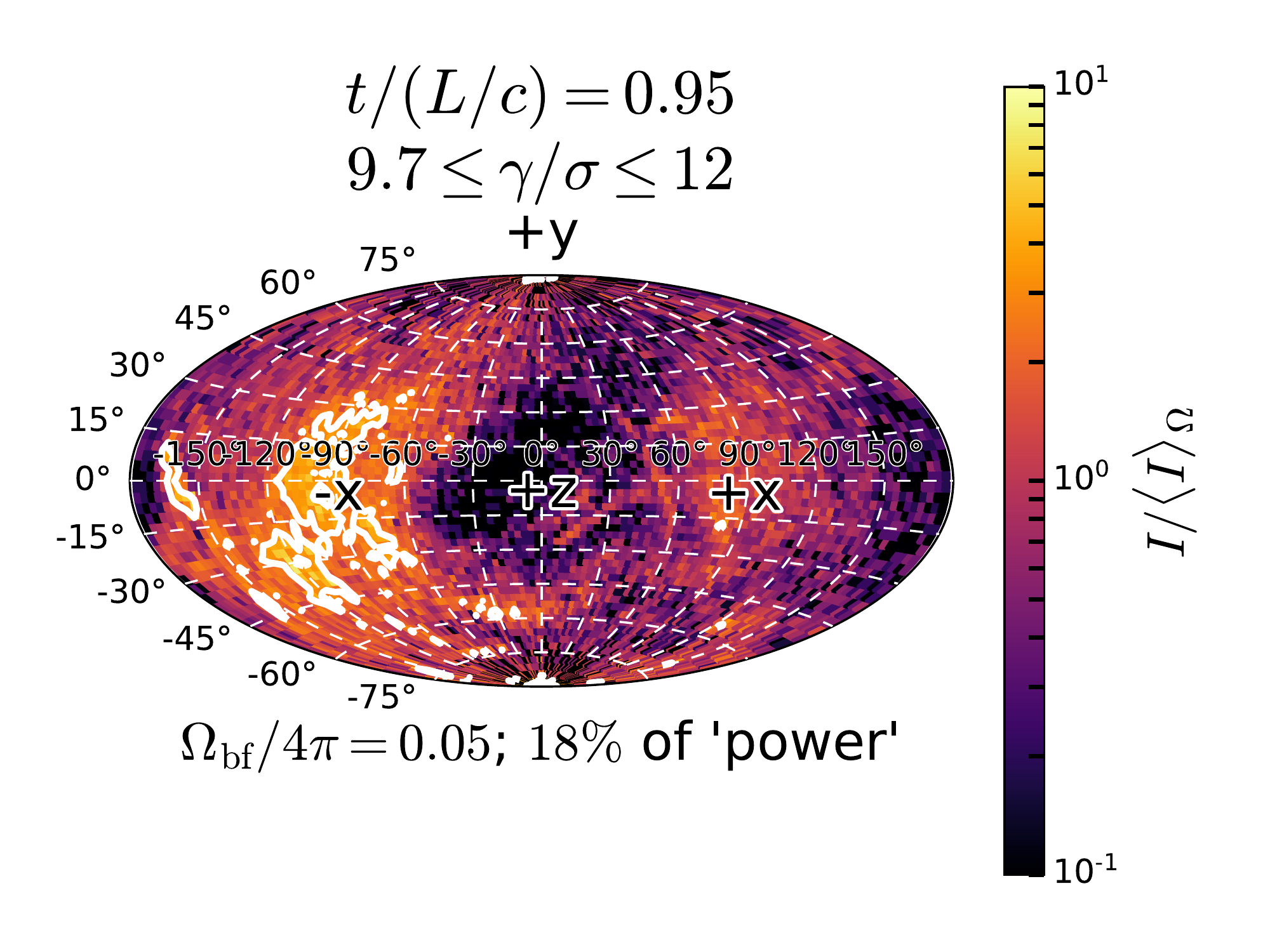}
  \end{subfigure}
  \begin{subfigure}{0.33\textwidth}
    \includegraphics[width=\linewidth]{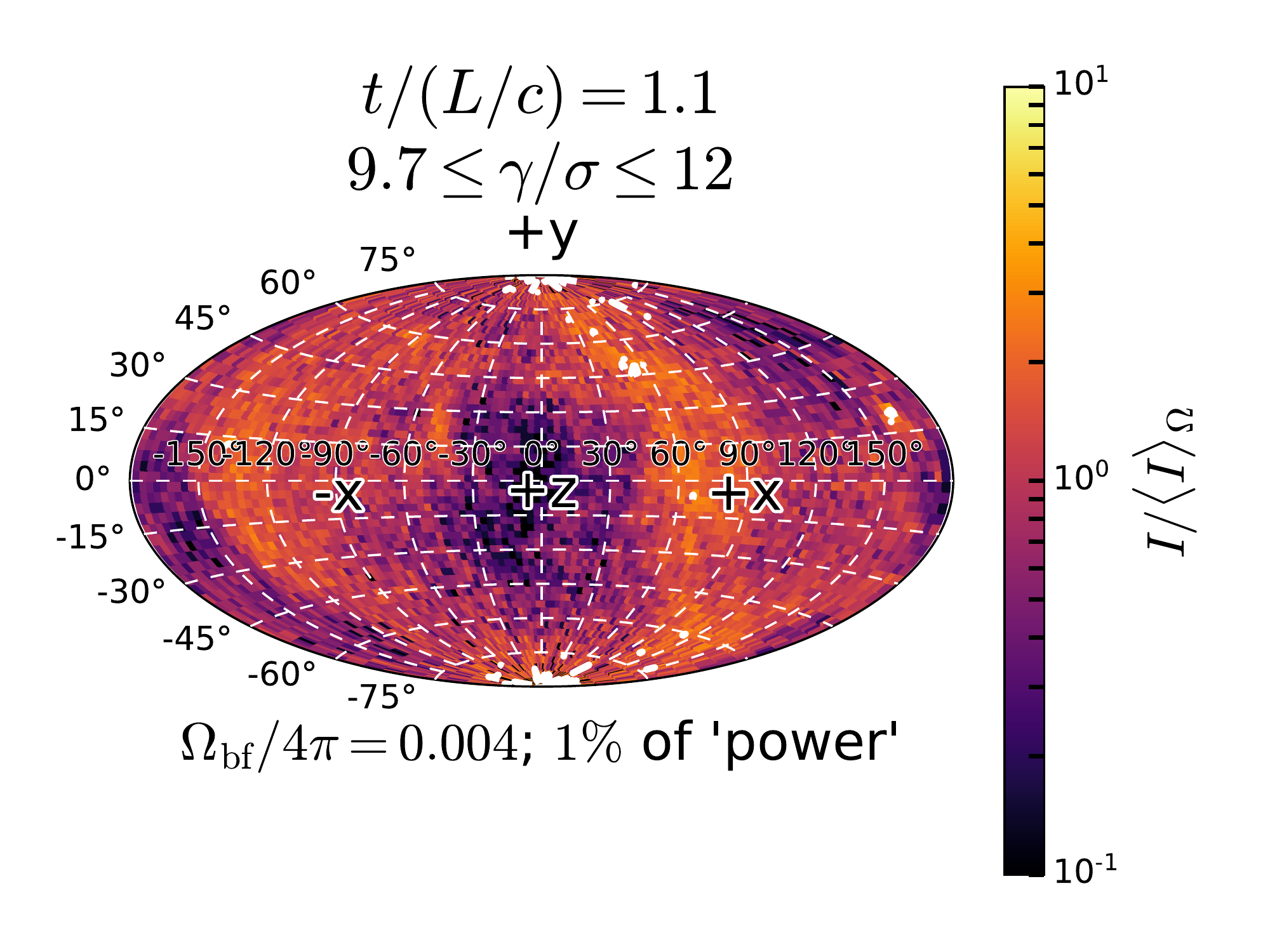}
  \end{subfigure}
  \caption{Intensity maps for an exemplary high-energy bin across several times for the~$\grad / \sigma = \infty$ simulation. Beamed fraction contours are in white and beamed fraction values are labelled below each map. Initially, the energy bin contains no particles; only around~$0.36L/c$ are particles first accelerated into the bin. Although initially beamed, in the absence of cooling, the particles maintain their energies and isotropize over time. Later, newly accelerated and collimated particles continue to appear in this energy band, but their contribution is drowned out by older, isotropized particles. The generic behaviour of the high-energy uncooled particle distribution may be summarized as follows: high-energy bands start empty, they become briefly beamed when they first acquire particles, but they then relax to an isotropic state.}
  \label{fig:heatmapintime}
\end{figure*}

Returning now to Fig.~\ref{fig:BeamFracVsT_norad}, one sees that at later times, beaming is quenched. After one \lc[,]both the~$bf$ and~$\Omega_{50}$ curves approach their isotropic values --~$0$ and~$0.5 \times 4 \pi$, respectively -- across nearly all particle energies. As discussed below, this occurs because, after their initial energization, particles quickly isotropize due to gyration about reconnected magnetic field lines. At first glance, it may appear that the highest Lorentz factors -- those near the cut-off in the particle distribution -- are exceptions to this rule, with dramatic beaming occurring even at late times.
This is not really a persistent effect, however, because beaming lasts only temporarily at any fixed Lorentz factor, beginning when the high-energy cut-off crosses (from below to above) that particular energy and ending shortly thereafter. Evidently, high-energy bands retain their beaming only until they may be populated by a significant number of particles. The reason for this is illustrated in Fig.~\ref{fig:heatmapintime}, and we discuss it here.

High-energy bands in the particle distribution begin the simulation empty. At some point, X-point energization begins to populate such a band with particles. These \quoted{young}(recently accelerated) particles are beamed in the same way that particles emerge beamed from the reconnection layer in the early part of the simulation \citep[$t < L/c$, as observed previously by][who confined their analysis to early times]{cwu12}. In the absence of radiative cooling, however, particles essentially remain in the energy band to which they are first accelerated. Meanwhile, they settle into plasmoids, where magnetic gyromotion isotropizes their momentum distribution. After a while, a given high-energy band is dominated by \quoted{old}particles whose angular spread has lost the beaming imprint left by X-points. Though a few young particles may still be injected into the energy band, their contribution to the angular intensity is washed out by the large number of older particles that has already accrued there. As a result, the high-energy parts of the particle distribution contain brief, intense beaming when they first acquire particles, but subsequently isotropize as older, increasingly isotropic particles begin to pile up.

We have simplified this explanation by considering particle acceleration to be dominated by the impulsive X-point mechanism. Alternative slower and more isotropic acceleration channels have been studied, for example, by \citet{ps18}, \citet{gld19}, and \citet{hps20}. However, these should be suppressed in the strongly radiative regime, unable to keep pace with the rapid cooling of the most energetic particles. Moreover, even when radiative losses are weaker (as in the present case~$\grad / \sigma = \infty$) and these slower mechanisms are more likely to operate, they can only serve to reduce the amount of beaming we measure, tending to swamp the highly anisotropic angular signatures produced at X-points. Practically speaking, this means that, while an isotropic angular map may be the combined result of dispersing beams \textit{and} intrinsically isotropic energization mechanisms, maps indicating strong kinetic beaming can only be attributed to X-point acceleration.

\subsection{Strong cooling: $\grad / \sigma = 1$}
We now turn to the case of strong radiative cooling~$\grad / \sigma \lesssim 1$, analysing our~$\grad / \sigma = 1$ simulation in detail. In this regime, the radiative cut-off $\grad$ is not far above the typical energy~($\sigma$) -- and well below the maximum energy~\citep[e.g. several~$\sigma$,][]{wuc16} -- that an energized particle would have in the absence of cooling. Thus,~$\grad$ chops off the part of the \nonthermal power-law tail that could otherwise extend to energies above~$\grad$ (see Fig.~\ref{fig:tdpdists} and the surrounding discussion).

\begin{figure}
  \centering
  \includegraphics[width=\columnwidth]{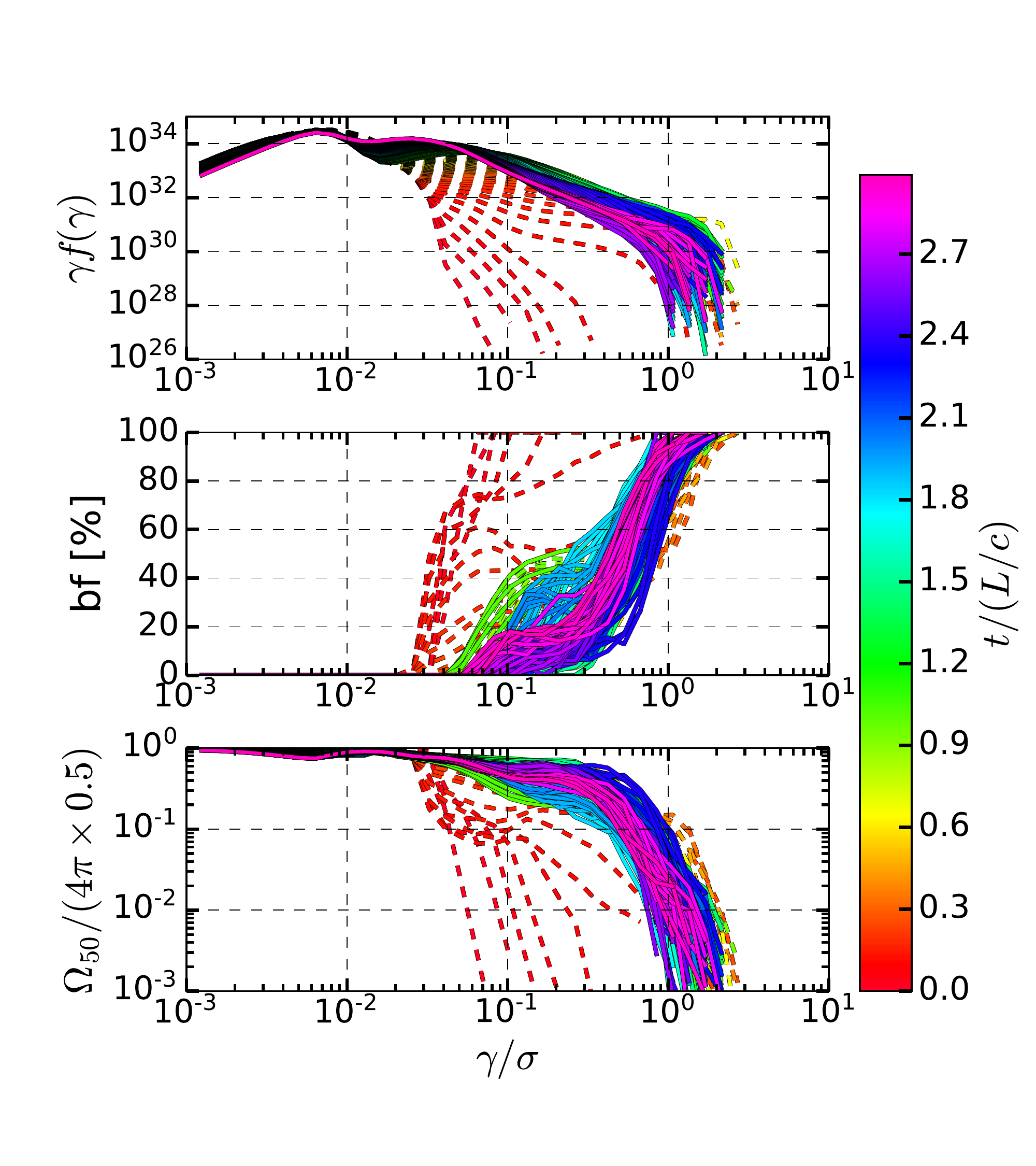}
  \caption{\revtext{(Figure updated/changed from original manuscript.)} The same as Fig.~\ref{fig:BeamFracVsT_norad} but for our simulation with~$\grad / \sigma = 1$ (strongly cooled). Unlike the non-radiative case, beaming is present across a moderate range of energies and persists well beyond~$t=L/c$. The envelope of the late-time beamed fraction curves indicates at least mild kinetic beaming across a decade in particle energies. The strongest beaming occurs over a somewhat smaller range, where the beamed fraction curves begin to rise steeply and the~$\Omega_{50}$ curves begin to turn over.}
  \label{fig:BeamFracVsT_rad}
\end{figure}

\begin{figure}
  \centering
  \includegraphics[width=\columnwidth]{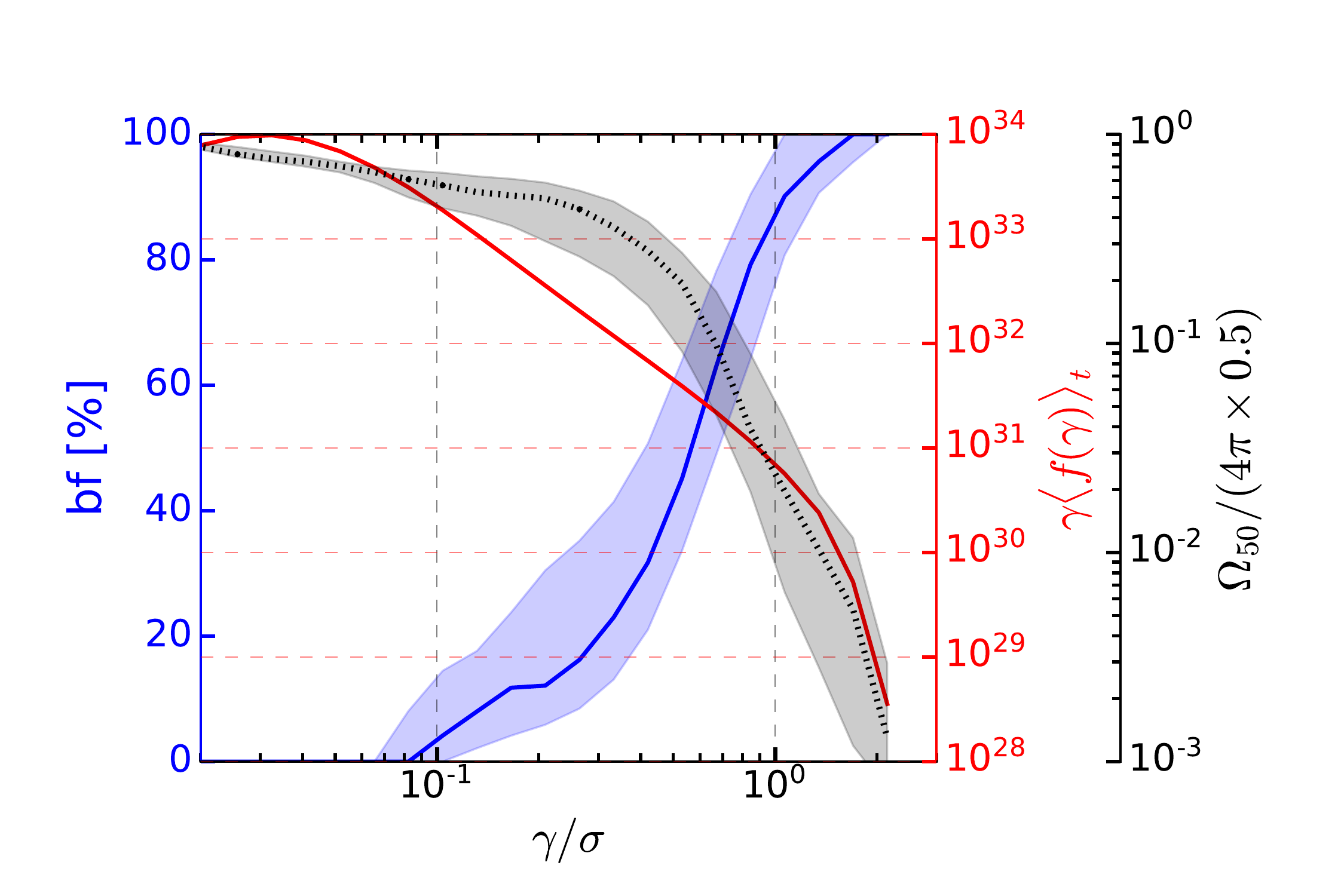}
  \caption{\revtext{(Figure updated/changed from original manuscript.)} The time-averaged electron distribution (red solid line), median electron beamed fraction (blue solid line), and median electron~$\Omega_{50}$ (black dotted line) as a function of particle energy for times between~$L/c$ and~$3L/c$ in our~$\grad / \sigma = 1$ simulation. The shaded beamed fraction and~$\Omega_{50}$ envelopes indicate the middle~$68 \rm \, per \, cent $ of the time series data at each particle energy. As discussed in Fig.~\ref{fig:BeamFracVsT_rad}, but somewhat more obvious here, the beamed fraction indicates at least weak beaming across a decade on the horizontal axis. Both beamed fraction and~$\Omega_{50}$ indicate strong beaming over a slightly narrower energetic range.}
  \label{fig:BeamFracAvgT_ptcl_1lc} 
\end{figure}

Paralleling our treatment of the non-radiative case, Fig.~\ref{fig:BeamFracVsT_rad} plots for the simulation with~$\grad / \sigma = 1$ what Fig.~\ref{fig:BeamFracVsT_norad} plots for the simulation with~$\grad / \sigma = \infty$. Here, strong cooling causes the electron distribution, beamed fraction, and~$\Omega_{50}$ to depart from their previous behaviour, where they essentially grew monotonically in horizontal or vertical extent.
Because this makes it hard to discern the time evolution in Fig.~\ref{fig:BeamFracVsT_rad}, we also supply Fig.~\ref{fig:BeamFracAvgT_ptcl_1lc}, which presents the post-one-light-crossing time-averaged particle distribution, median beamed fraction, and median~$\Omega_{50}$.
As evident from Fig.~\ref{fig:BeamFracAvgT_ptcl_1lc}, beaming persists, when cooling is strong, to late times across almost a decade in particle energy. In contrast, as we saw in the previous section, kinetic beaming is only transient when cooling is weak; it is present across a wide range of energies at early times and thereafter relegated to energies near the cut-off of the particle distribution. 

\begin{figure*}
  \centering
  \begin{subfigure}{0.33\textwidth}
    \includegraphics[width=\linewidth]{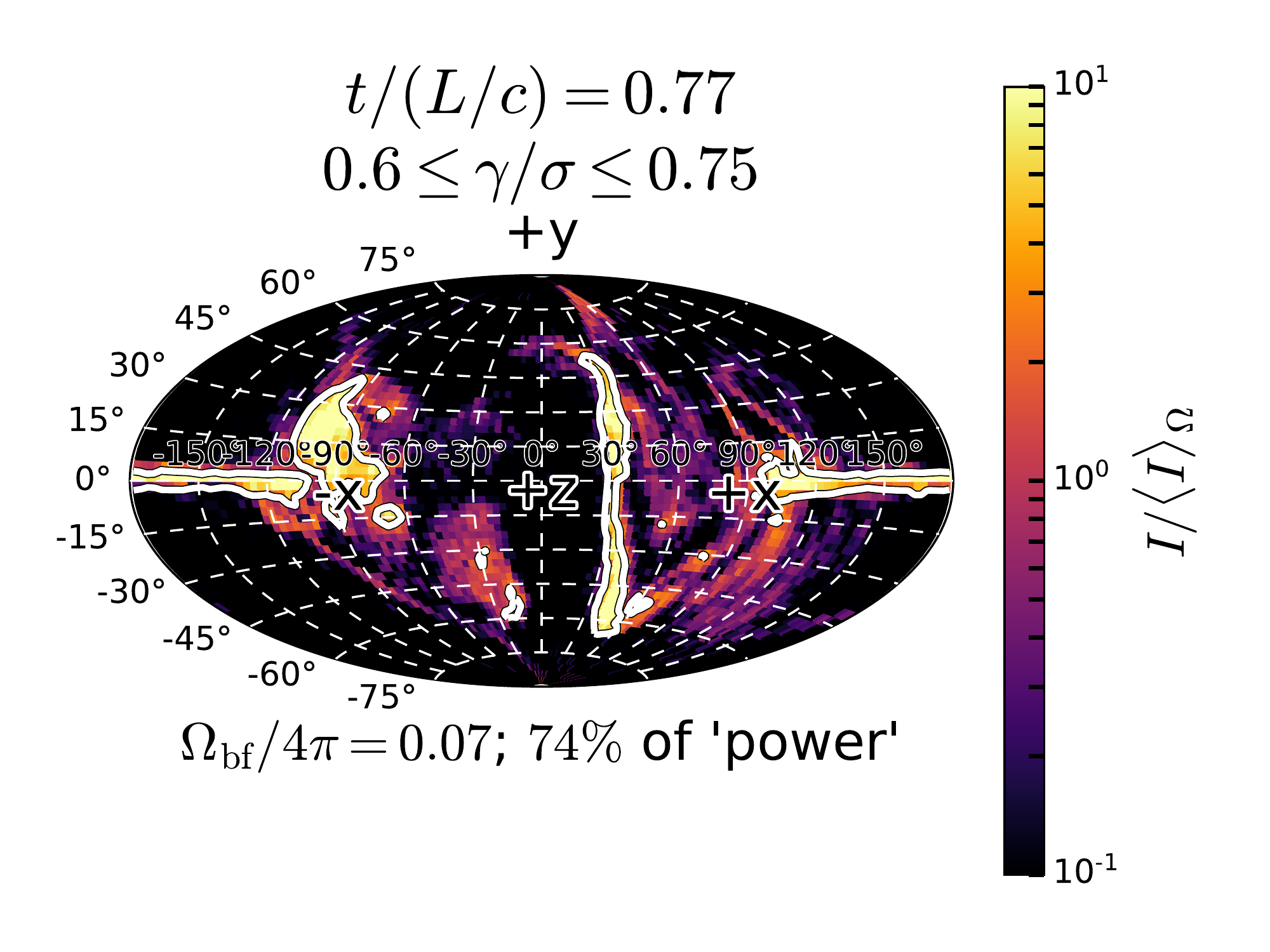}
  \end{subfigure}
  \begin{subfigure}{0.33\textwidth}
    \includegraphics[width=\linewidth]{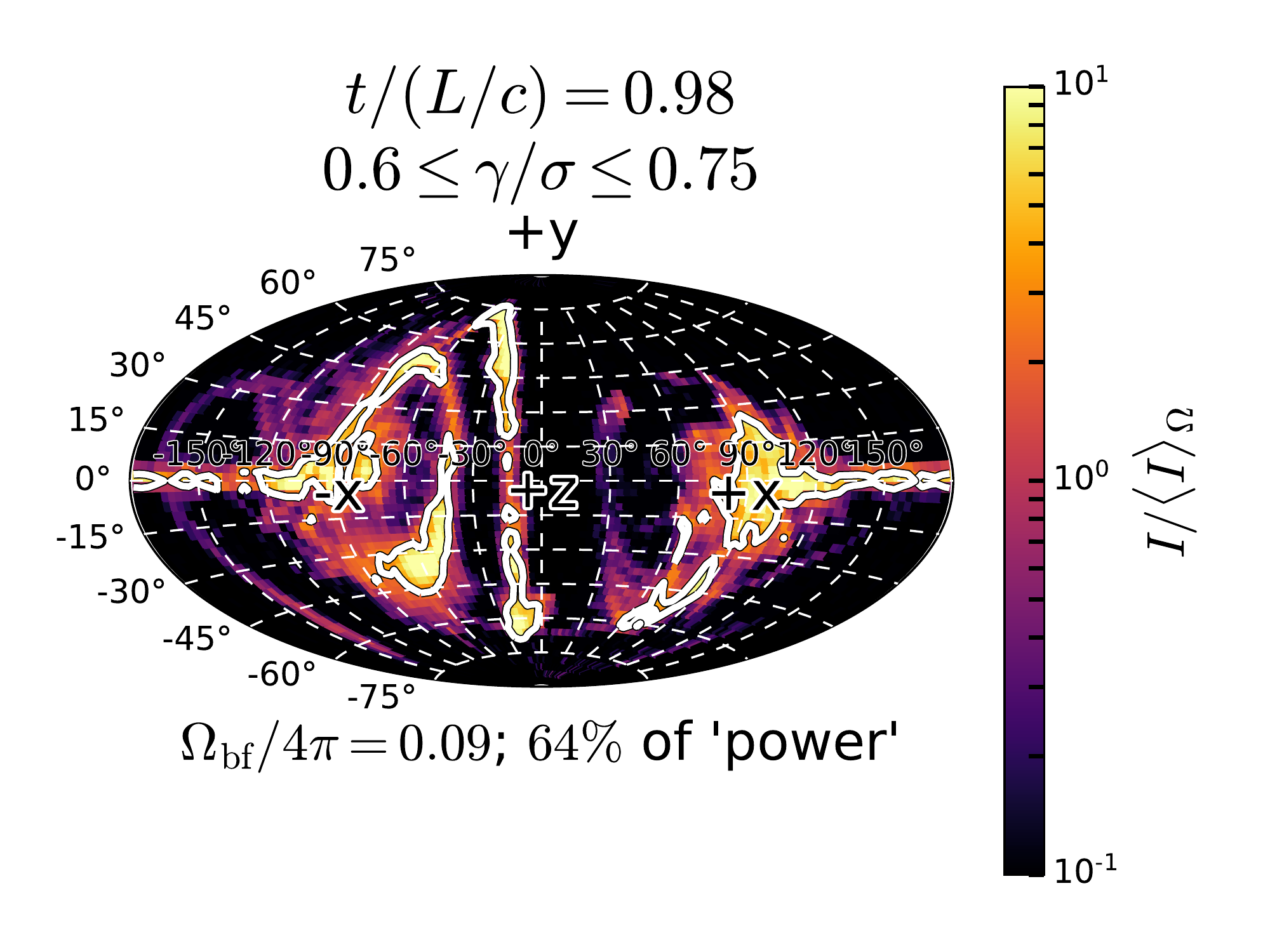}
  \end{subfigure}
  \begin{subfigure}{0.33\textwidth}
    \includegraphics[width=\linewidth]{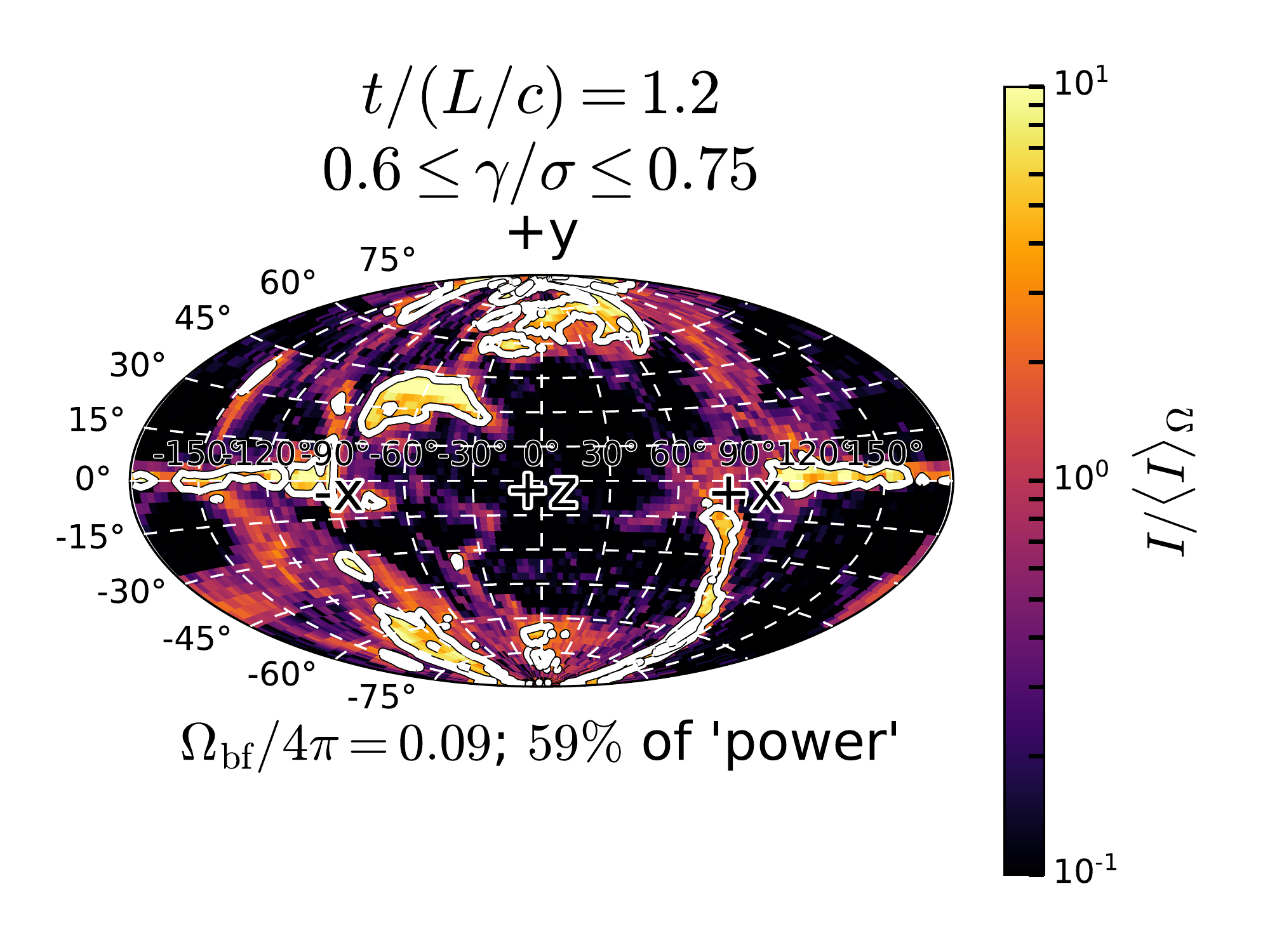}
  \end{subfigure} \\
  \begin{subfigure}{0.33\textwidth}
    \includegraphics[width=\linewidth]{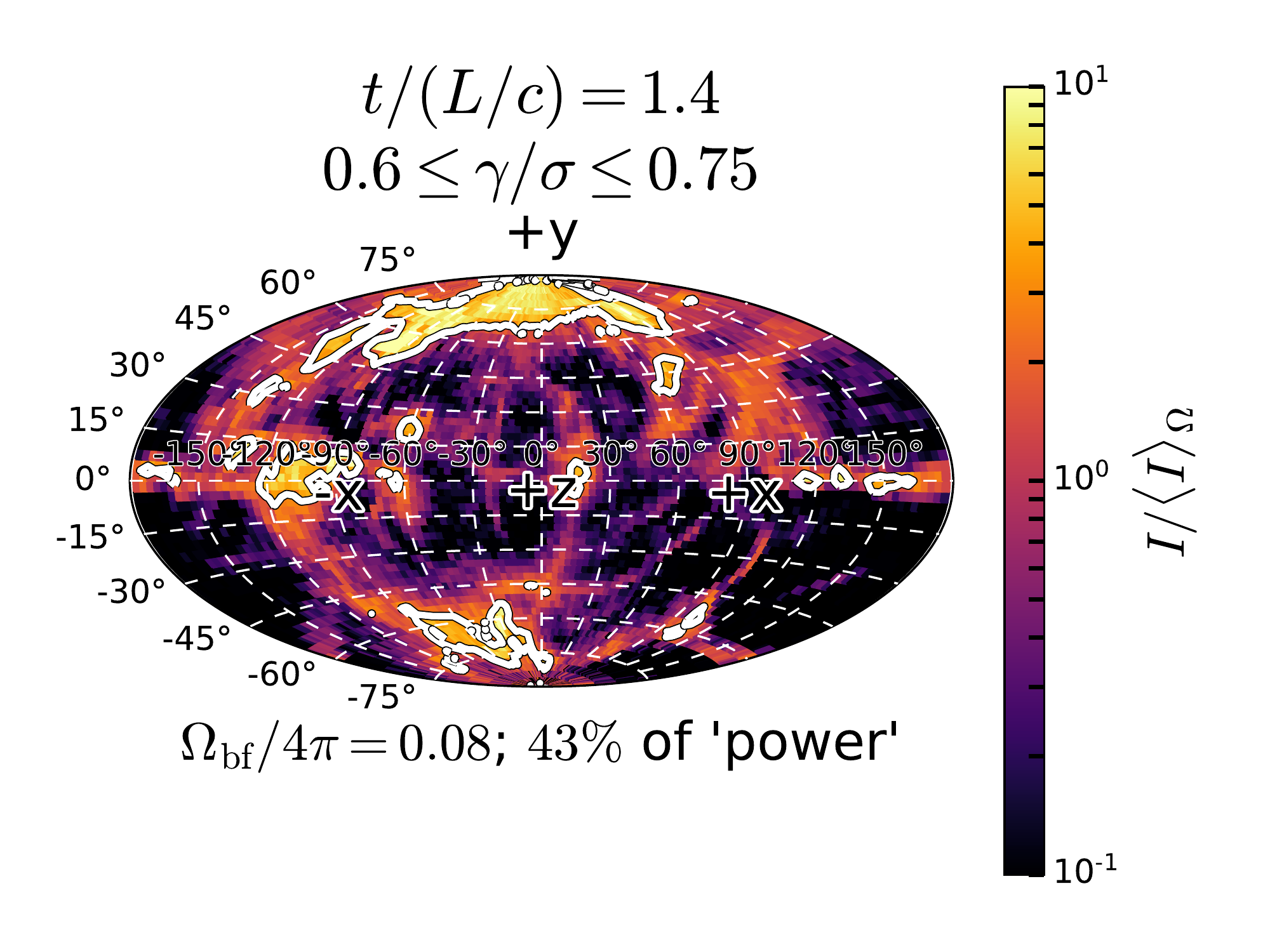}
  \end{subfigure}
  \begin{subfigure}{0.33\textwidth}
    \includegraphics[width=\linewidth]{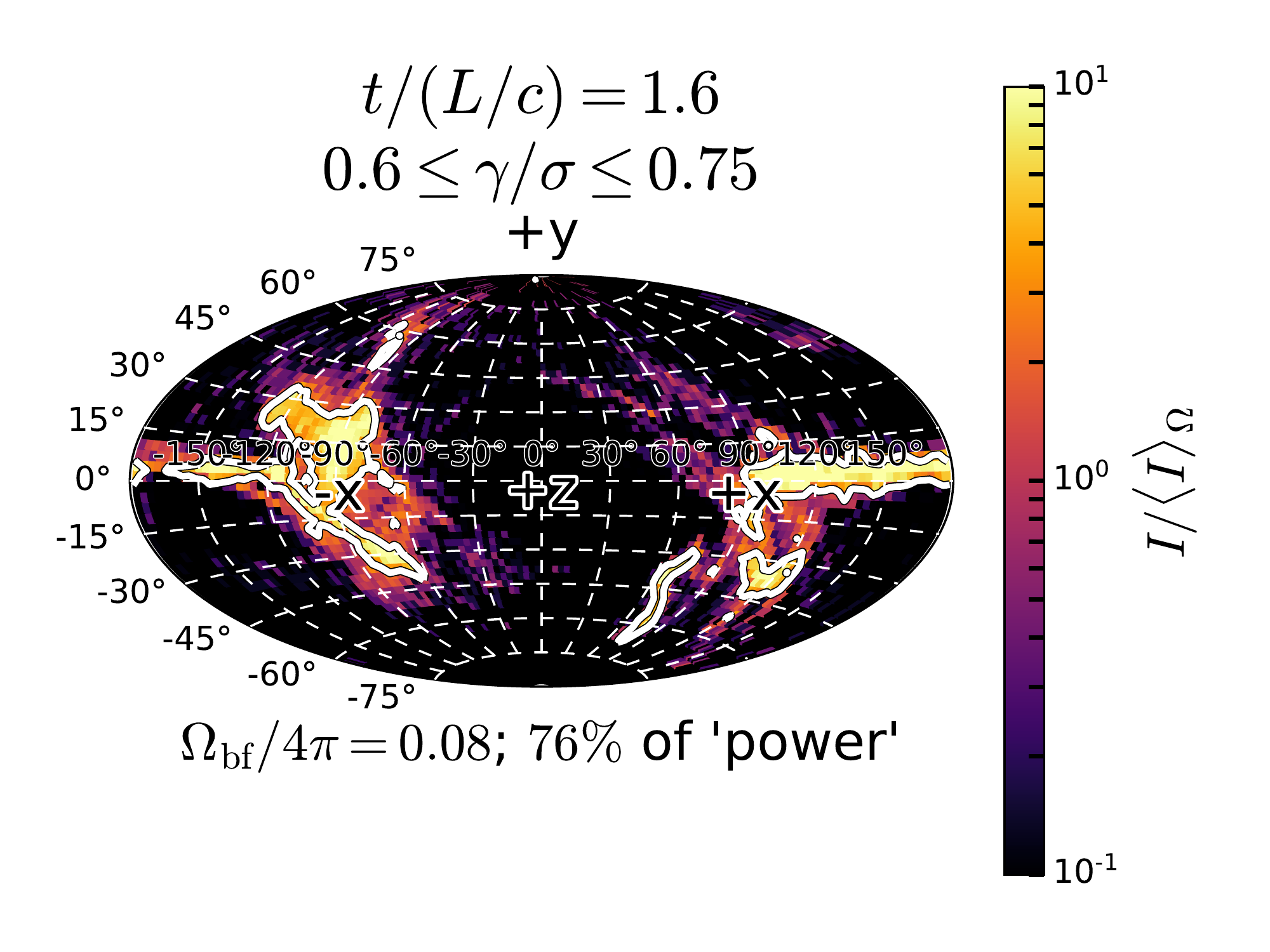}
  \end{subfigure}
  \begin{subfigure}{0.33\textwidth}
    \includegraphics[width=\linewidth]{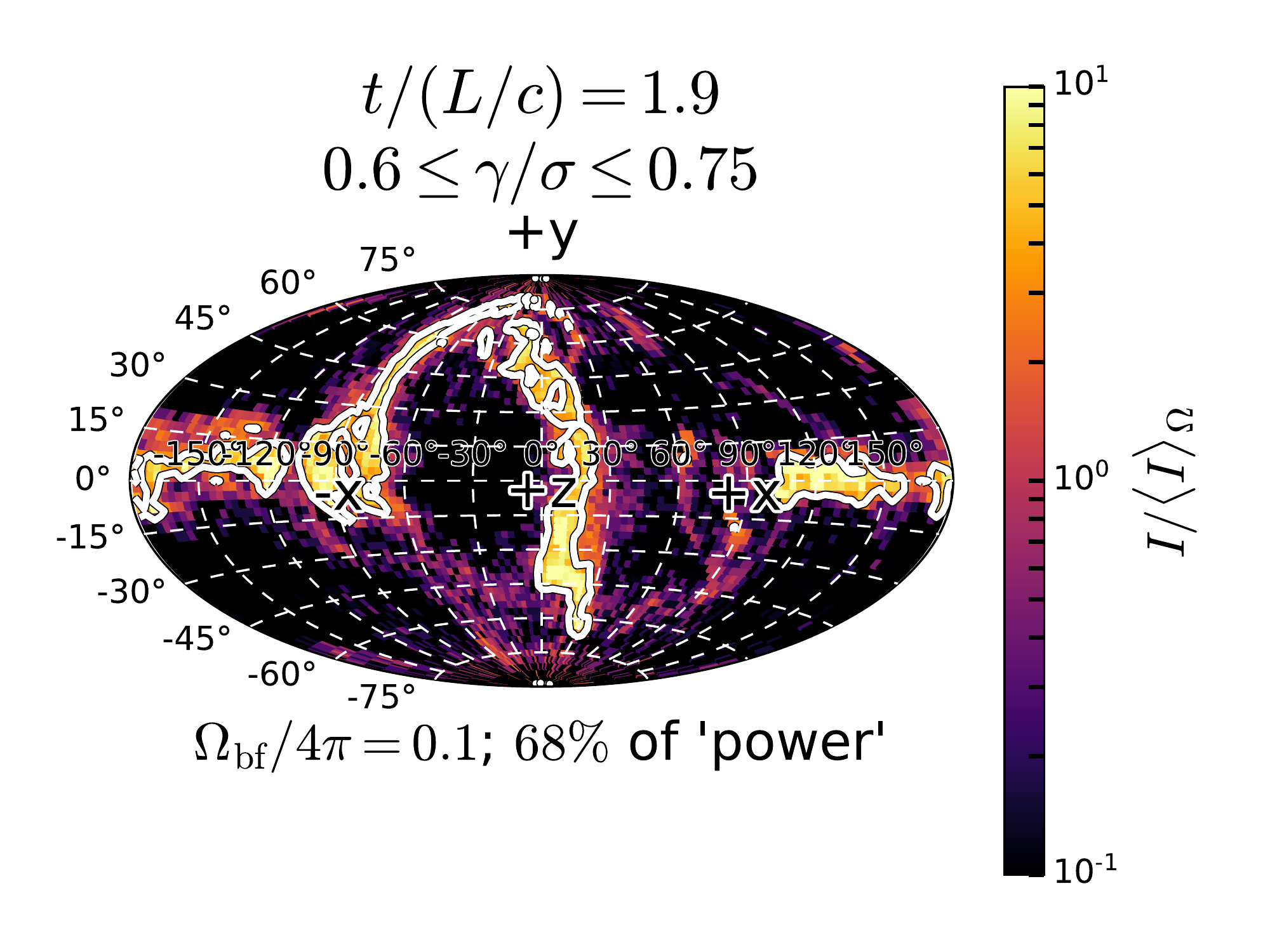}
  \end{subfigure}
  \caption{\revtext{(Figure updated/changed from original manuscript.)} The time-dependence of a high-energy heatmap for the~$\grad / \sigma = 1$ simulation. In the uncooled simulation, a similar set of figures (Fig.~\ref{fig:heatmapintime}) demonstrated the transience of beaming in that energy bin. But in this strongly cooled case, spectacular beaming patterns persist to times well past~$t = L/c$. Note that sometimes (particularly~$t = 1.6L/c$) X-points in the primary reconnection layer dominate the beaming, producing horizontal swaths of high intensity regions; at other times, the prominent vertical swaths indicate beaming produced at X-points between merging plasmoids (cf. Section~\ref{sec:angmaps} discussion).}
  \label{fig:heatmapintime_GradSigma1}
\end{figure*}
As is the case without radiative losses, acceleration from reconnection X-points preferentially collimates the more energetic particles into beams. The crucial difference with strong cooling is that particles radiate away most of their energy before they have had time to isotropize: the most energetic particles are always \quoted{young}(recently accelerated). As a result, the high-energy part of the particle distribution remains beamed at much later times (see Fig.~\ref{fig:heatmapintime_GradSigma1}). Beaming falls off with decreasing particle energy, however, because particles that have been cooling longer have also been isotropizing longer (they are \quoted[).]{older}

In principle, whether kinetic beaming persists at a given particle energy comes down to whether the isotropization \ts for those particles is longer or shorter than their cooling \ts[.] Suppose, for illustration, that the particle isotropization \ts[] is the gyration period~$t_{\rm iso} \sim \gamma / \omega_0$ where the nominal Larmor frequency is~$\omega_0 = e B_0 / m_{\rm e} c$. The cooling \ts[]for the same particles is~${t_{\rm IC} \sim \gamma m_{\rm e} c^2 / P_{\rm IC}(\gamma) \sim 10 \grad^2 / \omega_0 \gamma}$. One expects the smallest Lorentz factor $\gamma_{\rm iso}$ for which kinetic beaming persists to late times to be that for which these \ts[s]are equal:~$\gamma_{\rm iso} \sim \sqrt{10} \grad$. This picture is oversimplified, for it predicts~$\gamma_{\rm iso}$ to exceed (somewhat) the radiation-reaction limit~$\grad$, and it predicts $\grad / \gamma_{\rm iso} \sim \rm \, constant$, while we find (see Section~\ref{sec:kinbeamgrad}) a non-trivial scaling of~$\grad / \gamma_{\rm iso}$ with~$\grad$. What can be said for certain in the case~$\grad / \sigma = 1$ is that particles radiate more quickly than they isotropize over a considerable range of energies.

\subsection{Kinetic beaming as a function of radiative efficiency}
\label{sec:kinbeamgrad}
Having demonstrated that strong kinetic beaming persists in the presence of efficient radiative cooling but disappears after about~$t = L / c$ when cooling is negligible, we now analyse the transition between these regimes. In particular, we examine how sustained kinetic beaming weakens as the result of decreasing IC radiative efficiency. We also shift our focus from the angular particle distribution~$\dif N_t/ \dif \gamma \dif \Omega$ to the IC emission spectrum~$\dif P_t/ \dif \epsilon \dif \Omega$ where~$\epsilon$ and~$\Omega$ are the energy and direction, respectively, of IC photons.\footnote{The quantity~$\dif P_t/\dif \epsilon \dif \Omega$ is the instantaneous (at time~$t$), lower layer, volume-integrated IC emission coefficient~$j_{\rm IC}$, with~$j_{\rm IC}$ as defined by \citet{rl79}:~${\dif P_t/ \dif \epsilon \dif \Omega = \int_{y < L_y / 2} \dif x \dif y \, j_{\mathrm{IC}}(x, y, \epsilon, \pmb{\Omega}; t)}$.} This presents no challenge from an analysis standpoint, since the diagnostics we have been using so far (angular maps, beamed fraction,~$\Omega_{50}$) apply as well to photons as to particles. In fact, analysing the photons themselves rather than the emitting particles enables a more precise measurement of kinetic beaming, a point that we now briefly elaborate.

In this section, we shall be interested in the energetic extent of kinetic beaming: the range of (high) photon or particle energies across which a strong energy-dependent anisotropy is evident. Now, in the Thomson regime, photons Comptonized by a particle of Lorentz factor~$\gamma$ attain increased energies by the factor~$\gamma^2$ and are emitted along the particle's velocity vector within a cone of half-opening angle~$1/\gamma$. So, in the ultrarelativistic limit~($\gamma \gg 1$), not only is the angular distribution of emission nearly identical to that of the radiating particles, but kinetic beaming actually encompasses a wider range of energies in the former than in the latter. Therefore, treating the IC emission rather than the radiating particles directly allows us to measure more precisely the energetic extent of kinetic beaming as it becomes small.

\begin{figure*}
  \centering
  \begin{subfigure}{0.33\textwidth}
    \includegraphics[width=\linewidth]{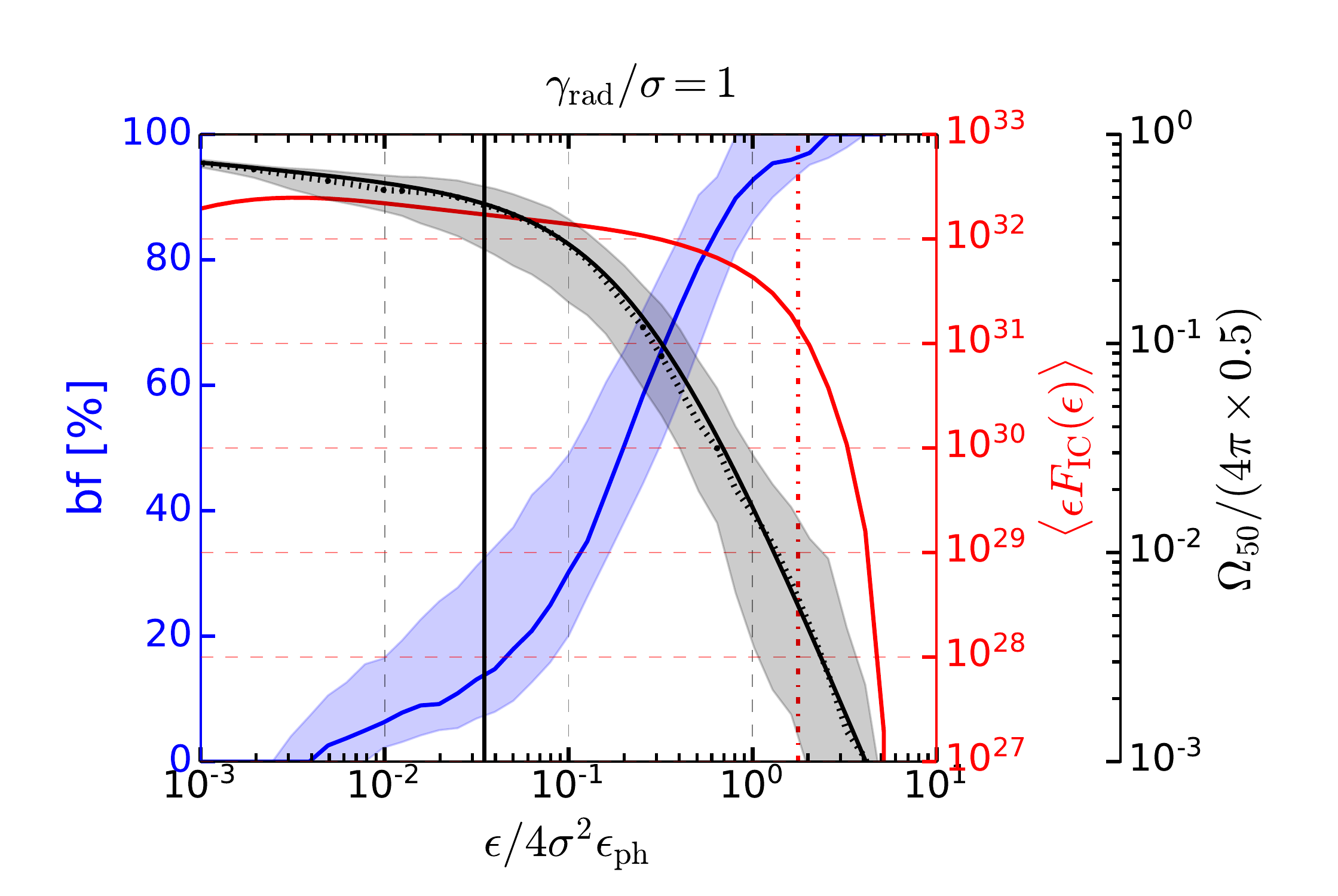}
  \end{subfigure}
  \begin{subfigure}{0.33\textwidth}
    \includegraphics[width=\linewidth]{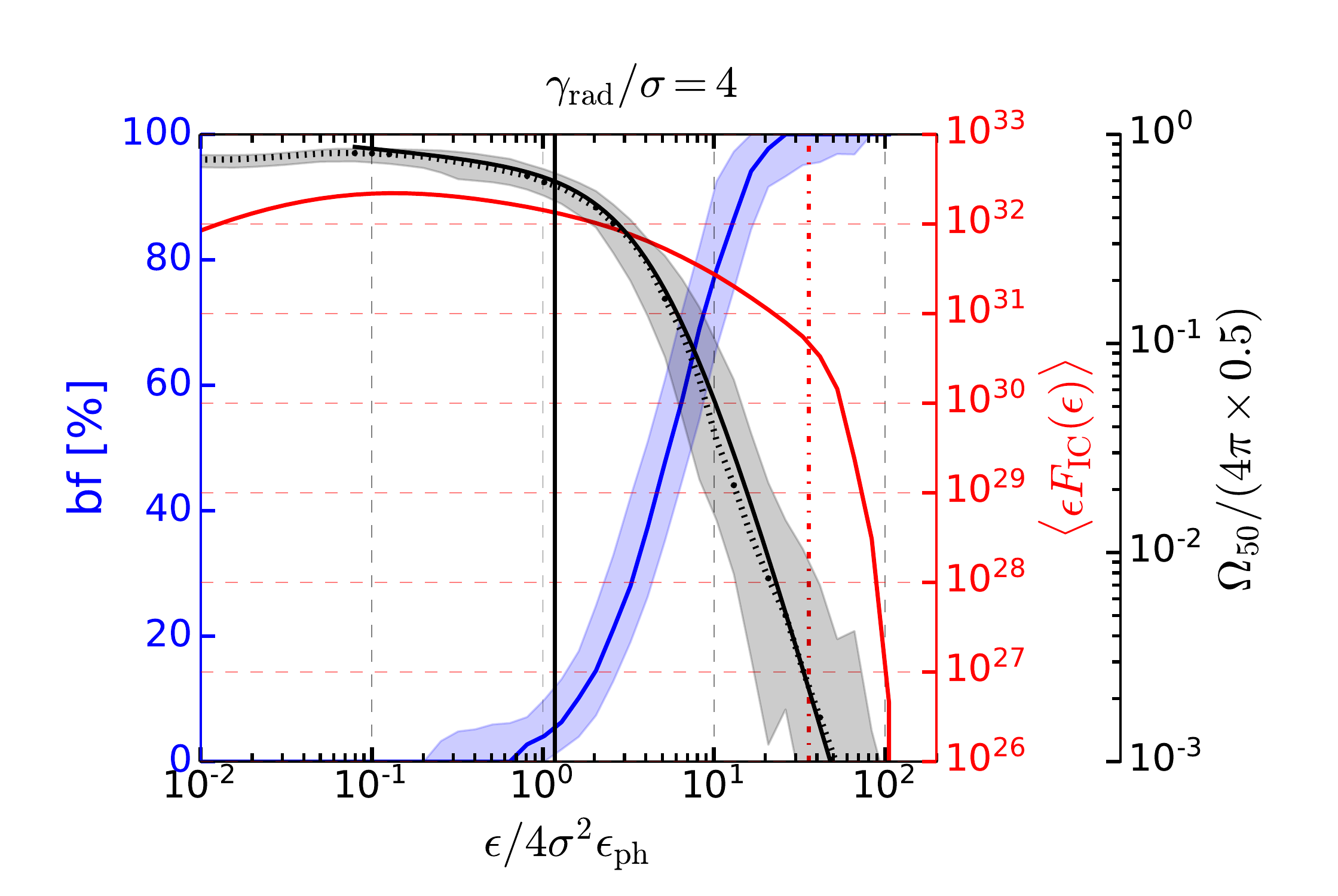}
  \end{subfigure}
  \begin{subfigure}{0.33\textwidth}
    \includegraphics[width=\linewidth]{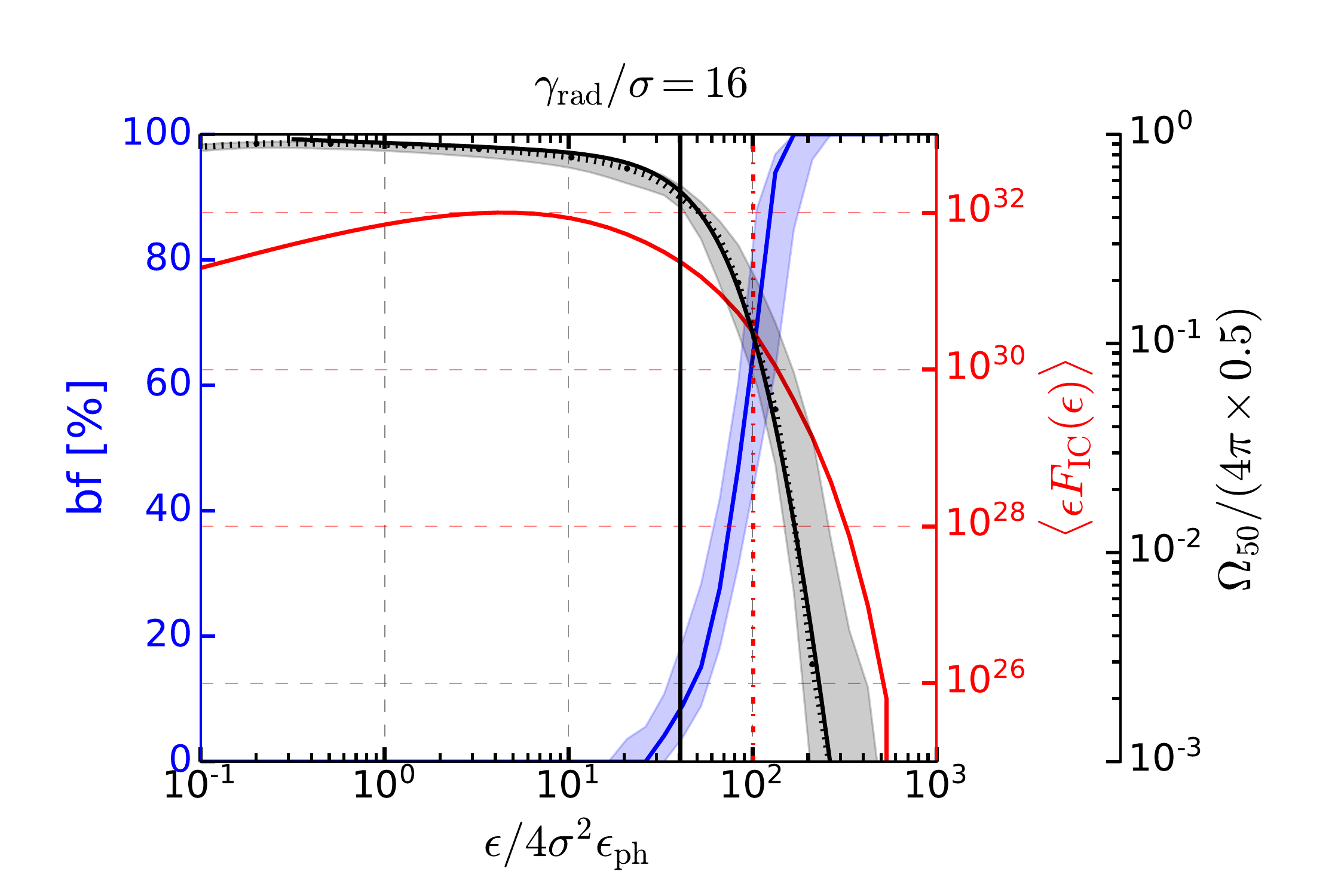}
  \end{subfigure}
  \caption{\revtext{(Figure updated/changed from original manuscript.)} The time-averaged (over~$1 \leq ct/L \leq 3$) electron IC emission spectrum (solid red), median beamed fraction~($bf$; solid blue), and median~$\Omega_{50}$ curve (dotted black) for the simulations with~$\grad / \sigma = 1$ (left),~$4$ (middle), and~$16$ (right). Shaded regions indicate, for~$\Omega_{50}$ and~$bf$ only, the middle~$68 \rm \, per \, cent $ of data. In each panel, the~$\Omega_{50}$ data (dotted black lines with shaded regions serving as error bars) are fit using a smoothly broken power law parametrized as in equation~(\ref{eq:sbpfit}). To avoid needing a more complicated fitting formula, only data following the last local maximum in the dotted~$\Omega_{50}$ curve are fit. The fit is drawn as a solid black line on top of the data that were used. Solid black vertical lines indicate the onset of the best-fitting spectral break~($\epsilon_{\rm iso}$ in the text); dot--dashed red vertical lines show the spectral cut-off~($\epsilon_{\rm c}$ in the text) in the displayed emission spectrum. The horizontal axis is normalized to the maximum photon energy~$4 \sigma^2 \eph$ to which a particle of Lorentz factor~$\sigma$ can upscatter $\eph$-energy photons.}
  \label{fig:omxplaw}
\end{figure*}

In Fig.~\ref{fig:omxplaw}, we illustrate beaming as a function of photon energy for three of our radiative runs~(${\grad / \sigma = 1, 4, \, \mathrm{and} \, 16}$). In that figure, one sees that kinetic beaming -- marked by rising~$bf$ and declining~$\Omega_{50}$ at the highest energies -- persists well beyond the first \lc[]time in all simulations. Also, the energy range across which beaming is kinetic widens for the more strongly radiative simulations.
Fig.~\ref{fig:omxplaw} quantifies these observations by displaying two characteristic photon energies. The first is the energy~$\epsilon_{\rm iso}$ above which beaming acquires pronounced spectral dependence -- where the~$\Omega_{50}$ curves begin to turn downward. The second is~$\epsilon_{\rm c}$, the cut-off in the IC emission spectrum. The ratio of the cut-off~$\epsilon_{\rm c}$ to the \quoted{isotropic}energy~$\epsilon_{\rm iso}$ characterizes the beamed range of photon energies.

\begin{figure}
  \centering
  \includegraphics[width=\columnwidth]{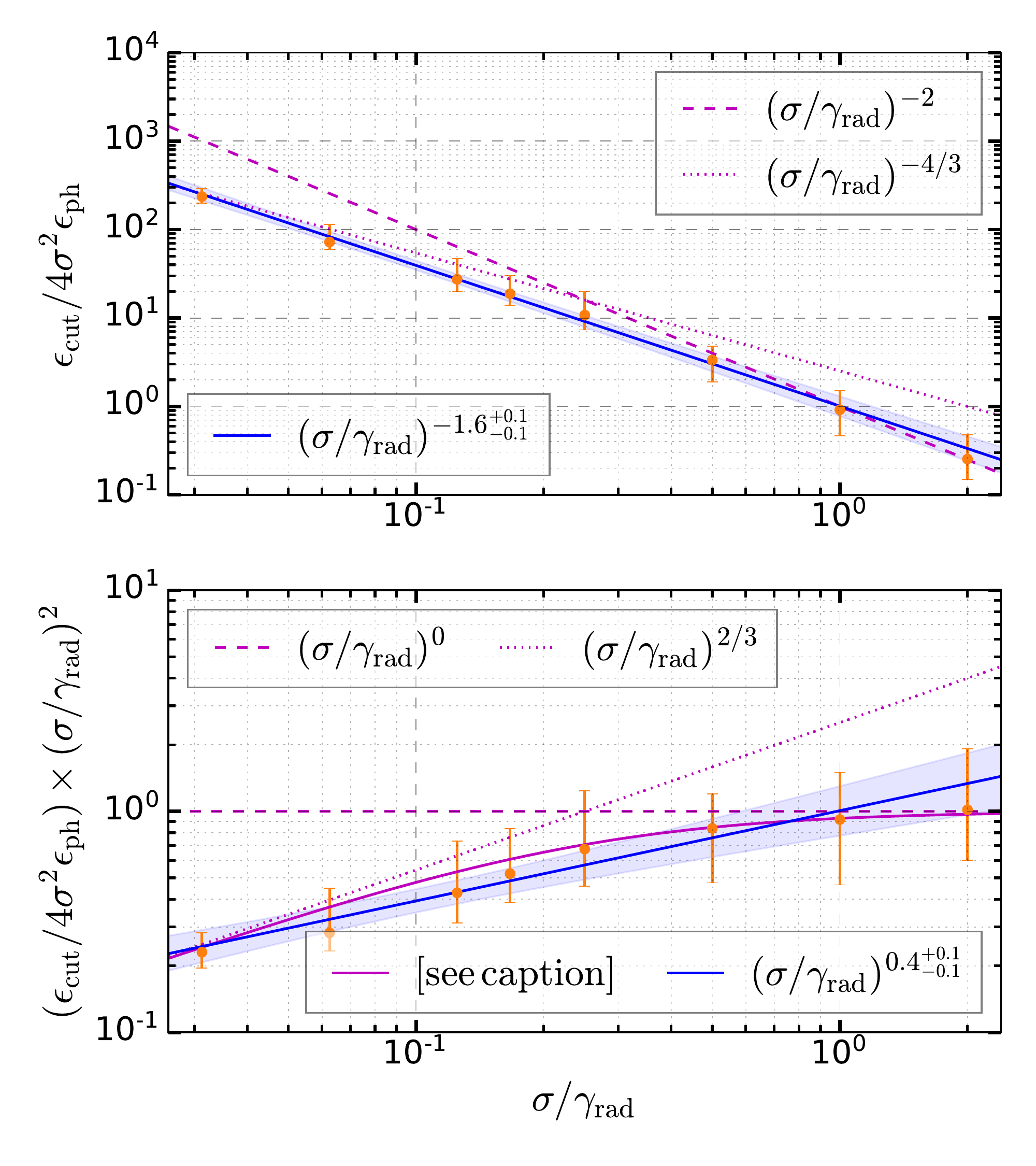}
  \caption{\revtext{(Figure updated/changed from original manuscript.)} The cut-off photon energy displayed as a function of IC cooling strength using two different normalizations. In the upper panel,~$\epsilon_{\rm c}$ is normalized to~$4 \sigma^2 \eph$; in the bottom panel, the normalization is~$4 \grad^2 \eph$. Each data point represents the median cut-off computed as a function of time (over~$1 \leq ct/L \leq 3$) for a given simulation. Error bars indicate the middle~$68 \rm \, per \, cent $ of data. A power-law fit to the data is presented in blue, with shaded blue envelope indicating the uncertainty in the fit. The expected {low-$\grad$} scaling~$\epsilon_{\rm c1} \sim \grad^{2}$ is displayed in dashed magenta and the {high-$\grad$} scaling~$\epsilon_{\rm c2} \sim \grad^{4/3}$ in dotted magenta. Additionally, the lower panel shows the empirical fitting formula~(\ref{eq:epscsmoothed}). A~$\chi^2$ goodness-of-fit test does not reject equations~(\ref{eq:epscpiecewise}) and~(\ref{eq:epscsmoothed}), or the single power law.}
  \label{fig:epscut}
\end{figure}

\begin{figure}
  \centering
  \includegraphics[width=\columnwidth]{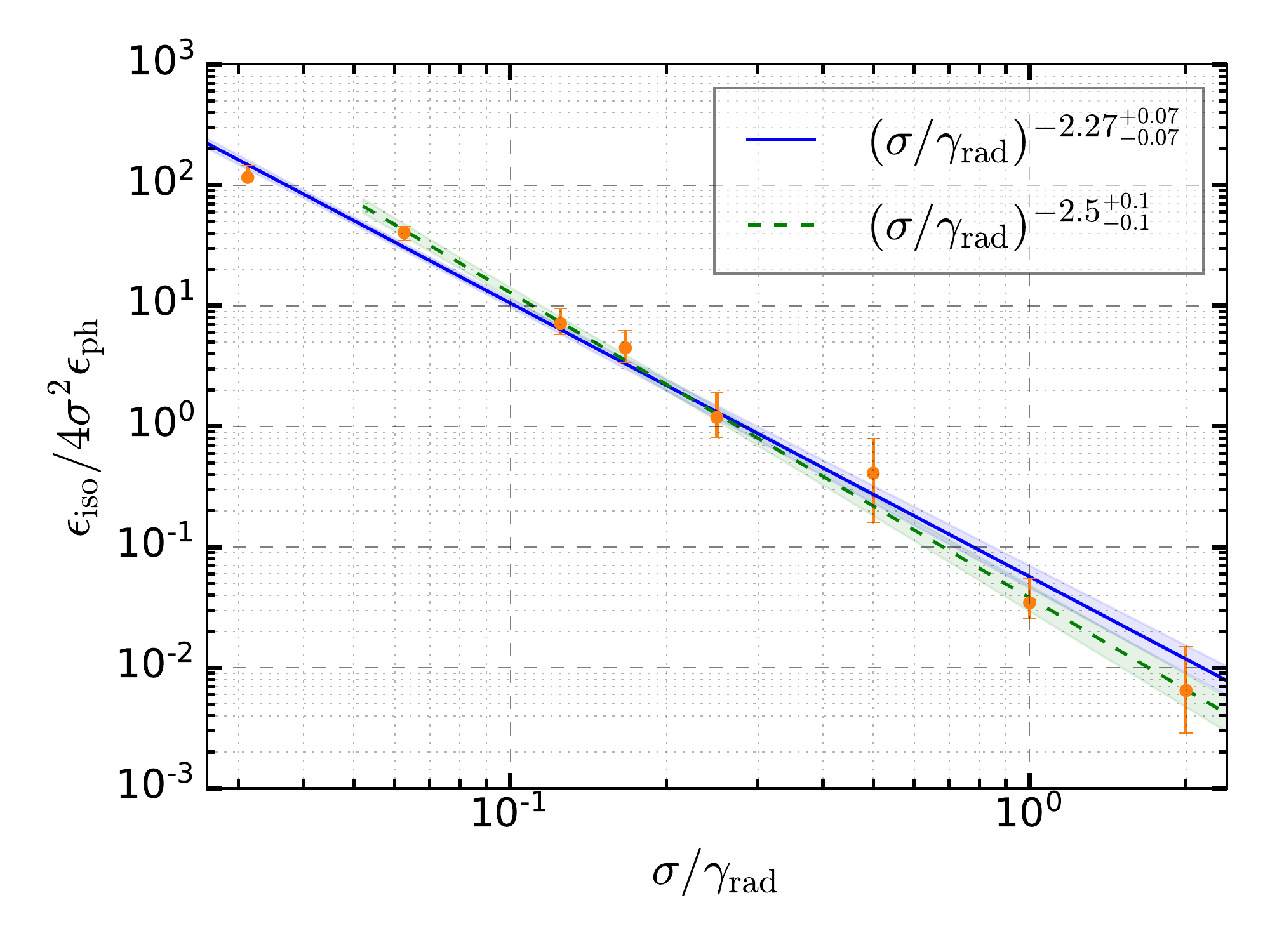}
  \caption{\revtext{(Figure updated/changed from original manuscript.)} The isotropic photon energy~$\epsilon_{\rm iso}$, which marks the transition to strong kinetic beaming in the~$\Omega_{50}$ curve, as a function of IC cooling strength. Error bars indicate~$68 \rm \, per \, cent $ confidence intervals given by Markov chain Monte Carlo fits \citep{fhg13} to each simulation's~$\Omega_{50}$ curve (see Fig.~\ref{fig:omxplaw}) using a smoothly broken power law [as parametrized in equation (\ref{eq:sbpfit})]. In this figure, the~$\epsilon_{\rm iso}$ data are fit with unbroken power laws across both the full range of data~$1/2 \leq \grad / \sigma \leq 32$ (solid blue with shaded error envelope) and a restricted range~$1/2 \leq \grad / \sigma \leq 16$ (dashed green with shaded error envelope). The restricted fit excludes the non-radiative asymptotic behaviour where~$\epsilon_{\rm iso}$ is expected to clamp to~$\epsilon_{\rm c}$.}
  \label{fig:epsiso}
\end{figure}

\begin{figure}
  \centering
  \includegraphics[width=\columnwidth]{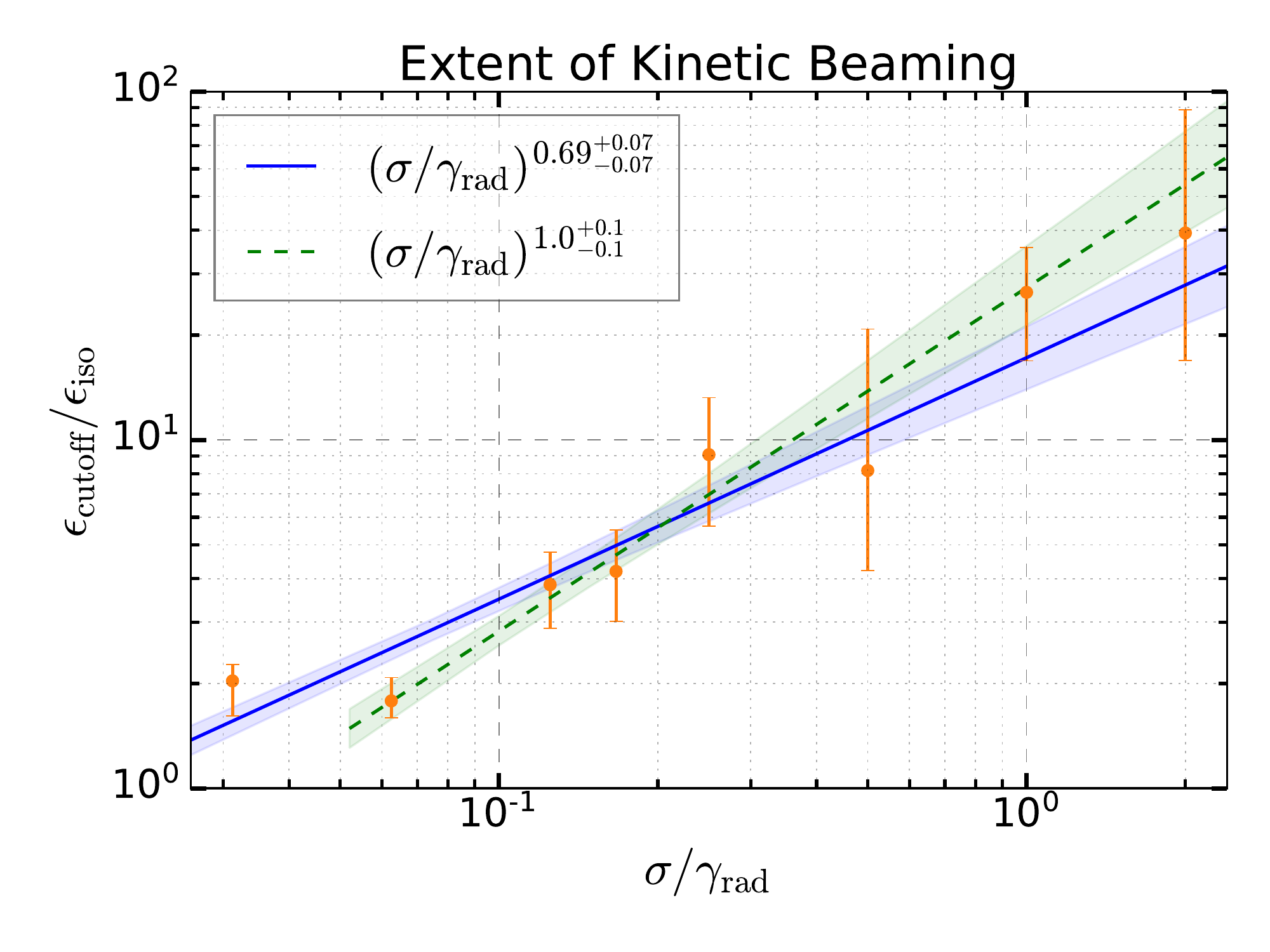}
  \caption{\revtext{(Figure updated/changed from original manuscript.)} The \quoted{range}of kinetic beaming as a function of~$\sigma / \grad$, expressed as the ratio of cut-off photon energy to photon energy of beaming onset:~$\epsilon_{\mathrm{c}} / \epsilon_{\mathrm{iso}}$. For each simulation, the photon energy~$\epsilon_{\mathrm{iso}}$ is measured as the break energy in a broken power-law fit to the~$\Omega_{50}$ curve, as in Fig.~\ref{fig:omxplaw}. Error bars presented are \quoted[,]{worst case}given by computing the ratio~$\epsilon_{\rm c} / \epsilon_{\rm iso}$ at the extreme ends of the confidence intervals in Fig.~\ref{fig:epscut} and Fig.~\ref{fig:epsiso}. The solid blue power-law fit uses the entire range of data~$1/2 \leq \grad / \sigma \leq 32$ and the dashed green power law uses the restricted range~$1/2 \leq \grad / \sigma \leq 16$. (As in Figs~\ref{fig:epscut} and~\ref{fig:epsiso}, shaded blue and green envelopes indicate fit uncertainties.) Fits omitting additional data from the weakly radiative end (high~$\grad$, low~$\sigma / \grad$) yield power-law scalings consistent with the dashed-green line, suggesting that~$\epsilon_{\rm c} / \epsilon_{\rm iso}$ may asymptote to a non-radiative limit of order unity for~$\grad \gtrsim 16 \sigma$.}
  \label{fig:epsCutDivEpsIso}
\end{figure}

These energy scales and their ratio will be critical to our eventual quantitative portrait of kinetic beaming as a function of radiative efficiency. Therefore, we will here expound upon the techniques we use to measure them as well as describe the trends in our measurements across our series of simulations. Let us begin with~$\epsilon_{\rm c}$. Following the method of equation~37 in \citet{bcs15} \citep[see also][]{sgp16, hps19}, we take
\begin{align}
  \label{eq:ecutdef}
  \epsilon_{\rm c} = \frac{\int \dif \epsilon \, \epsilon^n F_{\rm IC}(\epsilon)}{\int \dif \epsilon \, \epsilon^{n - 1} F_{\rm IC}(\epsilon)} \, ,
\end{align}
where $F_{\rm IC}(\epsilon) = \int \dif \Omega \, \dif P_t / \dif \epsilon \dif \Omega$ and $n$ is empirically determined. We use $n = 4$ (higher values do not change the power-law scaling of $\epsilon_{\rm c}$ with cooling strength in Fig.~\ref{fig:epscut}). The result of this calculation is displayed in Fig.~\ref{fig:omxplaw} for three reference simulations, and the functional dependence of~$\epsilon_{\rm c}$ on~$\grad$ is displayed, for our entire series of simulations, in Fig.~\ref{fig:epscut}. (Note in that figure, as well as in Figs~\ref{fig:epsiso} and~\ref{fig:epsCutDivEpsIso}, the horizontal axis is~$\sigma / \grad$, which \textit{increases} with stronger cooling.) We omit~$\grad / \sigma = 64, \infty$ from Figs~{\ref{fig:epscut}--\ref{fig:epsCutDivEpsIso}} because those simulations exhibit secular growth in~$\epsilon_{\rm c}$ throughout our analysis interval~$1 \leq ct / L \leq 3$, never reaching a quasi-steady state.

In Fig.~\ref{fig:epscut}, the {large-$\grad$} (weak radiation-reaction) scaling of~$\epsilon_{\rm c}$ with~$\gamma_{\rm rad}$ significantly deviates from~$\epsilon_{\rm c} = 4 \grad^2 \eph$ (recall here that~$\eph$ is the monochromatic energy of IC seed photons). This scaling is a special case of the more general result~$\epsilon_{\rm c} = 4 \gamma_{\rm c}^2 \eph$, which equals the maximum emitted photon energy from a particle at the cut-off Lorentz factor~$\gamma_{\rm c}$ in the particle distribution. At large~$\grad$, we measure~$\epsilon_{\rm c} < 4 \grad^2 \eph$. This means that~$\gamma_{\rm c} < \grad$ and, perhaps, that diminished radiative efficiency allows a slower particle acceleration mechanism to dominate the highest energies.\footnote{We have verified that~$\gamma_{\rm c} < \grad$, but do not present a corresponding plot.}

To see how this might work, let us suppose that such a mechanism operates and that the associated acceleration time~$t_{\rm slow}$ for a particle to double its Lorentz factor~$\gamma \to 2 \gamma$ scales as~$t_{\rm slow} \propto \gamma^\tidx$. Generally, we expect~$\tidx>1$ since the linear case~$\tidx=1$ corresponds to the fast \ts[~$t_{\rm X} \propto \gamma$]associated with direct X-point acceleration [to be exact,~$t_{\rm X}(\gamma) = \gamma m_e c^2 / c e \beta_{\rm rec} B_0 \simeq 10 \gamma / \omega_0$]. Furthermore, for self-consistency, this picture requires that the fast acceleration scaling~$t_{\rm X} \propto \gamma$ does not continue indefinitely to high energies; otherwise there would never be a large-$\gamma$ regime where~$t_{\rm slow} < t_{\rm X}$. We will therefore assume that X-point acceleration can only deliver particles up to a threshold Lorentz factor~$\gamma_{\rm X}$ and beyond that ceases to operate. (Formally,~$t_{\rm X} \propto \gamma$ for~$\gamma < \gamma_{\rm X}$ and~$t_{\rm X} = \infty$ otherwise.)

In this scenario, a good candidate Lorentz factor for~$\gamma_{\rm X}$ is~$\gamma_{\rm X} \sim 4 \sigma$. This has already been suggested by \citet{wuc16} as a natural limit set by the size of elementary current layers in the plasmoid hierarchy. Moreover,~$\grad = 4 \sigma$ is, quite suggestively, where the scaling~$\epsilon_{\rm c} \propto \grad^2$ appears to break down in Fig.~\ref{fig:epscut}. Let us therefore tentatively assign~$\gamma_{\rm X} = 4 \sigma$ (appropriate for our simulations, but, as discussed in Section~\ref{sec:kbsummary}, not necessarily the case in all astrophysical instances of reconnection).

Assuming radiative losses are weak enough that particles reach Lorentz factors exceeding~$\gamma_{\rm X} = 4 \sigma$~(i.e.~$\grad > 4 \sigma$), we are in a regime where~$t_{\rm slow} < t_{\rm X}$ at the highest energies. Equating, therefore,~$t_{\rm slow}$ to the IC cooling time~$t_{\rm IC}(\gamma) = 10 \grad^2 / \gamma \omega_0 \propto \grad^2 / \gamma$ gives an expected cut-off in the particle distribution~$\gamma_{\rm c} \propto \grad^{2/(\tidx+1)}$ decided not by the competition between radiative losses and X-point acceleration, but by that between radiation and the putative slower acceleration mechanism. Importantly, the corresponding photon energy cut-off~$\epsilon_{\rm c} = 4 \gamma_{\rm c}^2 \eph \propto \grad^{4/(\tidx+1)}$ scales more gently than~$\propto \grad^2$, which is expected only if~$\grad \leq \gamma_{\rm X} = 4 \sigma$.

The above considerations are more than just an abstract hypothetical exercise. In fact, a slower energization process with~$\tidx=2$ -- in which the Lorentz factors of high-energy particles generally follow~$\gamma(t) \propto \sqrt{t}$ -- has been identified by \citet{ps18} and recently elaborated by \citet{hps20}. In contrast to direct acceleration at reconnection X-points, this mechanism operates on particles inside plasmoids. There, particles are accelerated gradually due to conservation of their magnetic moments in the presence of a slowly growing magnetic field. The resultant scaling~$\epsilon_{\rm c} \propto \grad^{4/3}$ is not far from the apparent weak radiation-reaction scaling in Fig.~\ref{fig:epscut}.

Let us now use these ideas to construct a theoretical model that explains all of our~$\epsilon_{\rm c}$ measurements. In this effort, we regard the cut-offs~$\gamma_{\rm c}$ and~$\epsilon_{\rm c}$ as dependent functions of the independent variable~$\grad$. The particular value~$\grad = \gamma_{\rm X} = 4 \sigma$ is special, because we assume that, for~$\grad > \gamma_{\rm X}$, X-point acceleration is subdominant. Instead, particles are primarily accelerated by a process similar to that of \citet{ps18} and \citet{hps20}, which operates on a \ts[]~$t_{\rm slow} = C \gamma^2$. To fix the proportionality constant~$C$, we require that the slow and fast acceleration \ts[s,~$t_{\rm slow}$ and~$t_{\rm X}$,]give equal cut-off Lorentz factors~$\gamma_{\rm c}(\grad) = \grad$ at the transition value~$\grad = \gamma_{\rm X}$. This can be expressed as the condition~$C \gamma_{\rm c}^2 = t_{\rm slow} = t_{\rm X} = t_{\rm IC} = 10 \grad^2 / \gamma_{\rm c} \omega_0$, and yields, upon inserting~$\gamma_{\rm c} = \grad = 4 \sigma$, the result~$t_{\rm slow} = 5 \gamma^2 / 2 \sigma \omega_0$~($C = 5/2 \sigma \omega_0$). For~$\grad < \gamma_{\rm X}$, the cut-off~$\gamma_{\rm c}$ is set by the competition between X-point acceleration and radiative cooling and is given by~$t_{\rm X}(\gamma_{\rm c}) = t_{\rm IC}(\gamma_{\rm c})$; for larger~$\grad$, cooling balances the slower acceleration mechanism and~$\gamma_{\rm c}$ can be found from the condition~$t_{\rm slow}(\gamma_{\rm c}) = t_{\rm IC}(\gamma_{\rm c})$. The IC photon cut-off energy~$\epsilon_{\rm c} = 4 \gamma_{\rm c}^2 \eph$ is then a broken power law in~$\grad$:
\begin{align}
    \epsilon_{\rm c} = \begin{cases}
        \epsilon_{\rm c1} = 4 \grad^2 \eph & \grad \leq \gamma_{\rm X} = 4 \sigma \\
        \epsilon_{\rm c2} = 4 \grad^2 \eph (4 \sigma / \grad)^{2/3} & \mathrm{otherwise} \, .
    \end{cases}
    \label{eq:epscpiecewise}
\end{align}
Should one wish to smooth the transition between~$\epsilon_{\rm c} = \epsilon_{\rm c1}$ and~$\epsilon_{\rm c} = \epsilon_{\rm c2}$, we find that the empirical formula
\begin{align}
    \frac{1}{\epsilon_{\rm c}^2} = \frac{1}{\epsilon_{\rm c1}^2} + \frac{1}{\epsilon_{\rm c2}^2}
    \label{eq:epscsmoothed}
\end{align}
describes our~$\epsilon_{\rm c}(\grad)$ data quite well. Both this smoothed form and~$\epsilon_{\rm c1}$ and~$\epsilon_{\rm c2}$ individually are displayed in Fig.~\ref{fig:epscut}. Also shown is a power-law fit~$\epsilon_{\rm c} \propto \grad^{1.6}$ exhibiting a scaling intermediate between~$\epsilon_{\rm c1}$ and~$\epsilon_{\rm c2}$. Formally, equations~(\ref{eq:epscpiecewise}) and~(\ref{eq:epscsmoothed}), and a single power law all acceptably reproduce our~$\epsilon_{\rm c}(\grad)$ data, but, based on our theoretical considerations, we suspect that a broken power law more accurately reflects the underlying physics.

Let us now move on to describe our other important beaming-related energy scale~$\epsilon_{\rm iso}$. As we did with~$\epsilon_{\rm c}$, we will first describe how we measure this quantity, interpreting our measurements thereafter. To calculate~$\epsilon_{\rm iso}$, the first step is to fit a smoothly broken power law of the form
\begin{align}
  \label{eq:sbpfit}
  \Omega_{50}(\epsilon) = A \left( \frac{\epsilon}{\epsilon_{\rm br}} \right)^{-p_1} \left\{ \frac{1}{2} \left[ 1 + \left( \frac{\epsilon}{\epsilon_{\rm br}} \right)^{1 / \Delta} \right] \right\}^{(p_1 - p_2) \Delta}
\end{align}
to the~$\Omega_{50}$ curve \citep[\texttt{astropy.modeling.powerlaws.Smoothly BrokenPowerLaw1D},][see Fig.~\ref{fig:omxplaw}]{astropy18}. The parameters~$A$,~$p_1$,~$p_2$, and~$\epsilon_{\rm br}$ are the scale, power-law indices, and spectral break of the fit. The parameter~$\Delta$ controls the width of the break in the sense that equation~(\ref{eq:sbpfit}) constitutes a pure power law with index~$p_1$~($p_2$) at energies below~$\epsilon_{\rm br} / 10^\Delta$ (above~$\epsilon_{\rm br} 10^\Delta$).

Once the parameters in~(\ref{eq:sbpfit}) are determined, we take~$\epsilon_{\rm iso} = \epsilon_{\rm br} / 10^\Delta$. This definition is empirically motivated from two observations. First, our fits generally produce~$p_1$ nearly flat and~$p_2$ steep~($|p_1| \leq 0.2$ and~$p_2 \geq 1.5$ across all fits). Second, in the (low-energy)~$p_1$ segment of the curve,~$\Omega_{50}$ hovers near isotropy~[$\Omega_{50}(\epsilon) \simeq 0.5 \times 4 \pi$]. Hence, at energies below~$\epsilon_{\rm br} / 10^\Delta$,~$\Omega_{50}$ is both energy-independent and isotropic (i.e. kinetic beaming is absent), but above~$\epsilon_{\rm br} / 10^\Delta$,~$\Omega_{50}$ begins to turn over, eventually declining precipitously with photon energy. Thus, our choice~$\epsilon_{\rm iso} = \epsilon_{\rm br} / 10^\Delta$ provides a good description for when beaming starts becoming kinetic, as intended.\footnote{Because we wish to flag the onset of kinetic beaming, we choose~$\epsilon_{\rm iso} = \epsilon_{\rm br} / 10^\Delta$ rather than~$\epsilon_{\rm iso} = \epsilon_{\rm br}$. This means that, intentionally, the measured~$\epsilon_{\rm iso}$ values indicated in Fig.~\ref{fig:omxplaw} are often just before the~$\Omega_{50}$ curves turn over, rather than in the middle of the spectral break.}

In order to build confidence in our~$\epsilon_{\rm iso}$-extraction method, and to illustrate the utility of having two metrics of beaming~($\Omega_{50}$ and~$bf$), we will discuss one subtlety associated with our procedure. Namely, because~$10^\Delta$ lies between~$2$ and~$8$ across all our fits (except for one broad transition~$10^\Delta = 20$ in our~$\grad / \sigma = 2$ simulation), and because~$\epsilon_{\rm br}$ is rather large to begin with, the energy~$\epsilon_{\rm br} 10^\Delta$ signalling the end of the spectral break often falls near the rightmost edge of the~$\Omega_{50}$ data (or, for~$\grad / \sigma = 2$, well beyond it). This means that~$p_2$ and, to some extent,~$\Delta$ and~$\epsilon_{\rm br}$ are not necessarily well constrained. We deal with this difficulty in two ways. First, we do not rigorously study the~$p_2$ measurements. We only report the lowest value~$p_2 = 1.5$ (see above) to generally indicate the pronounced energy-dependence acquired by~$\Omega_{50}$ beyond~$\epsilon_{\rm iso}$. Secondly, for our fitted~$\epsilon_{\rm br}$ and~$\Delta$ values, which are used directly in our definition of~$\epsilon_{\rm iso}$, we provide the following sanity check using our second metric of beaming: the beamed fraction. In particular, for each simulation, the location~$\epsilon_{\rm iso} = \epsilon_{\rm br} / 10^\Delta$ -- despite being entirely determined from the~$\Omega_{50}$ data -- roughly coincides with photon energies where the beamed fraction slope increases most rapidly [where~$\dif^{\, 2}(bf)/\dif \epsilon^2$ is peaked]. This qualitative agreement between beaming metrics suggests that~$\epsilon_{\rm iso}$ flags a real feature in the~$\Omega_{50}$ curves, and is not merely an artefact of truncated high-energy~$\Omega_{50}$ information. Fig.~\ref{fig:epsiso} displays the dependence of~$\epsilon_{\rm iso}$ on~$\sigma / \grad$.

The kinetic beaming range~$\epsilon_{\rm c} / \epsilon_{\rm iso}$, measured from actual PIC simulations using the above techniques, allows us to quantify the energetic extent of kinetic beaming versus cooling strength. This is done in Fig.~\ref{fig:epsCutDivEpsIso}, where, for each of our radiative simulations, we present the value of~$\epsilon_{\rm c} / \epsilon_{\rm iso}$ as a function of~$\sigma / \grad$. In both Figs~\ref{fig:epsiso} and~\ref{fig:epsCutDivEpsIso}, we supply power-law fits to our data. This is not meant to indicate a robust theoretical description, but merely to characterize how quickly these quantities change from the weakly radiative~($\grad \gg \sigma$) to the strongly radiative~($\grad \lesssim \sigma$) regime.

Fig.~\ref{fig:epsCutDivEpsIso} demonstrates a clear dependence of the kinetic beaming range on IC cooling strength. Moreover, this dependence may be stronger than is suggested by naively fitting a single power law to our entire data set~$1/2 \leq \grad / \sigma \leq 32$. This is because~$\epsilon_{\rm iso}$ can never exceed~$\epsilon_{\rm c}$: particles and photons can be neither isotropic nor beamed at energies where none of them exist. Thus, the ratio~$\epsilon_{\rm c} / \epsilon_{\rm iso}$, although strongly dependent on~$\grad$ when~$\grad$ approaches the strong cooling regime, must ultimately asymptote to unity in the non-radiative limit. According to our data, this occurs closer to~$\grad = 16 \sigma$ than to~$\grad = 32 \sigma$. To demonstrate this, we conduct a series of fits to the data in Fig.~\ref{fig:epsCutDivEpsIso}, of which we display only the first two. In each successive iteration, we remove the most weakly radiative simulation~(that is, we keep all the data for the first fit, omit~$\grad = 32 \sigma$ for the second, omit~$\grad = 32 \sigma$ and~$\grad = 16 \sigma$ for the third, etc.). The power law becomes insensitive to this procedure once we are restricted to~$\grad \leq 16 \sigma$ and at that point exhibits nearly linear scaling.

This suggests that our simulation series captures an important transition in the range of beamed photon energies. For the mildest radiative cooling~($\grad \gtrsim 16 \sigma$), kinetic beaming is unobservable, manifesting itself nowhere in the distribution of emitted photons~($\epsilon_{\rm iso} \sim \epsilon_{\rm c}$). However, once the radiative efficiency is increased, kinetic beaming suddenly appears, and persists at late times across a sizeable range of photon energies. This range increases throughout the entire set of~$\grad$ that is numerically accessible to us, and even surpasses one decade when~$\grad \lesssim 4 \sigma$.

Finally, we note that the simulation with~$\grad = \sigma / 2$, which (as noted in Section~\ref{sec:sims}) is problematic from the standpoint of cooling in the upstream plasma, is not essential to any of the findings in this section. In particular, excluding it leaves all best-fitting power-law scalings essentially unchanged. Because this simulation does not modify any of the overall trends, and indeed appears to fall in line with those trends, we have included it in Figs~{\ref{fig:epscut}--\ref{fig:epsCutDivEpsIso}}.

\subsection{System size dependence}
\label{sec:lscan}
Before placing our numerical results in an astrophysical context, let us briefly explore how those results depend on the size of our computational box~$L_x = L$. To do so, we report on a small series of simulations with fixed radiation-reaction strength~$\grad = 4 \sigma$ and varying~$L / \sigma \rho_0 \in [80, 120, 160, 240, 320]$. All other parameters are the same as described in Section~\ref{sec:sims} and summarized in Table~\ref{table:params}.

We have chosen~$\grad = 4 \sigma$ for these simulations because, on the one hand, our results in the previous section indicate that this radiative efficiency is strong enough to yield a substantial range of beamed photon energies. On the other hand, the radiation is weak enough that the upstream plasma does not cool at all during the simulation~[$t_{\rm cool}/(3L/c) = 22$ in equation~(\ref{eq:upstreamcool})]. In our~$\grad$ scan, we tolerated a larger degree of upstream cooling because the expected error imparted to our main measured quantities --~$\epsilon_{\rm iso}$,~$\epsilon_{\rm c}$, and~$\epsilon_{\rm c} / \epsilon_{\rm iso}$ -- was both within our measurement error and swamped by the pronounced observed~$\grad$-dependence. However, here we need to be more strict. Our goal is to demonstrate system-size insensitivity of the same beaming quantities, and, hence, we need to eliminate any upstream cooling effects that could selectively come into play at larger~$L$.

\begin{figure}
    \centering
    \includegraphics[width=\columnwidth]{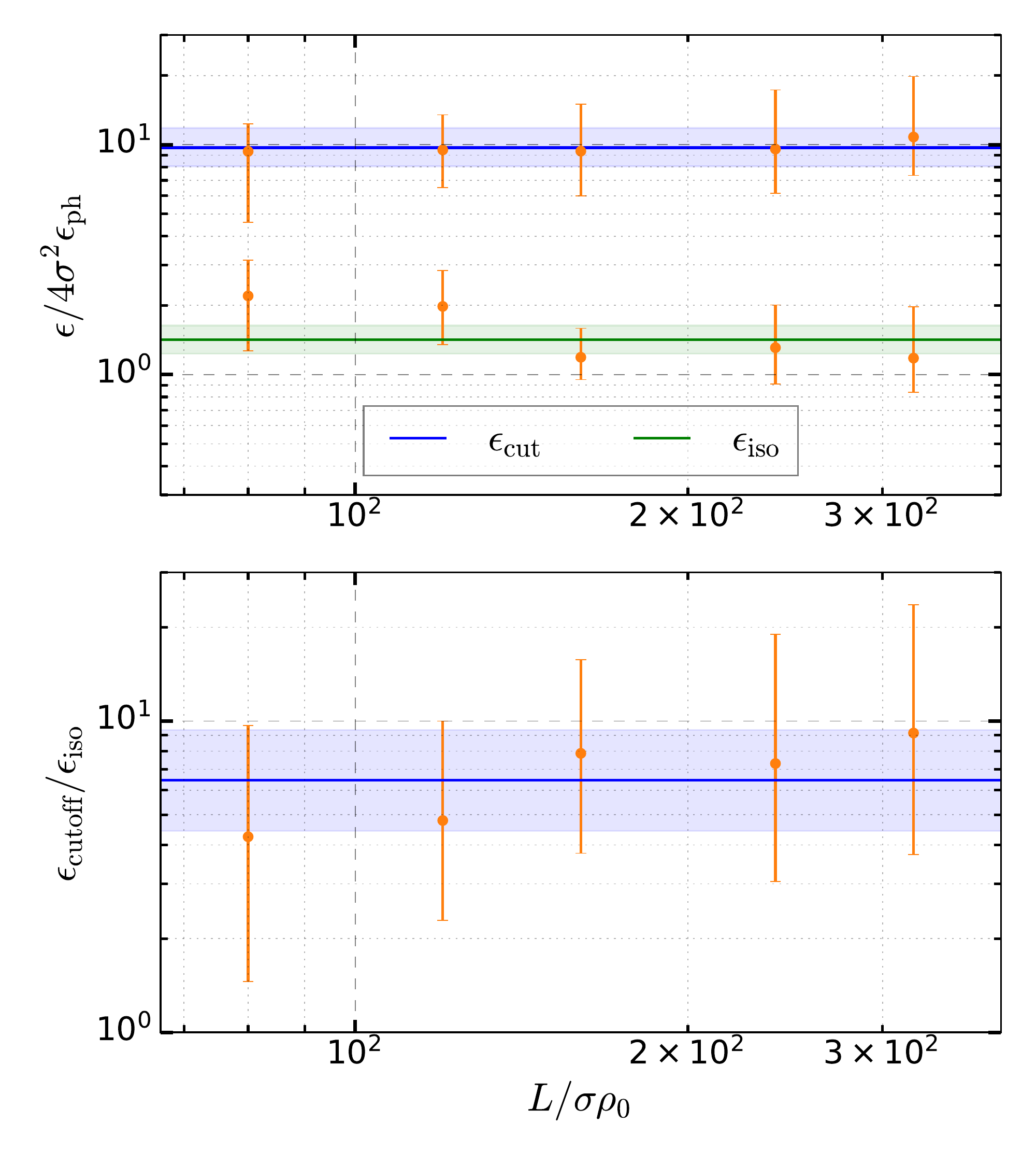}
    \caption{\revtext{(Figure added in revised manuscript.)} Top panel:~$\epsilon_{\rm c}$ and~$\epsilon_{\rm iso}$ measured using the same techniques as for Fig.~\ref{fig:epscut} and Fig.~\ref{fig:epsiso} but for a series of simulations at uniform~$\grad = 4 \sigma$ and varying~$L / \sigma \rho_0 = [80, 120, 160, 240, 320]$. Bottom panel: the kinetic beaming range~$\epsilon_{\rm c} / \epsilon_{\rm iso}$ for the same series of simulations. Solid lines indicate constant fits to each quantity, with shaded regions displaying~$68$ per cent confidence intervals. A~$\chi^2$ goodness-of-fit test does not reject any of these flat lines, showing that, given the error bars, the data are consistent with~$L$-independence.}
    \label{fig:lscan}
\end{figure}

For this series of simulations, Fig.~\ref{fig:lscan} displays the values of~$\epsilon_{\rm c}$,~$\epsilon_{\rm iso}$, and the kinetic beaming range~$\epsilon_{\rm c} / \epsilon_{\rm iso}$ calculated by the methods described in Section~\ref{sec:kinbeamgrad}.
If kinetic beaming were to weaken with system size, presumably because of a diminished importance of kinetic scale phenomena, then one would expect the kinetic beaming range~$\epsilon_{\rm c} / \epsilon_{\rm iso}$ to exhibit a downward trend with~$L$. Instead, our data show that this quantity is consistent with being constant with~$L$. This is in fact true not only of the ratio~$\epsilon_{\rm c} / \epsilon_{\rm iso}$ but also of~$\epsilon_{\rm c}$ and~$\epsilon_{\rm iso}$ individually. If there is any non-constant trend at all, although this is not statistically significant, the kinetic beaming range \textit{increases} with~$L$ (owing primarily to a decrease in~$\epsilon_{\rm iso}$).

In this system-size scan, we have not explored larger~$L$ because~(1) our existing results establish a compelling trend and~(2) the cost of larger boxes is prohibitive: a factor of~2 increase leads to a factor of~$2^3$ additional computational expense. Pending a future study that more thoroughly elucidates system-size effects -- both by going to larger~$L$ and by trialing additional values of~$\grad$ -- we will move forward assuming that the basic picture of kinetic beaming is captured by our main simulation series (with constant~$L = 320 \sigma \rho_0$ and varying~$\grad$). We will, in particular, assume that those results can be extrapolated to astrophysically large systems, as suggested by the $L$-invariance apparent in this system-size investigation.

\subsection{Summary of kinetic beaming}
\label{sec:kbsummary}
In this section, we would like to collect and summarize what we have learned so far, particularly from Section~\ref{sec:kinbeamgrad} (the fundamental features of which, as suggested in Section~\ref{sec:lscan}, may apply even to astrophysically large systems), with an eye towards extrapolating these findings to astrophysical situations. These remarks will be mostly at a general level, independent of any particular class of astrophysical objects, but they will set the stage for our specific application to TeV blazar flares in Section~\ref{sec:fsrqflares}.

Assuming the connection between beaming and rapid light-curve variability \citep[described in the earlier works of][]{cwu12, cwu13, cwu14b, cwu14a}, our main result is that kinetic beaming requires a high degree of radiative efficiency in order to leave an observational signature. From our analysis of the kinetic beaming range~$\epsilon_{\rm c} / \epsilon_{\rm iso}$ (Fig.~\ref{fig:epsCutDivEpsIso}), we have quantitatively discovered that beaming-induced variability should only exist in the highest energy spectral bands, where the emitting particles are near their radiatively imposed cut-off energy. An implication in the context of gamma-ray (e.g. blazar) flares is that, if high-cadence observations could be made in both bands, TeV flares should not have similar \ts[]GeV counterparts (unless multiple radiative processes enable strongly cooled particles to emit in several bands): kinetic beaming does not appear to be sustained over such a broad energy range.

Our findings further suggest that, when radiative cooling is weak, kinetic beaming does not necessarily explain the total duration of rapid flares. While the transient initial phase of beaming seen in our weakly cooled simulations could influence the rising part of a flare, subsequent isotropization would likely prevent, in that case, the fast-rise fast-decay pattern characteristic of a collimated beam crossing the line of sight. On the other hand, kinetic beaming may well shape the entire temporal profile of rapid outbursts when radiative cooling is efficient. Then the energetic particle beams emit corresponding photon beams before diverging. Near the spectral cut-off, beaming is pronounced and kinetic, and the light curves in this band are expected to exhibit increasingly dramatic variability at higher and higher energies.

One point worth emphasizing is that our usage of the terms \quoted{strong}and \quoted{efficient}cooling is not necessarily the same as that in other works. Often in astrophysics, radiative cooling is said to be efficient if particles cool faster than some macroscopic system \ts[.]Here, the definition of efficient radiative cooling is at least as strong, but (and depending on the particular system) often much stronger than these more conventional notions. For us, strong cooling is \textit{microscopically} strong. A particle with Lorentz factor close to its radiatively imposed limit~$\grad$ has a cooling time matching its \textit{acceleration} time through the reconnection layer, or, equivalently, has a cooling length (the distance it travels in one cooling time) of order its Larmor radius \citep[see][]{u16}. Either scale may potentially be much smaller than any macroscopic system scale.

Finally, we would like to state a conjecture that may broaden the scope of our results, enhancing the potential variety of astrophysical sources for which kinetic beaming may explain rapid flares. This conjecture concerns the range of~$\grad / \sigma$ for which kinetic beaming extends across an appreciable span of energies~(for which~$\epsilon_{\rm c} / \epsilon_{\rm iso}$ is sizable). Fig.~\ref{fig:epsCutDivEpsIso} suggests that this range only exceeds about a decade when~$\grad \leq 4 \sigma$. This, however, does not necessarily mean that sources for which~$\grad \gg \sigma$ should not exhibit kinetically beamed emission. In particular, we believe that the figure of merit for a kinetic beaming scenario is not whether~$\grad$ is of order several~$\sigma$ or less, but whether the actual energies achieved by particles are close to~$\grad$. As suggested by our results (Fig.~\ref{fig:epscut}), particles are only expected to reach Lorentz factors~$\sim \grad$ if they are accelerated by the fast X-point mechanism. Slower acceleration channels radiatively saturate at energies less than~$\grad$. Hence, if particles are somehow able to reach~$\grad$ even when~$\grad \gg \sigma$, then they must have been accelerated and, consequently, beamed near an X-point.

Thus, what kinetic beaming really depends on is not whether~$\grad \lesssim (\mathrm{several}) \sigma$, but whether~$\grad$ is comparable to or less than the maximum Lorentz factor~$\gamma_{\rm X}$ achievable due to X-point acceleration. As hypothesized by this and prior studies \citep{wuc16, ps18, hps20},~$\gamma_{\rm X}$ may be of the order of~$4 \sigma$ in reconnection set-ups like the one employed by us in this work, meaning that the requirement~$\grad \leq \gamma_{\rm X}$ in our case simplifies to~$\gamma_{\rm rad} \leq 4 \sigma$. However, in alternative and more astrophysical situations, it may be possible for~$\gamma_{\rm X}$ to circumvent this~$4 \sigma$ limit. For example, many of the sources (e.g. pulsar wind nebulae and active galactic nuclei) for which kinetic beaming nicely explains a number of aspects of observed flares are also highly \nonthermal emitters, even in their quiescent states. This suggests that the upstream plasma is itself \nonthermal[,]possessing a long tail of already high-energy particles -- very different from the thermal upstream conditions in our (and almost all other) simulations. Injected into the reconnection layer near an X-point, these high-energy particles may not be limited to Lorentz factors~$4 \sigma$, and could indeed already exceed those Lorentz factors \textit{before} even experiencing the reconnection electric field.

Such particles could, in principle, reach Lorentz factors all the way up to the Hillas limit~$e E_{\rm rec} l / m_e c^2$ where~$E_{\rm rec} \simeq 0.1 B_0$ is the reconnection electric field, and~$l$ is its potentially macroscopic coherence length. Importantly, the effective~$l$ should be larger for particles that are more energetic upon entering the reconnection region. Indeed, \citet{wuc16} found that the characteristic~$\gamma_{\rm X} \simeq 4 \sigma$ limit arises because cold thermal particles only experience direct linear acceleration in elementary current layers between the smallest-scale plasmoids. Before their energy can grow too large, these particles become magnetized and trapped inside small plasmoids flanking the elementary layer where they were originally accelerated. However, higher energy particles have much larger Larmor radii and therefore sample larger scale fields. They may potentially traverse multiple acceleration regions (spanning several elementary current layers) before finally becoming trapped inside a necessarily large (and therefore rare) plasmoid \citep[cf.][]{cwu13}.

These remarks motivate a future systematic study of the maximum Lorentz factors achievable by X-point acceleration, and of kinetic beaming, in the presence of alternative upstream conditions. However, when it comes to our astrophysical discussion below, we will simply assume that it is possible to achieve Lorentz factors~$\gamma \sim \grad$ (i.e.~$\grad \leq \gamma_{\rm X}$) even if~$\grad \gg \sigma$, and we will not require~$\grad \leq 4 \sigma$ as a necessary condition for kinetic beaming.

At this point, it is clear that the astrophysical relevance of our findings is predicated on whether they survive under a number of non-trivial generalizations (e.g. to larger systems, to more realistic upstream conditions, and even to~3D), most of which are beyond the scope of this work. With that in mind, one should read our specific astrophysical remarks below not as predictions made by a robust and fully-fledged theory, but as provocative inferences that can be made should the fundamental character of our findings be preserved in real astrophysical systems. It is to those inferences that we now turn. Focusing specifically on TeV blazar flares, we examine whether kinetic beaming -- as understood within the simplified framework of this study -- can reasonably explain the extreme variability observed in these events in a manner that is consistent with, and possibly constrains, blazar radiative environments.

\section{Rapid TeV flares in FSRQs}
\label{sec:fsrqflares}
We now shift our discussion towards a concrete astrophysical application of our numerical results: rapid TeV blazar (specifically FSRQ) flares. Before our analysis, we provide some brief background concerning blazars.

Blazars comprise a class of active galactic nuclei (AGNs) with a relativistic jet pointed towards us. Their observed spectra are generally quite broad, extending from the radio band to the gamma-rays, and characterized by two \nonthermal humps. In models where the emission is leptonic, the lower energy (optical/UV/X-ray) hump is thought to stem from synchrotron radiation and the higher energy (gamma-ray) component from IC process, whereby soft ambient photons are upscattered by relativistic particles \citep{bfr08, brs13, ms16}. The photons seeding IC scattering are typically supplied either by synchrotron emission from within the jet itself \citep[synchrotron self-Compton, or SSC, models; e.g.][]{mgc92, bm96} or by various external sources \citep[external IC models; e.g.][]{bs87, mk89, sbr94}. The most common view is that internal synchrotron emission seeds Compton scattering in BL Lacs -- blazars characterized by a lack of strong emission lines -- whereas external photons dominate the ambient radiation field in FSRQs, which are more luminous, showing strong emission lines and thermal radiation attributed to an accretion disc \citep{tbg11, ms16}.

Here, we will follow in the footsteps of many prior studies \citep[e.g.][]{gub09, ngb11, g13, spg15, pgs16, wub18, cps19, cps20, on20}, positing relativistic magnetic reconnection as the driving mechanism behind blazar flares. Because it is difficult for a reconnection layer to sustain internal radiation energy density larger than the upstream (unreconnected) magnetic energy density \citep{b17, cps19}, the most natural emission model for reconnection-powered Compton-dominated flares -- for which the IC spectral component dominates the synchrotron emission -- is external IC \citep[e.g.][]{cps20}. Conveniently, the objects with the most extreme observed Compton dominance, FSRQs, are also those objects that come pre-equipped with rich external radiation environments.\footnote{A notable exception is the BL Lac PKS 2155-304, which produced a Compton-dominated flare in 2006 \citep{hess12}.} Not only is an externally illuminated reconnection region precisely the set-up addressed in our simulations, but in the case of the first sub-hour TeV FSRQ flare ever observed -- that from PKS~1222+21 on~2010 June~17 \citep{magic11, tst11} -- a strong case has already been made by \citet{nbc12} that kinetic beaming was at play. For these reasons, we will devote the main part of our analysis to understanding rapid TeV FSRQ flares, concentrating on the prototypical PKS~1222+21 outburst.

Two prominent sources of external background radiation in FSRQs like PKS~1222+21 are the broad-line region (BLR) and dusty torus (also called the hot dust region; HDR). The BLR contains gas subject to ionizing radiation from the AGN accretion disc, and it reprocesses this light into UV line emission \citep[most prominently~Ly~$\alpha$;][]{tg08}. The HDR is made of dust clouds radiantly heated by the AGN and producing thermal emission predominantly in the IR \citep{nsi08a, nsi08b}.

As discussed by \citet{nbc12} \citep[see also][]{magic11, tbg11}, the very high-energy (VHE;~${\gtrsim 0.1 \, \rm TeV}$) radiation detected from PKS~1222+21 must have been produced beyond the BLR, at least~$\sim 0.5 \, \rm pc$ from the AGN. Otherwise, it would have been absorbed while traversing the intense BLR radiation fields. At that distance, the extremely rapid variability \ts[]requires the VHE flare to be fed by an unrealistically high energy density packed into a small fraction of the jet's cross section. However, via kinetic beaming, magnetic reconnection can achieve the same variability \ts[]in a much larger space \citep{cwu13, cwu14b, cwu14a}. This, combined with the highly collimated emitting particles \citep[this study and, originally,][]{cwu12}, relaxes the necessary energy density, enabling the flare to be fuelled at the parsec-scale on a reasonable energy budget \citep[the full details of this argument are presented by][]{nbc12}.

Our numerical results may be used to constrain further this general picture of kinetic beaming in reconnection-powered VHE FSRQ flares. Namely, one may stipulate that the putative beaming operates in the regime of strong radiative cooling (as defined in this work), and examine what new astrophysical insight may be derived from this requirement. Let us make this idea more quantitive. In our simulations, kinetic beaming was apparent in the late-time distribution of particles only for strong cooling~($\grad \lesssim 4 \sigma$) and for emitting particle Lorentz factors~$\gamma_{\rm emit}$ above the isotropization threshold~$\gamma_{\rm iso}$. In the case of particularly efficient radiation~($\grad = \sigma$),~$\gamma_{\rm iso}$ was nearly an order of magnitude smaller than the radiative cut-off, well-approximated by~$\grad$ (see Fig.~\ref{fig:BeamFracVsT_rad} and Fig.~\ref{fig:BeamFracAvgT_ptcl_1lc} and surrounding discussion). Thus, a necessary condition for efficiently cooled kinetic beaming is that the emitting particles bear energies within a fairly narrow band given by
\begin{align}
  \label{eq:kbresptcl}
  \gamma_{\rm rad} \geq \gamma_{\rm emit} \geq \gamma_{\rm iso} \sim \frac{\gamma_{\rm rad}}{10} \, .
\end{align}
As already discussed (Section~\ref{sec:kbsummary}), despite that it was only in our simulations with~$\grad \lesssim 4 \sigma$ that we measured an appreciable range of beamed particle and photon energies, we refrain from employing~$\grad \lesssim 4 \sigma$ as a requirement for kinetic beaming in addition to~(\ref{eq:kbresptcl}). Rather, we assume that it is possible for X-points to accelerate particles up to the radiative limit~$\grad$ even if~$\grad \gg \sigma$, provided some particles in the reconnection inflow already possess relatively high Lorentz factors. This could be the case, for example, if the upstream plasma is highly \nonthermal[,]as may reasonably be expected from \nonthermal[]quiescent blazar spectra. 

We will now present simple estimates to check whether equation~(\ref{eq:kbresptcl}) is satisfied by an external IC model for the PKS~1222+21 VHE flare. Here, unprimed quantities are evaluated in the observer's frame and primed quantities in the frame of the VHE emitting region, which is the assumed rest frame of the reconnection layer. (However, we leave particle Lorentz factors~$\gamma$ unprimed, though they are always evaluated in the reconnection frame.) These frames are connected by the emitting region bulk Lorentz factor~$\Gamma$. For simplicity, we ignore the source redshift~$z \simeq 0.4$, and assume the angle~$\theta_{\rm obs}$ between the emitting region bulk velocity and the line of sight to be such that the Doppler factor~$\delta = \{ \Gamma [1 - (1 - 1 / \Gamma^2)^{1/2} \cos \theta_{\rm obs} ] \}^{-1}$ is approximately equal to~$\Gamma$. We follow \citet{nbc12}, adopting~$\Gamma = 40$ (sufficient to render external IC radiation more efficient than SSC) and a fiducial comoving (unreconnected) magnetic field strength~$B_0' = 0.1 \, \rm G$ typical at the parsec-scale.

Beyond the BLR, the likely dominant source of external photons illuminating the jet is the dusty torus. In the observer's frame, the torus radiation is approximately uniform and isotropic, with energy density~$U_{\rm HDR} \simeq 9 \times 10^{-5} \, \rm erg \, cm^{-3}$ and typical photon energy~$\epsilon_{\rm HDR} \simeq 0.3 \, \rm eV$ \citep{nsi08a, nsi08b, ssm09, mmj11, tbg11, nbc12}. Particles upscattering these photons to the characteristic observed energy~$\epsilon_{\rm obs} = 100 \, \rm GeV$ \citep{magic11} have approximate Lorentz factors
\begin{align}
  \label{eq:gemithdr}
  \gamma_{\rm emit,HDR} &\sim \sqrt{\frac{\epsilon_{\rm obs}'}{\epsilon_{\rm HDR}'} } \sim \sqrt{\frac{\epsilon_{\rm obs} / \Gamma}{\epsilon_{\rm HDR} \Gamma}} \notag \\
  &\sim 1 \times 10^4 \left( \frac{\Gamma}{40} \right)^{-1} \left( \frac{\epsilon_{\rm obs}}{100 \, \rm GeV} \right)^{1/2} \left( \frac{\epsilon_{\rm HDR}}{0.3 \, \rm eV} \right)^{-1/2} \, .
\end{align}
We note that the Comptonization occurs in the marginal Klein--Nishina regime, since, in the rest frames of the scattering particles, the seed photon energies are close to the electron rest mass [cf. equation (\ref{eq:kndef})]:
\begin{align}
  \label{eq:knemithdr}
  \frac{\gamma_{\rm emit,HDR} \epsilon_{\rm HDR}'}{m_{\rm e} c^2} &\sim \sqrt{\frac{\epsilon_{\rm obs} \epsilon_{\rm HDR}}{(m_{\rm e} c^2)^2}} \notag \\
  &\sim 0.3 \left( \frac{\epsilon_{\rm obs}}{100 \, \rm GeV} \right)^{1/2} \left( \frac{\epsilon_{\rm HDR}}{0.3 \, \rm eV} \right)^{1/2} \, .
\end{align}
Klein--Nishina effects are even more important for hypothetical particles at the much-higher upper-limit Lorentz factor imposed by the HDR, which can be estimated via equation~(\ref{eq:graddef}) as
\begin{align}
  \label{eq:gradhdr}
  \gamma_{\rm rad,HDR} &= \sqrt{\frac{0.3 e B_0'}{4 \sigma_{\rm T} U_{\rm HDR}'} } \sim \sqrt{\frac{0.3 e B_0'}{4 \sigma_{\rm T} U_{\rm HDR} \Gamma^2}} \notag \\
  &\sim 6 \times 10^6 \left( \frac{ \Gamma }{40} \right)^{-1} \left( \frac{B_0'}{0.1 \, \rm G} \right)^{1/2} \left( \frac{ U_{\rm HDR} }{9 \times 10^{-5} \, \rm erg \, cm^{-3}} \right)^{-1/2} \, .
\end{align}
Even though our numerical study was confined to Thomson IC radiation, with our definition of~$\grad$ even relying on that fact, let us momentarily maintain equation~(\ref{eq:graddef}) as a definition and suppose that our result~(\ref{eq:kbresptcl}) also holds in the deep Klein--Nishina regime. Then, because~$\gamma_{\rm rad,HDR}$ and~$\gamma_{\rm emit,HDR}$ are widely separated, the emitting particles are far below our expected isotropization threshold:
\begin{align}
  \label{eq:grathdr}
  \gamma_{\rm emit} \sim \frac{\gamma_{\rm rad,HDR}}{400} \ll \frac{\gamma_{\rm rad,HDR}}{10} \sim \gamma_{\rm iso,HDR} \, .
\end{align}

Equation~(\ref{eq:grathdr}) suggests that the IC(HDR) process does not impose sufficient radiative losses for kinetic beaming to imprint itself upon the emitted photons. Rather, the radiating particles are expected to emit isotropically. Admittedly, the fact that the VHE photons are produced in the marginal Klein--Nishina regime challenges the applicability of our numerical results -- and, indeed, we plan to study kinetic beaming using a fully Klein--Nishina Compton cross section in a future work. However, we do not expect this to extend kinetic beaming to a broader range of particle energies. This is because Klein--Nishina effects suppress radiative cooling, likely lengthening a particle's cooling time relative to its isotropization time.

We therefore see that, on the one hand, radiatively efficient kinetic beaming appears strained to fit into the picture of Compton-dominated flares seeded by dusty torus photons. On the other hand, kinetic beaming solves an important and challenging energy budget problem for parsec-scale FSRQ flares independently of the underlying radiative mechanism \citep{nbc12}. Rather than abandon the kinetic beaming framework, we submit that the new insight gleaned in this work -- that kinetic beaming requires efficient radiative cooling to manifest itself observationally -- hints that a more elaborate emission model may be appropriate.

Let us therefore conduct our analysis in the opposite direction. Rather than model the flare's radiative environment, testing afterward whether it is consistent with strongly cooled kinetic beaming, let us start by assuming that beaming and efficient radiation operate together and see what this implies about the background photon population. In that spirit, we consider the properties of a hypothetical radiation field, characterized by its (assumed narrowly distributed) photon energy~$\epsilon_{\rm ph}$ and energy density~$U_{\rm ph}$, that satisfies our main requirement~$\gamma_{\rm emit,ph} \geq \gamma_{\rm iso,ph} \sim \gamma_{\rm rad,ph} / 10$ in equation~(\ref{eq:kbresptcl}). This requirement can be recast, using equations~(\ref{eq:graddef}) and~(\ref{eq:gemithdr}), as the following inequality involving~$\epsilon_{\rm ph}'$,~$U_{\rm ph}'$,~$B_0'$, and~$\epsilon_{\rm obs}'$: 
\begin{align}
  \label{eq:ephuphineq}
  10 &\geq \frac{\gamma_{\rm rad,ph}}{\gamma_{\rm emit,ph}} \simeq \sqrt{\frac{9}{40} \frac{B_0'}{B_{\rm c}} \frac{U_{\rm c}}{U_{\rm ph}'} \frac{\epsilon_{\rm ph}'}{\epsilon_{\rm obs}'}} \, ,
\end{align}
where~$B_{\rm c} = 8 \pi e / 3 \sigma_T = 6.0 \times 10^{15} \, \rm G$ is the classical critical field and~$U_{\rm c} = B_{\rm c}^2 / 8 \pi$. For a fixed~$B_0'$ and~$\epsilon_{\rm obs}'$, saturation of this inequality defines a~1D space of radiation fields~$U_{\rm ph}'(\epsilon_{\rm ph}') \propto \epsilon_{\rm ph}'$ for which~$\gamma_{\rm emit,HDR}$ is at the expected isotropization threshold. To pinpoint one candidate combination of~$\epsilon_{\rm ph}'$ and~$U_{\rm ph}'$, we require that~$U_{\rm ph}' \geq U_{\rm HDR}'$, necessary for Comptonization of~$U_{\rm ph}$-photons to dominate those from the dusty torus, and implying~$\gamma_{\rm rad,ph} \leq \gamma_{\rm rad,HDR}$. In turn, this yields a smallest permissible emitting particle Lorentz factor [via~(\ref{eq:gradhdr}) and~(\ref{eq:ephuphineq})] of~$\gamma_{\rm emit,min} \sim \gamma_{\rm rad,ph} / 10 \leq \gamma_{\rm rad,HDR} / 10 \sim 6 \times 10^5$. The corresponding Compton seed photons have characteristic energies~$\epsilon_{\rm ph}' \sim \epsilon_{\rm obs}' / \gamma_{\rm emit,min}^2 \geq 7 \times 10^{-3} \, \rm eV$, which are small enough that the IC emission takes place safely in the Thomson regime:~$\gamma_{\rm emit,min} \epsilon_{\rm ph}' / m_{\rm e} c^2 \sim 8 \times 10^{-3}$.

Now that we know what kind of seed photon population (i.e. combination of~$\epsilon_{\rm ph}$ and~$U_{\rm ph}$) is required for efficiently cooled kinetic beaming, we ask whether such a population can be realized in nature. As an affirmative plausibility argument, we briefly consider the possibility of a structured jet. However, since a detailed global flare model is beyond the scope of our present study, we discuss only a subset of the possible parameters.

In particular, we consider a spine-sheath configuration, where the transverse jet structure consists of two regions: a central, fast-moving spine surrounded by a slower-moving sheath \citep{gtc05, tg16, t17, srb16}. Photons produced in the sheath are blueshifted to the frame of the spine where they seed reconnection-powered Compton radiation.\footnote{Strictly speaking, we could equally well consider an emitting blob plowing through an otherwise unstructured jet; the important part is the relative motion.} We suggest that the sheath emission mechanism is synchrotron -- perhaps due to a simultaneous but less luminous reconnection event -- but, to maintain a simple and general discussion, avoid explicitly invoking this fact. We will merely suppose that the spine, which contains the VHE-producing magnetic reconnection site, inherits the (fast) bulk Lorentz factor~$\Gamma_{\rm >} = \Gamma = 40$, while the sheath moves at more typical (slower) speeds:~$\Gamma_{\rm <} = 10$. The relative Lorentz factor between the two regions is~$\Gamma_{\rm r} \simeq \Gamma_{\rm >} / 2 \Gamma_{\rm <} = 2$. To generalize our prior convention, primed quantities continue to refer to the reconnection (i.e. spine) rest frame and unprimed quantities (save particle Lorentz factors~$\gamma$) to the observer's frame; we will not write anything down in the sheath frame itself.

In this set-up, the photon energy~$\epsilon_{\rm ph}' \sim 7 \times 10^{-3} \, \rm eV$ corresponds to an observed seed photon energy~$\epsilon_{\rm ph} \sim \epsilon_{\rm ph}' \Gamma_{\rm <} / \Gamma_{\rm r} \sim 0.03 \, \rm eV$. This lies on the part of the broad-band spectral energy distribution presented by \citet{tbg11} attributed to the dusty torus \citep{mmj11, tbg11}, and hence is consistent with observed spectral features. Let us see whether the inferred energy density~$U_{\rm ph}'$ -- which, again, is expected to be larger than~$U_{\rm HDR}'$ in order for its Comptonization to dominate the flare -- is also consistent with observations. To that end, we suppose the sheath luminosity peaks at~$L_{\rm ph} = 10^{46} \, \rm erg \, s^{-1}$, similar to that observed in the broad-band spectrum near~$\epsilon_{\rm ph}$ (\citealt{tbg11}; however, these data are not simultaneous with the VHE flare). The spine-frame seed photon energy density is then~$U_{\rm ph}' \sim \Gamma_{\rm r}^2 L_{\rm ph} / 4 \pi c \Gamma_{\rm <}^4 R_{\rm sh}^2 \sim 0.2 \, \rm erg \, cm^{-3}$, and, importantly, exceeds~$U_{\rm HDR}' \sim \Gamma_{\rm >}^2 U_{\rm HDR} \sim 0.1 \, \rm erg \, cm^{-3}$, as required. In this estimate, we have assumed that the transverse size~$R_{\rm sh}$ of the sheath photon source is comparable to that of the VHE-emitting region,~$R_{\rm sh} \simeq R' \sim 10 c t_{\rm var} \Gamma_{\rm >} \simeq 2 \times 10^{-3} \, \rm pc$, implied by the TeV variability \ts[~$t_{\rm var} = 10 \, \rm min$]\citep{magic11} and enlarged by a factor of~$10$ due to kinetic beaming \citep[cf.][]{cwu12, nbc12, fermi16a}. We have checked that the sheath thickness~$R_{\rm sh}$ can be relaxed without substantial change to the model (reducing~$U_{\rm ph}'$ much less severely than the naive expectation~$U_{\rm ph}' \propto R_{\rm sh}^{-2}$).

\begin{figure}
  \centering
  \includegraphics[width=\columnwidth]{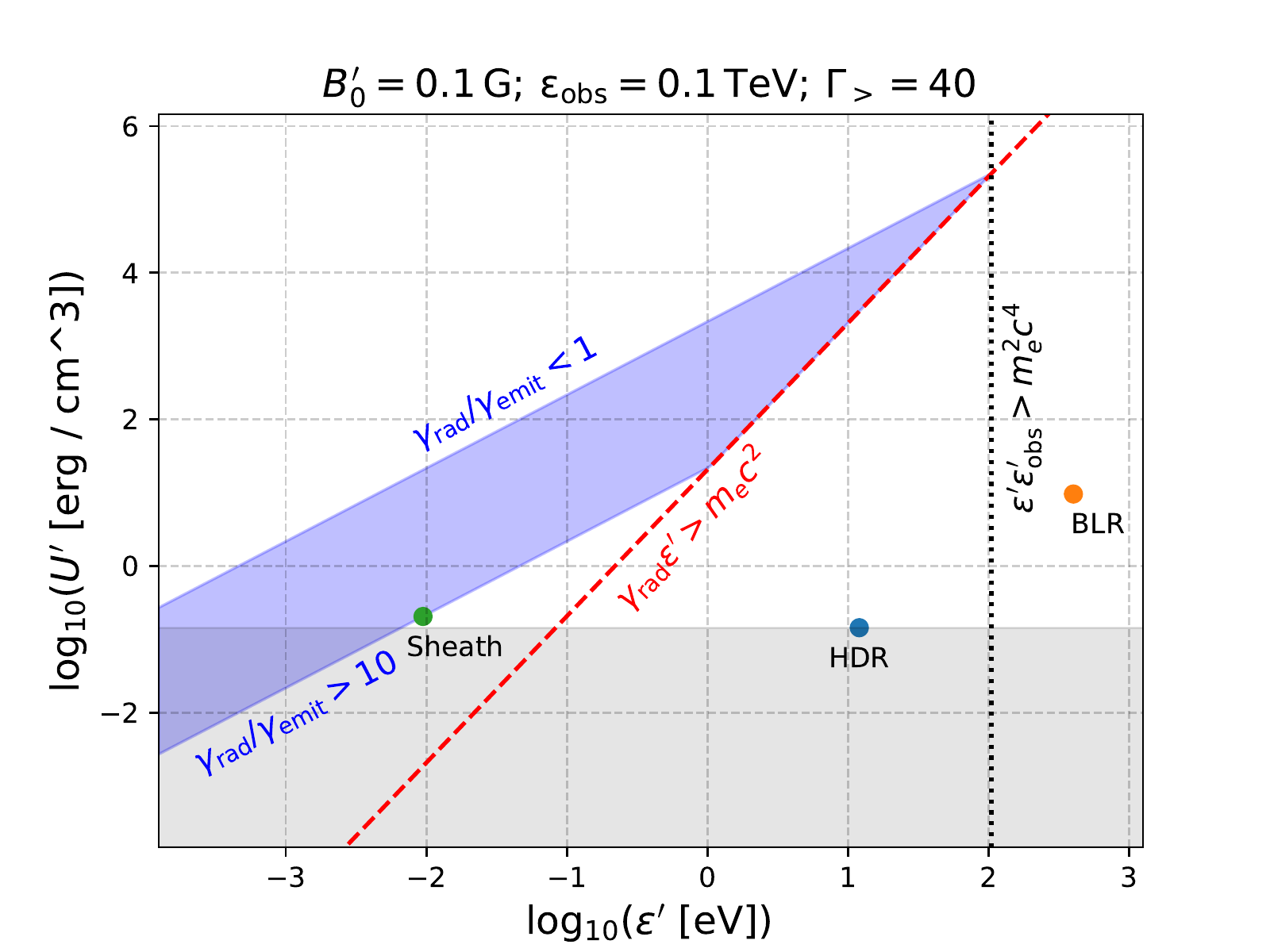}
  \caption{Radiative phase diagram for the PKS~1222+21 VHE flare. We project an otherwise high-dimensional parameter space onto the~$U'-\epsilon'$ plane by fixing the parameters~$\Gamma = \Gamma_{\rm >} = 40$,~$\Gamma_{\rm <} = 10$,~$B_0' = 0.1 \, \rm G$, and~$\epsilon_{\rm obs} = 100 \, \rm GeV$. Individual radiation fields are assumed to be monochromatic, occupying single points. To the right of the dotted vertical line, VHE photons pair-produce with their seed population. This excludes broad-line region illumination (see text), which we illustrate by adopting a characteristic BLR radiation field:~$U_{\rm BLR} = 6 \times 10^{-3} \, \rm erg \, cm^{-3}$ and~$\epsilon_{\rm BLR} = 10 \, \rm eV$ \citep{ssm09, tbg11, nbc12}. Right of the dashed red line, particles scatter photons in the Klein--Nishina~($\epsilon' \grad \geq m_{\rm e} c^2$) regime with suppressed efficiency. In contrast, the blue band indicates efficient but not unphysical~($\grad \geq \gamma_{\rm emit} \geq \gamma_{\rm iso} \sim \grad / 10$) Thomson~($\epsilon' \grad < m_{\rm e} c^2$) IC cooling, and its lower border is given by saturating inequality~(\ref{eq:ephuphineq}). Lying outside this band, HDR illumination is probably unable to mediate observable kinetic beaming. A spine-sheath radiation field is more viable, and is above the shaded grey zone, where fields more tenuous than~$U_{\rm HDR}'$ yield potentially unobservable IC output.}
  \label{fig:ercphasespace}
\end{figure}

Kinetic beaming is more viable in this spine-sheath model not so much because~$U_{\rm ph}' > U_{\rm HDR}'$, but because~$\epsilon_{\rm ph}' \ll \epsilon_{\rm HDR}'$. That is, the sheath photons appear much softer than the HDR photons in the spine frame, requiring higher energy particles for Comptonization to the VHE band -- particles that are then quite strongly cooled. This is illustrated in Fig.~\ref{fig:ercphasespace}, which presents the main results of this section through a radiative \quoted[.]{phase diagram}The blue band in the figure, with lower border given by saturating inequality~(\ref{eq:ephuphineq}), designates seed photon populations conducive to efficiently radiative kinetic beaming. The sheath photons occupy this band, and the HDR photons do not, mostly because of the large energetic disparity between the two populations. This is accentuated by the sheath motion, which enlarges the energy gap to~$\epsilon_{\rm HDR}' / \epsilon_{\rm ph}' = (\Gamma_{\rm >} \Gamma_{\rm <} / \Gamma_{\rm r}) \epsilon_{\rm HDR} / \epsilon_{\rm ph} = 200 \epsilon_{\rm HDR} / \epsilon_{\rm ph}$ in the spine frame.

Hence, purely by invoking relative motion between a VHE-emitting spine and a seed-photon-emitting sheath, one may reconcile the seed photon population required by radiatively efficient kinetic beaming with one that may plausibly be realized during an actual flare. Although we leave a detailed model to future work, we view the above remarks as illustrating the potential utility of a kinetic beaming framework that includes the new ingredient of strong radiative cooling. Whereas kinetic beaming on its own has previously been used to balance the energy budget in VHE FSRQ flares \citep{nbc12}, we now see that the added radiative requirement may constrain possible emission mechanisms. Surprisingly, the most appropriate radiative model for rapid FSRQ flares may be one that does not rely on external structures at all, and, therefore, presents a potentially universal mechanism for the most rapid TeV flares in all blazars, even BL Lacs.

\section{Conclusions}
\label{sec:conclusions}
In this paper, we present the first systematic investigation of the role radiative cooling plays in the kinetic (energy-dependent) beaming of particles and their emission in collisionless relativistic magnetic reconnection. In agreement with prior studies \citep{cwu12, knp16}, we measure definite and pronounced kinetic beaming during the early stages of all our simulations, independent of cooling strength (see Fig.~\ref{fig:BeamFracVsT_norad} and Fig.~\ref{fig:BeamFracVsT_rad}). When radiation is inefficient, kinetic beaming fades at later times to a nearly isotropic distribution of particles and emission, as anticipated \citep{knp16, sgp16, ynz16}. However, as we demonstrate explicitly, kinetic beaming remains persistently observable when radiative cooling is strong (see again Fig.~\ref{fig:BeamFracVsT_norad} and Fig.~\ref{fig:BeamFracVsT_rad}), and may then extend across more than an order of magnitude in photon energies. Moreover, enhanced radiative efficiency increases the beamed range of photon energies (Fig.~\ref{fig:omxplaw} and Fig.~\ref{fig:epsCutDivEpsIso}). In every case, late-time kinetic beaming is apparent only when the emitting particles have energies that are moderately close to (within an order of magnitude of) the radiatively imposed cut-off~$\grad$.

The underlying picture is a competition of \ts[s:]that over which the radiating particles cool and that over which they isotropize. As a generic side effect of impulsive X-point acceleration, particles are always initially beamed \citep{ucb11, cub12, cwu12}. However, in the limit of weak radiative losses, they produce most of their radiation after they have isotropized, and their initial collimation leaves no observable remnant. When cooling is strong, the opposite situation occurs: high-energy radiation comes only from beamed, recently accelerated particles. Reconnection focuses particles in both circumstances, but only in one is this focusing imprinted on the high-energy emission, manifesting itself as rapid variability along a particular observer's line of sight.

Our findings have important consequences for rapid high-energy (HE;~$\sim \rm GeV$) and very high-energy~(VHE;~$\sim \rm TeV$) astrophysical flares. At a very general level, we predict that rapid flares observed in a given spectral band (e.g. VHE) should not exhibit similar variability at energies that are lower by more than a couple orders of magnitude (e.g. HE), barring counterparts produced by the same particles simultaneously shining via multiple radiative processes. At a more detailed level, a kinetic beaming framework may constrain emission models in specific flaring systems, and we examine TeV FSRQ flares as an example. Analysing the~2010 June~17 flare of PKS~1222+21, we find that a kinetic beaming origin of the rapid variability seems at odds with the picture of IC-scattered dusty torus photons dominating the TeV outburst \citep[we do not consider models invoking Comptonization from inside the broad-line region, which are precluded by pair-production considerations; e.g.][]{magic11}. We postulate that an alternative spine-sheath model, wherein reconnection-energized particles upscatter photons originating in the outer fringes of the jet, may be viable. Importantly, this could potentially operate in both FSRQs and BL Lacs. Thus, our results hint that the same physical mechanism may underlie rapid TeV flares from all blazars, regardless of their class.

This study opens the door to a wide scope of future work. It remains to be seen, for example, how our numerical results regarding the effects of radiative cooling on kinetic beaming extend to: different magnetizations~$\sigma$,~3D, larger systems, different guide field strengths, a \nonthermal upstream plasma, and electron--ion reconnection. Indeed, our main astrophysical results depend on whether the fundamental picture of kinetic beaming described here survives in the presence of a number of more realistic physical set-ups, which future studies may test. Additionally, our astrophysical analysis motivates the incorporation of more exotic physics into future simulations. For example, as we have seen in our study of the~2010 June~17 flare of PKS~1222+21, if Comptonization of dusty torus photons were primarily responsible for the outburst, then the IC emission would have taken place in the marginal Klein--Nishina limit. Although we do not expect radiative cooling to be strong enough to mediate kinetic beaming in this case, a dedicated study of kinetic beaming with fully Klein--Nishina Compton cooling may yield surprising results. We intend to carry out such a study in the future.

Finally, we stress that, although we have specialized to rapid FSRQ flares, the general framework presented here of kinetic beaming mediated by radiative cooling and facilitating short \ts[]variability is quite general; it requires only that relativistic magnetic reconnection take place in its radiative regime.
As illustrated by our particular application to FSRQs, a kinetic beaming hypothesis places powerful constraints on a flaring system without introducing many free parameters, and, if the case of blazars is any indication, may help to refine our understanding of reconnection-powered outbursts from other types of astrophysical systems.
There appears, then, great potential for the kinetic beaming paradigm -- as has already been applied to pulsar wind nebulae \citep{cwu12, cwu13, cwu14b, cwu14a} and now to blazars \citep[see also][]{nbc12, zuw20} -- to find fruitful application in attempts to explain many of the most extreme and diversely sourced flares in astrophysics.

\section*{Acknowledgements}
This work is supported by NASA, NSF, and the DOE, grant numbers NASA ATP NNX16AB28G, NASA ATP NNX17AK57G, NSF AST 1411879, NSF AST 1903335, and DE-SC0008409. This work used the Extreme Science and Engineering Discovery Environment (XSEDE) Stampede2 at the Texas Advanced Computing Center through allocation TG-PHY140041.


\section*{Data Availability}

The simulation data underlying this article were generated at the XSEDE/TACC Stampede2 supercomputer and are archived at the TACC/Ranch storage facility. As long as the data remain in the archive, they will be shared on reasonable request to the corresponding author.



\bibliographystyle{mnras}
\bibliography{ref}







\bsp	
\label{lastpage}
\end{document}